\pdfoutput=1
\newcommand*{\ATLASLATEXPATH}{latex/}
\documentclass[cernpreprint,texlive=2016,txfonts,UKenglish]{\ATLASLATEXPATH atlasdoc}
\usepackage[backend=bibtex]{\ATLASLATEXPATH atlaspackage}
\usepackage{\ATLASLATEXPATH atlasbiblatex}
\usepackage{\ATLASLATEXPATH atlascontribute}
\usepackage{\ATLASLATEXPATH atlasphysics}
\usepackage{\ATLASLATEXPATH atlasprocess}

\graphicspath{{logos/}{figures/}}

\usepackage{paper-defs}
\usepackage{lscape}
\usepackage{multirow}
\usepackage{rotating}
\usepackage{ amssymb }

\AtlasTitle{Evidence for the $H \to b\bar{b}$ decay with the ATLAS detector}
\author{The ATLAS Collaboration}
\AtlasRefCode{HIGG-2016-29}
\PreprintIdNumber{CERN-EP-2017-175}
\AtlasJournal{JHEP}
\AtlasAbstract{
A search for the decay of the Standard Model Higgs boson into a $b\bar{b}$ pair when produced in association with a $W$ or $Z$ boson is performed with the ATLAS detector. The analysed data, corresponding to an integrated luminosity of 36.1~\fb, were collected in proton--proton collisions in Run 2 of the Large Hadron Collider at a centre-of-mass energy of 13 \TeV. Final states containing zero, one and two charged leptons (electrons or muons) are considered, targeting the decays \znn, \wln\ and \zll.  For a Higgs boson mass of 125~\GeV, an excess of events over the expected background from other Standard Model processes is found with an observed significance of 3.5 standard deviations, compared to an expectation of 3.0 standard deviations. This excess provides evidence for the Higgs boson decay into $b$-quarks and for its production in association with a vector boson.
The combination of this result with that of the Run 1 analysis yields a ratio of the measured signal events to the Standard Model expectation equal to $0.90 \pm 0.18 \mathrm{(stat.)} ^{+0.21}_{-0.19} \mathrm{(syst.)}$. 
Assuming the Standard Model production cross-section, the results are consistent with the value of the Yukawa coupling to $b$-quarks in the Standard Model.
}

\hypersetup{pdftitle={ATLAS document},pdfauthor={The ATLAS Collaboration}}

\begin{document}

\maketitle

\tableofcontents

\clearpage

\section{Introduction}
\label{sec:intro}
The Higgs boson, predicted more than 50 years ago~\autocite{Englert:1964et,Higgs:1964ia,Higgs:1964pj,Guralnik:1964eu}, 
was discovered in 2012 by the ATLAS and CMS 
Collaborations~\autocite{HIGG-2012-27,CMS-HIG-12-028}, analysing the results of 
proton--proton ($pp$) collisions produced by the Large Hadron Collider (LHC)~\autocite{Evans:2008zzb}. The 
properties of the discovered particle have been measured using the Run~1 dataset, 
collected at centre-of-mass energies of 7~\TeV\ and 8~\TeV, and were found to be compatible with those
predicted by the Standard Model (SM) within uncertainties, typically of the order of $\pm$20\%~\autocite{HIGG-2013-02,HIGG-2013-01,CMS-HIG-12-036,HIGG-2015-07}. 
The Run~2 dataset at an energy of 13 \TeV\ provides an opportunity to increase the 
precision of such measurements, and to challenge theory predictions further.
While analyses of Higgs bosons decaying into vector bosons are entering an era of detailed differential measurements, direct evidence for the
coupling of the Higgs boson to fermions was established only via the observation of the decay into $\tau$-leptons through the combination of 
ATLAS and CMS Run~1 results~\autocite{HIGG-2015-07}, and, more recently, through the combination of CMS Run~1 and Run~2 results~\autocite{Sirunyan:2017khh}. Although the gluon--gluon fusion production mode provides indirect evidence for the coupling of the Higgs boson to top quarks, there is currently no direct observation of the coupling of the Higgs boson to quarks.

The decay of the SM Higgs boson into pairs of $b$-quarks is expected to have a 
branching ratio of 58\% for $\mh=125$~\GeV~\autocite{Djouadi:1997yw}, 
the largest among all decay modes. Accessing \Hbb\ decays is therefore 
crucial for constraining, under fairly general assumptions~\autocite{Lafaye:2009vr,Heinemeyer:2013tqa}, 
the overall Higgs boson decay width. 
At the LHC, the very large backgrounds arising from multi-jet production make an 
inclusive search extremely challenging. The most sensitive production 
modes for probing  \Hbb\ decays are those where the Higgs boson 
is produced in association with a $W$ or $Z$ boson~\cite{Glashow:1978ab};
their leptonic decay modes lead to clean signatures that can be efficiently triggered on, while rejecting most of the multi-jet backgrounds.

Searches for a Higgs boson in the $b\bar{b}$ decay mode were 
conducted at the Tevatron by the CDF and D0 Collaborations. 
They reported an excess of events in $VH$ associated production 
(where $V$ is used to denote $W$ or $Z$) in the mass range of 120~\GeV\ to 135~\GeV, with a global significance of 3.1 standard deviations, and a local significance of 2.8~standard deviations at a mass of 125~\GeV~\cite{TEV-HIGGS-COMB-12}.
ATLAS and CMS reported results from Run~1 each using 
approximately 25~fb$^{-1}$ of integrated luminosity
~\cite{HIGG-2013-23,CMS-HIG-13-012}.
Excesses of events consistent with a Higgs boson with a mass of 
125~\GeV\ were observed in $VH$ associated production 
with significances of 1.4 and 2.1 standard deviations by ATLAS and CMS, respectively. 
Searches for the Higgs boson decay into $b\bar{b}$ have been also performed for
 the vector-boson fusion (VBF)~\autocite{HIGG-2014-12,CMS-HIG-14-004} and 
$t\bar{t}H$~\autocite{HIGG-2013-27,CMS-HIG-13-029} production 
modes, but with sensitivities smaller than 
for $VH$ production. The combination of the Run~1 ATLAS and CMS 
analyses resulted in observed and expected significances
of 2.6 and 3.7 standard deviations for the \Hbb\ decay mode, respectively~\cite{HIGG-2015-07}.

This article reports on the search for the SM Higgs boson in the 
$VH$ production mode and decaying into a $b\bar{b}$ pair
with the ATLAS detector in Run~2 of the LHC, using 
an integrated luminosity of 36.1~fb$^{-1}$. Three main signatures are explored, $ZH \to \nu\nu b\bar{b}$, 
$WH \to \ell \nu b \bar{b}$ and $ZH \to \ell\ell b\bar{b}$. The 
respective analysis categories that target these decay modes are referred to 
as the 0-, 1- and 2-lepton channels, based on the number of selected charged leptons.
In this article, the term "lepton", unless modified by a qualifier, refers to electron and muon.
A $b$-tagging algorithm is used to identify the jets consistent with originating from a $H \to b\bar{b}$ decay. 
In order to maximise the sensitivity to the Higgs boson signal, a set of observables encoding 
information about event kinematics and topology is combined into a multivariate discriminant. A binned maximum-likelihood fit, referred to as the global likelihood fit, is applied to data simultaneously across the three channels in multiple analysis regions. The likelihood fit uses the multivariate discriminant as the main fit observable, in order to extract the signal yield and normalisations of the main backgrounds. The signal extraction method is validated with two other analyses: the \textit{dijet-mass analysis}, where the signal yield is extracted using the mass of the dijet system of $b$-tagged jets as the main fit observable, and the \textit{diboson analysis}, where the nominal multivariate analysis is modified to extract the $(W/Z)Z$ diboson process, with the $Z$ boson decaying into $b\bar{b}$.
The combination of the results of the Higgs boson search with those of the previously published analysis of the Run~1 
dataset~\autocite{HIGG-2013-23} is also presented. 
 

\section{ATLAS detector}
\label{sec:detector}
ATLAS~\cite{PERF-2007-01} is a general-purpose particle detector covering nearly the entire solid angle\footnote{ATLAS uses a right-handed coordinate system with its origin at the nominal interaction point (IP) in the centre of the detector and the $z$-axis coinciding with the axis of the beam pipe.  The $x$-axis points from the IP towards the centre of the LHC ring, and the $y$-axis points upward. Cylindrical coordinates ($r$,$\phi$) are used in the transverse plane, $\phi$ being the azimuthal angle around the $z$-axis. The pseudorapidity is defined in terms of the polar angle $\theta$ as $\eta = - \ln \tan(\theta/2)$. The distance in ($\eta$,$\phi$) coordinates, $\Delta R = \sqrt{(\Delta\phi)^2+(\Delta\eta)^2}$, is also used to define cone sizes. Transverse momentum and energy are defined as $\pt=p\sin\theta$ and $\et=E\sin\theta$, respectively.}
 around the collision point. It consists of an inner tracking detector surrounded by a thin superconducting solenoid, electromagnetic and hadronic calorimeters,
and a muon spectrometer incorporating three large superconducting toroidal magnets.

The inner tracking detector (ID or inner detector in the rest of the article), located within a 2~T axial magnetic field generated by the superconducting solenoid, is used to measure the trajectories and momenta of charged particles. The inner layers, consisting of high-granularity silicon pixel detectors, instrument a pseudorapidity
range $|\eta| < 2.5$. 
A new innermost silicon pixel layer, the insertable B-layer~\cite{Capeans:1291633} (IBL), was added to the detector between Run~1 and Run~2. The IBL improves the ability to identify displaced vertices and thereby significantly improves the $b$-tagging performance~\cite{ATL-PHYS-PUB-2015-022}. 
Silicon strip detectors covering $|\eta| < 2.5$ are located beyond the pixel detectors. 
Outside the strip detectors and covering $|\eta | < 2.0$, there are straw-tube tracking detectors, which also provide measurements of transition radiation that are used in electron identification. 

The calorimeter system covers the pseudorapidity range $|\eta| < 4.9$.
Within the region $|\eta|< 3.2$, electromagnetic calorimetry is provided by barrel ($|\eta| < 1.475$) and 
endcap ($1.375 < |\eta| < 3.2$) high-granularity lead/liquid-argon (LAr) electromagnetic calorimeters,
with an additional thin LAr presampler covering $|\eta| < 1.8$ to correct for energy loss in material upstream of the calorimeters.
Hadronic calorimetry is provided by a steel/scintillator-tile calorimeter, segmented into three barrel structures within $|\eta| < 1.7$, and two copper/LAr hadronic endcap calorimeters extend the coverage to $|\eta|=3.2$.
The solid angle coverage for $|\eta|$ between 3.2 and 4.9 is completed with copper/LAr and tungsten/LAr calorimeter modules optimised for electromagnetic and hadronic measurements, respectively.

The outermost part of the detector is the muon spectrometer, which measures the curved trajectories of muons in the field of three large air-core toroidal magnets. High-precision tracking is performed within the range $|\eta| < 2.7$ and there are chambers for fast triggering within the range $|\eta| < 2.4$. 

A two-level trigger system~\cite{Aaboud:2016leb} is used to reduce the recorded data rate. 
The first level is a hardware implementation that makes use of only a subset of the total available information to make fast decisions to accept or reject an event, aiming to reduce the rate to approximately 100~kHz, and the second level is the software-based high-level trigger that provides the remaining rate reduction to approximately 1~kHz.

\newpage

\section{Dataset and simulated event samples}
\label{sec:samples}
The data used in this analysis were collected at a centre-of-mass energy of 13~\TeV\ during the 2015 and 2016 running periods, and correspond to integrated luminosities of $3.2\pm 0.1$~\fb\ and $32.9 \pm 1.1$~\fb, respectively~\cite{Aaboud:2016hhf}. They were collected using missing transverse momentum (\met) triggers for the 0- and 1-lepton channels and single-lepton triggers for the 1- and 2-lepton channels. 
Events are selected for analysis only if they are of good quality and if all the relevant detector components are known to be in good operating condition. In the combined dataset, the recorded events have an average of 25 inelastic $pp$ collisions (the collisions other than the hard scatter are referred to as pile-up).

Monte Carlo (MC) simulated events are used to model the SM background and $VH$, $H \rightarrow b\bar{b}$ signal processes. All simulated processes are normalised using the most accurate theoretical predictions  currently available for their cross-sections. Data-driven methods 
are used to estimate the multi-jet background from strong interactions (QCD) for the 1-lepton channel, as discussed in Section \ref{sec:multijet}.
This background is negligible in the other channels, as a result either of the high \met\ requirement and dedicated selection criteria (0-lepton channel) or of the two lepton selection (2-lepton channel).
   
   All samples of simulated events were passed through the ATLAS detector simulation~\cite{SOFT-2010-01} based on \textsc{GEANT~4}~\cite{geant}
   and are reconstructed with the standard ATLAS reconstruction software. The effects of pile-up from multiple interactions in the same and nearby bunch crossings were modelled by overlaying minimum-bias events, simulated using the soft QCD processes of \textsc{Pythia 8.186}~\cite{Pythia8} with the A2~\cite{ATLAS:2012uec} set of tuned parameters (tune) and \textsc{MSTW2008LO}~\cite{Martin:2009iq} parton distribution functions (PDF). For all samples of simulated events, except for those generated using \textsc{Sherpa}~\cite{sherpa}, the \textsc{EvtGen v1.2.0} program~\cite{Lange:2001uf} was used to describe the decays of bottom and charm hadrons.
A summary of all the generators used for the simulation of the signal and background
processes is shown in Table \ref{tab:samples}.
\newcommand{\hsp}{\hspace*{0.3cm}}
\begin{table}[tb!]
\begin{center}{\fontsize{7}{7.9}\selectfont
\caption
{The generators used for the simulation of the signal and background
processes. If not specified, the order of the cross-section calculation refers to the expansion in the strong coupling constant ($\alphas$).
The acronyms ME, PS and UE stand for matrix element, parton shower and underlying event,  respectively.
$(\star)$ The events were generated using the first PDF in the NNPDF3.0NLO set and subsequentially reweighted to PDF4LHC15NLO set~\cite{Butterworth:2015oua} using the internal algorithm in \textsc{Powheg-Box v2}.
$(\dagger)$ The NNLO(QCD)+NLO(EW) cross-section calculation for the $pp \to ZH$ process already includes the $gg\to ZH$ contribution.
The $qq\to ZH$  process is normalised using the NNLO(QCD)+NLO(EW) cross-section for the $pp \to ZH$ process, after subtracting the $gg\to ZH$ contribution.
\protect\label{tab:samples}}
\begin{tabular}{lllllll} \hline\hline
\hsp Process & ME generator & ME PDF &PS and & UE model & Cross-section \hspace{2.5cm}\\
& & & Hadronisation & tune & order\\ \hline
\hline
\multicolumn{6}{l}{Signal} \\
\hline
\hsp $qq\to WH$  &\textsc{Powheg-Box v2}~\cite{Alioli:2010xd} +& NNPDF3.0NLO$^{(\star)}$~\cite{Ball:2014uwa} &\textsc{Pythia8.212}~\cite{Pythia8} & AZNLO~\cite{AZNLO:2014} & NNLO(QCD)+  \\
\hspace*{0.5cm} $\to\ell\nu b\bar{b}$&   \textsc{GoSam}~\cite{arXiv:1111.2034} + \textsc{MiNLO}~\cite{arXiv:1206.3572,Luisoni:2013kna}  & & & &NLO(EW)~\cite{Ciccolini:2003jy,Brein:2003wg,Ferrera:2011bk,Brein:2011vx,Ferrera:2013yga,Ferrera:2014lca,Campbell:2016jau} \\
\hsp $qq\to ZH$ &\textsc{Powheg-Box v2} + & NNPDF3.0NLO$^{(\star)}$ &\textsc{Pythia8.212} & AZNLO & NNLO(QCD)$^{(\dagger)}$+  \\
\hspace*{0.5cm}$\to\nu\nu b\bar{b}/\ell\ell b\bar{b}$&   \textsc{GoSam} + \textsc{MiNLO}  & & & &NLO(EW) \\
\hsp $gg  \to ZH$ &\textsc{Powheg-Box v2} & NNPDF3.0NLO$^{(\star)}$ &\textsc{Pythia8.212} & AZNLO & NLO+  \\
\hspace*{0.5cm}$\to\nu\nu b\bar{b}/\ell\ell b\bar{b}$&  & & & &NLL~\cite{Altenkamp:2012sx,Hespel:2015zea,ggzhnll,Harlander:2013mla,Brein:2012ne}\\
\hline\hline
\multicolumn{6}{l}{Top quark}  \\
\hline
\hsp $t\bar{t}$ &\textsc{Powheg-Box v2}~\cite{Frixione:2007nw} &  NNPDF3.0NLO &\textsc{Pythia8.212} &A14~\cite{ATL-PHYS-PUB-2014-021}& NNLO+NNLL~\cite{Czakon:2011xx} \\
\hsp $s$-channel &\textsc{Powheg-Box v1}~\cite{Alioli:2009je} & CT10~\cite{Lai:2010vv}&\textsc{Pythia6.428}~\cite{Pythia6} &P2012~\cite{Skands:2010ak} & NLO~\cite{PhysRevD.81.054028} \\
\hsp $t$-channel &\textsc{Powheg-Box v1}~\cite{Alioli:2009je} & CT10f4&\textsc{Pythia6.428} &P2012& NLO~\cite{Kidonakis:2011wy} \\
\hsp $Wt$ &\textsc{Powheg-Box v1}~\cite{Re:2010bp} & CT10&\textsc{Pythia6.428} &P2012& NLO~\cite{PhysRevD.82.054018} \\
\hline\hline
\multicolumn{6}{l}{Vector boson + jets} \\
\hline
\hsp $W\to\ell\nu$ & \textsc{Sherpa 2.2.1}~\cite{sherpa,Cascioli:2011va,Gleisberg:2008fv} & NNPDF3.0NNLO &  \textsc{Sherpa 2.2.1}~\cite{Schumann:2007mg,Hoeche:2012yf} & Default & NNLO~\cite{Catani:2009sm}  \\
\hsp $Z/\gamma^{*}\to\ell\ell$ & \textsc{Sherpa 2.2.1} & NNPDF3.0NNLO &  \textsc{Sherpa 2.2.1} & Default & NNLO    \\
\hsp $Z\to\nu\nu$ & \textsc{Sherpa 2.2.1} & NNPDF3.0NNLO &  \textsc{Sherpa 2.2.1} & Default & NNLO    \\
\hline\hline
\multicolumn{6}{l}{Diboson} \\
\hline
\hsp $WW$ & \textsc{Sherpa 2.1.1} & CT10 &  \textsc{Sherpa 2.1.1} & Default & NLO  \\
\hsp $WZ$ & \textsc{Sherpa 2.2.1} & NNPDF3.0NNLO &  \textsc{Sherpa 2.2.1} & Default & NLO  \\
\hsp $ZZ$ & \textsc{Sherpa 2.2.1} & NNPDF3.0NNLO &  \textsc{Sherpa 2.2.1} & Default & NLO  \\
\hline\hline
\end{tabular}}
\end{center}
\end{table}

Simulated events for $qq \rightarrow VH$ plus zero or one jet production at next-to-leading order (NLO) were generated with the \textsc{Powheg-Box v2 + GoSam + MiNLO} generator~\cite{Alioli:2010xd,arXiv:1111.2034,arXiv:1206.3572,Luisoni:2013kna} (named \textsc{Powheg+MiNLO} in the rest of the article).
The contribution from $gg \rightarrow ZH$ (gluon-induced) production was simulated using the leading-order (LO) \textsc{Powheg-Box v2} matrix-element generator. 
An additional scale factor is applied to the $qq \rightarrow VH$ processes as a function of the vector boson's transverse momentum to account for electroweak (EW) corrections at NLO. This makes use of the $VH$ differential cross-section computed with \textsc{Hawk}~\cite{Denner:2011id,Denner:2014cla}. The samples of simulated events include all final states where the Higgs boson decays into $b\bar{b}$ and the vector boson to a leptonic final state, including those with a $\tau$-lepton. 
The analysis has only a small acceptance for other Higgs boson production and decay modes which are therefore neglected.
The mass of the Higgs boson was fixed at 125~\GeV\ and the $H \to b\bar{b}$ branching fraction was fixed at 58\%.  
The inclusive $pp\to VH$ cross-sections~\cite{Ciccolini:2003jy,Brein:2003wg,Ferrera:2011bk,Brein:2011vx,Ferrera:2013yga,Ferrera:2014lca,Campbell:2016jau} were calculated at next-to-next-to-leading order (NNLO)~(QCD) and NLO~(EW). Electroweak corrections include the photon-induced contributions, which are of the order of 5\% for the $WH \to \ell \nu b \bar{b}$ process and 1\% for the $ZH \to \ell\ell b\bar{b}$ process. 
For the gluon-induced $ZH$ production, the cross-section is calculated at next-to-leading order and next-to-leading-logarithm accuracy~(NLO+NLL) in QCD~\cite{Altenkamp:2012sx,Hespel:2015zea,ggzhnll,Harlander:2013mla,Brein:2012ne}. This is then subtracted from the inclusive $pp\to ZH$ production cross-section to estimate the quark-induced contribution to the cross-section. 

For the generation of $\ttb$ at NLO, the \textsc{Powheg-Box v2} generator~\cite{Frixione:2007nw} was used. 
Single top quark events in the $Wt$-, $s$- and $t$-channels were generated using the \textsc{Powheg-Box v1} generator~\cite{Alioli:2009je,Re:2010bp}. 
The top quark mass was set to 172.5~\GeV.  Events were filtered such that at least one $W$ boson in each event decays leptonically. 
The overall yield predicted for the $\ttb$ process is rescaled according to the NNLO cross-section, including the resummation of soft gluon emission at next-to-next-to-leading-logarithm accuracy (NNLL) as available in \textsc{Top++2.0}~\cite{Czakon:2011xx}. The overall yields predicted for single top quark production in the $s$-, $t$-, and $Wt$-channels are rescaled according to their respective NLO cross-sections~\cite{PhysRevD.81.054028,Kidonakis:2011wy,PhysRevD.82.054018}. 

Events containing $W$ or $Z$ bosons with jets (\vjets) were simulated using the \textsc{Sherpa 2.2.1} generator. Matrix elements were calculated for up to two partons at NLO and four partons at LO using the \textsc{OpenLoops}~\cite{Cascioli:2011va} and \textsc{Comix}~\cite{Gleisberg:2008fv} matrix-element generators.
The number of expected \vjets\ events is rescaled using the NNLO cross-sections \cite{Catani:2009sm}.

Diboson $WZ$ and $ZZ$ (referred to as $VZ$) processes were generated using \textsc{Sherpa 2.2.1}, which calculates up to one additional parton at NLO and up to three additional partons at LO.  The $WW$ process was generated using \textsc{Sherpa 2.1.1}, which calculates the inclusive production at NLO, and up to three additional partons at LO. 
The cross-section from \textsc{Sherpa} at NLO are used to normalise the events.

Samples produced with alternative generators are used to estimate systematic uncertainties in the event modelling, as described in Section \ref{sec:syst}.

\newpage

\section{Object and event selection}
\label{sec:selection}

Events with two jets tagged as containing $b$-hadrons and with either zero, 
one or two charged leptons (electrons or muons) are selected 
in this analysis. In the following, the physics objects
and the event selection for each channel are described.

\subsection{Object reconstruction}
\label{subsec:objectReco}

Interaction vertices are reconstructed ~\autocite{Aaboud:2016rmg} from tracks measured by the inner detector. 
The vertex with the highest sum of squared transverse momenta of all associated tracks 
is selected as the primary vertex, whereas all others are considered to be pile-up vertices.

Electrons are reconstructed~\autocite{Aaboud:2016vfy,ATLAS-CONF-2016-024}  by applying a sliding-window algorithm to noise-suppressed clusters of energy deposited in the calorimeter  and matching to a track in the inner detector.
Their energy calibration is based primarily on a data sample of $Z \to e^+e^-$ events~\autocite{PERF-2013-05}.  
Electron candidates are required to satisfy criteria for the 
shower shape, track quality and track-to-cluster match, corresponding to either the \textit{loose} or \textit{tight} 
likelihood-based requirements, denoted ``LooseLH'' and ``TightLH'' in Ref.~\cite{Aaboud:2016vfy}. 
All electrons are required to have $\pt>7$~\GeV\ and $|\eta|<2.47$. Non-prompt and pile-up tracks are rejected by 
requiring small transverse ($\mathrm{IP}_{{r\phi}}$) and longitudinal ($\mathrm{IP}_{{z}}$) impact parameters, 
defined with respect to the primary vertex position:\footnote{When computing impact parameters, the beam line is used to approximate the primary vertex position in the transverse plane.}  tracks must have
$|\mathrm{IP}_{r\phi}|/\sigma_{\mathrm{IP}_{r\phi}}<5$ and $|\mathrm{IP}_{{z}}|<0.5$~mm, with 
$\sigma_{\mathrm{IP}_{{r\phi}}}$ representing the uncertainty in the transverse impact parameter. 
A loose isolation requirement is applied: the electron track must be isolated from other tracks 
reconstructed in the inner detector, based on a variable cone size with $\Delta R_{\textrm{max}}=0.2$, with a requirement 
that is tuned to yield a constant 99\% efficiency as a function of electron \pt. Tight electrons are also 
required to pass a more stringent calorimeter-based isolation requirement, where the sum of the transverse energy 
of all the clusters of calorimeter 
cells, not associated with the electron candidate but found within a cone of $\Delta R=0.2$ around the electron track,
is required to be below 3.5~\GeV.

Muons are reconstructed~\autocite{PERF-2014-05,PERF-2015-10} as tracks in the inner detector matched 
to tracks in the muon spectrometer up to $|\eta|=2.5$. Some  acceptance is gained up to $|\eta|=2.7$ 
using the muon spectrometer alone, and within the region $|\eta|<0.1$ of limited muon-chamber acceptance, 
using tracks reconstructed in the inner detector that do not have a full matching track in the muon spectrometer, 
but have deposited energy in the calorimeter that is consistent with the energy loss of a muon. 
Two selection categories are defined: \textit{loose} and \textit{medium}, based on the respective muon identification criteria defined in Ref.~\autocite{PERF-2015-10}. 
All muon candidates are required to have $\pt>7$~\GeV, and not to be matched to 
an inner detector track that is likely to arise from a non-prompt muon or from pile-up, 
by applying impact parameter requirements similar to those for the electron selection:
$|\mathrm{IP}_{{r\phi}}|/\sigma_{\mathrm{IP}_{{r\phi}}}<3$ and $|\mathrm{IP}_{{z}}|<0.5$~mm.
A loose isolation requirement is applied, based on the momenta of tracks
in the inner detector which lie within a variable-size cone, with $\Delta R_{\textrm{max}}=0.3$, around the muon track;
analogously to the electron case, the requirement is tuned to yield a 99\%  efficiency for any value of  \pt.
For medium quality muons, a stringent track-based isolation requirement is applied, where 
the sum of the \pt\ of all the tracks found within a cone of $\Delta R=0.2$ around the muon track 
is required to be below 1.25~\GeV.

Jets are reconstructed from noise-suppressed energy clusters in the 
calorimeter~\autocite{ATL-LARG-PUB-2008-002} with the \antikt\ algorithm~\autocite{Cacciari:2008gp,Cacciari:2011ma} with radius parameter $R=0.4$.
The energies of the jets are calibrated using a jet energy scale correction (JES) derived from 
both simulation and \textit{in situ} calibration using data~\autocite{Aad:2014bia,Aaboud:2017jcu}. Jet cleaning 
criteria are applied to find jets arising from non-collision sources  or noise in the calorimeters and any event 
containing such a jet is removed~\autocite{ATLAS-CONF-2012-020,ATLAS-CONF-2015-029}. Jets with \pt\ below 60~\GeV\ and with $|\eta|<2.4$ have to pass a requirement on the \textit{jet vertex tagger} (JVT)~\autocite{Aad:2015ina}, 
a likelihood discriminant that uses track and vertex information in order to suppress jets originating from
pile-up activity. Jets in the central region ($|\eta|<2.5$) are required to have $\pt>20$~\GeV. For jets in the forward region ($2.5 \leq |\eta| < 4.5$), thus outside the acceptance of the inner detector, a stricter requirement of $\pt>30$~\GeV\ is applied in order to suppress jets from pile-up activity.

Jets in the central region 
can be tagged as containing $b$-hadrons by using
a multivariate discriminant (MV2c10) ~\autocite{PERF-2012-04,ATL-PHYS-PUB-2016-012} that combines information from an impact-parameter-based algorithm, from the explicit reconstruction of a secondary vertex and from a multi-vertex fitter that attempts to reconstruct the full $b$- to $c$-hadron decay chain. A significantly improved algorithm, which also profits 
 from the addition of the IBL detector, was developed for Run 2~\autocite{ATL-PHYS-PUB-2016-012}. At the chosen working point, the improved algorithm provides nominal light-flavour ($u$,$d$,$s$-quark and gluon) and $c$-jet misidentification efficiencies of 0.3\% and 8.2\%, respectively, for an average 
70\% $b$-jet tagging efficiency, as estimated from simulated \ttbar\ events for jets with $\pt>20$~\GeV\ and $|\eta|<2.5$.
The flavour tagging efficiencies in simulation are corrected separately for $b$-, $c$- and light-flavour jets, based on the respective data-based calibration analyses. 
The ratio of the efficiencies in data and simulation
is close to unity for $b$-jets, while more significant corrections are needed for $c$- and light-flavour jets, up to $\approx 1.4$ and $\approx 2$, respectively.

Simulated jets are labelled according to which hadrons with $\pt>5$~\GeV\ are found 
within a cone of size $\Delta R=0.3$ around the jet axis.
If a $b$-hadron is found the jet is labelled as a $b$-jet. If no $b$-hadron is found, but a $c$-hadron is present, then the jet is labelled as a $c$-jet. Otherwise the jet is labelled as a light (i.e., $u$,$d$,$s$-quark, or gluon) jet. Simulated $V$~+~jets events are categorised depending on the generator-level  \textit{truth} labels of the jets in the event that are selected to form the Higgs boson candidate: $V+bb$, $V+bc$, $V+cc$, $V+bl$, $V+cl$, $V+ll$ where $b$, $c$, $l$ stand for $b$-jet, $c$-jet and light-jet respectively. An inclusive $V$~+~heavy flavour ($V$~+~HF) category is defined as containing the first four: $V+bb$, $V+bc$, $V+cc$, $V+bl$. The $V+bb$ component is dominant: its fraction ranges from 70\% to 90\% of \vhf\ events, depending on the channel and analysis region.

Hadronically decaying $\tau$-leptons are reconstructed~\autocite{PERF-2014-06,ATLAS-CONF-2017-029} as jets from noise-suppressed energy clusters, using the \antikt\ algorithm with radius parameter $R=0.4$. They are required to have exactly one or three matching tracks in the inner 
detector within a cone of size $\Delta R=0.2$ around the jet axis,
to have $\pt >20$~\GeV\ and $|\eta|<2.5$, and to be 
outside the transition region between the barrel and endcap calorimeters ($1.37 < |\eta| < 1.52$). To reject jets 
being reconstructed and identified as $\tau$-leptons, a multivariate approach using boosted decision trees is employed,  
based on information from the calorimeters and from the tracking detectors; and the \textit{medium} 
quality criteria described in Ref.~\autocite{ATLAS-CONF-2017-029} are applied. 
Hadronically decaying $\tau$-leptons are only used in the analysis in the overlap removal procedure described at the end of this subsection.
This has an impact on the determination of the event's  jet multiplicity.

The uncertainty in the expected number of events depends on the size of the samples of simulated events. 
The combination of processes with large production cross-section and small selection efficiencies can make the production of 
samples exceeding the integrated luminosity of the data challenging.
For cases where the small selection efficiency is due to the high rejection achieved by the application of $b$-tagging, a method called \textit{parameterised tagging} is applied. 
Unlike when explicitly applying the $b$-tagging algorithm (\textit{direct tagging}), in parameterised tagging all jets are kept but the event is weighted by the expected probability for a jet with a certain flavour label ($b$, $c$ or light) to be tagged as a $b$-jet. These probabilities are parameterised as a function of jet kinematics (\pt\  and $\eta$) based on a large sample of simulated $t\bar{t}$ events. Parameterised tagging is used for the $V+cc$, $V+cl$, $V+ll$ and $WW$ samples, which simulate small background contributions  ($<2$\% of the total background). For all other samples, direct tagging is applied.

In addition to the JES correction, two more corrections are applied to $b$-tagged jets. The \textit{muon-in-jet} correction is applied when a medium quality muon with $\pt>5$~\GeV\ is found within $\Delta R=0.4$ of a jet, to account for the presence of $b$- and $c$-hadron decays into muons 
which do not deposit their full energy in the calorimeter. 
Unlike in the lepton selection introduced previously, no isolation criteria are applied. When more than one muon is found, the one closest to the jet axis is chosen. The muon four-momentum is added to that of the jet, and the energy deposited by the muon in the calorimeter is removed. To further improve the jet response, a second correction, denoted \textit{PtReco}, is applied as a function of jet \pt. This correction is based on the residual difference in jet response expected from the signal simulation between the reconstructed $b$-jets (with all corrections previously applied) and the corresponding \textit{truth jets} (formed by clustering final-state particles taken from the Monte Carlo truth record, including muons and neutrinos). 
This correction increases the energy of jets with  $\pt \sim 20$~\GeV\ by 12\% and the energy of those with $\pt > 100$~\GeV\ by 1\%. 
A larger correction is applied in case a muon or electron is identified within $\Delta R=0.4$ of the jet axis, to account for the missing neutrino energy.

\begin{figure}[h]
  \centering
  \begin{tabular}{cccc}
    \includegraphics[width=0.6\linewidth]{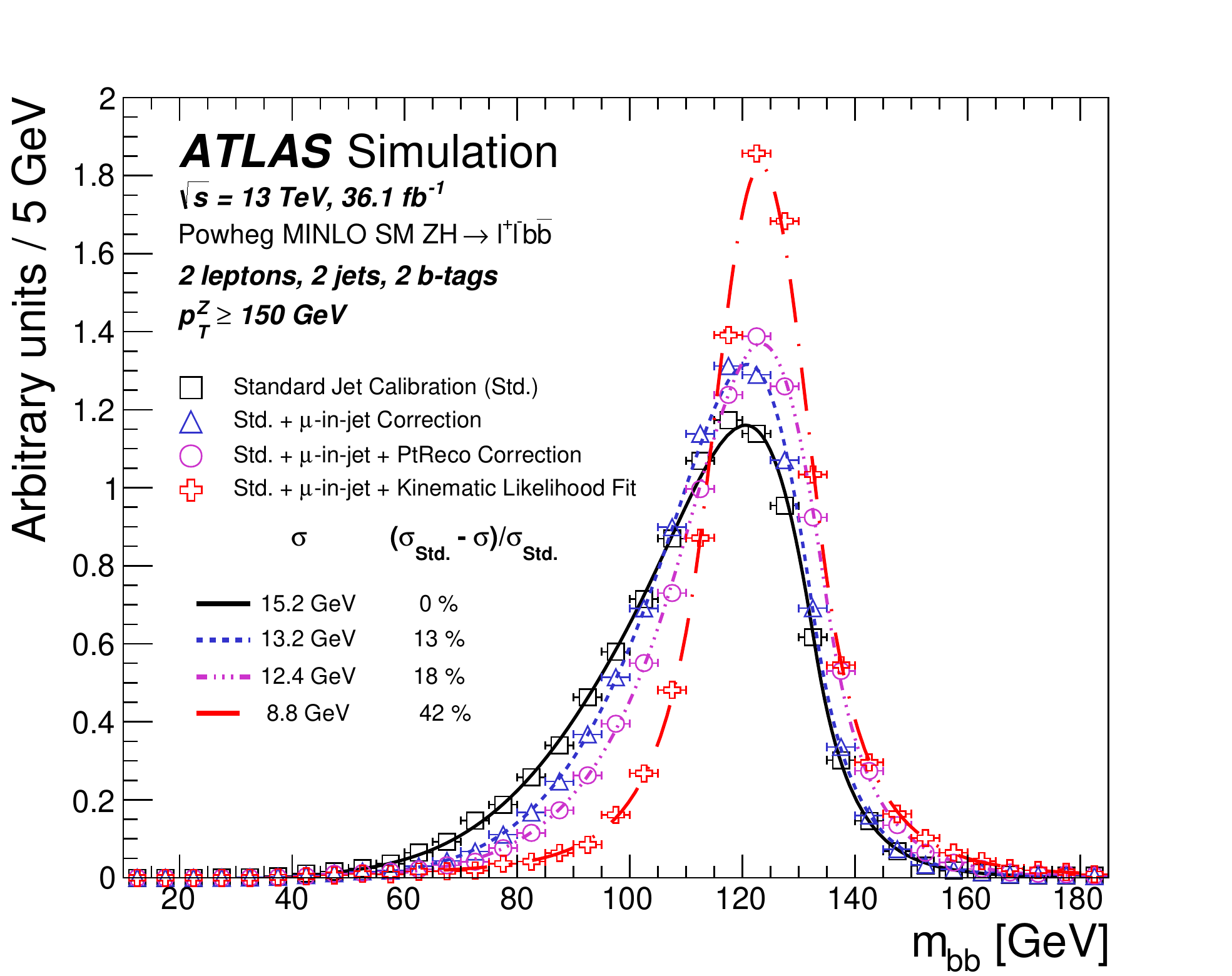}
  \end{tabular}
  \caption{Comparison of the \mbb\ distributions as additional corrections are applied to the jet energy scale, shown for simulated events in the  2-lepton channel in the 2-jet and $\ptz\ >150$~\GeV\ region. A fit to a Bukin function~\autocite{2007arXiv0711.4449B} is superimposed on each distribution, and the resolution values and improvements are reported in the legend.}
    \label{fig:massreso}
\end{figure}

In the 2-lepton channel, where the $ZH \to \ell\ell b\bar{b}$ event kinematics can be fully reconstructed, a per-event kinematic likelihood fit, 
 described in more detail in Ref.~\cite{HIGG-2013-23},
is used to improve the estimate of the energy of the two $b$-jets, in place of the PtReco correction. These corrections result in an improved \mbb\ mass distribution in the region of the Higgs boson signal, as illustrated in Figure~\ref{fig:massreso}; the central value is moved closer to its nominal value, and the resolution is improved by up to about 40\%.

The presence of neutrinos can be inferred by measuring the momentum imbalance in the event. This is measured by the missing transverse momentum $\vec{E}_{\mathrm{T}}^{\mathrm{miss}}$, defined as the negative vector sum of the transverse momenta of electrons, muons and jets\footnote{Hadronically decaying $\tau$-leptons are treated as jets in the measurement of the \vecmet .} associated with the primary vertex. A soft term~\autocite{PERF-2011-07,ATL-PHYS-PUB-2015-023,ATL-PHYS-PUB-2015-027} is added  to include well-reconstructed tracks matched to the primary vertex that are not already matched to any of the physics objects.

The object reconstruction and identification algorithms do not always result in unambiguous identifications. An overlap removal 
procedure is therefore applied, with the following actions taken in sequence. 
Any hadronically decaying $\tau$-lepton reconstructed closer than $\Delta R = 0.2$ to an electron or muon is removed, except in cases where the muon is deemed to be of low quality.
If a reconstructed muon shares an electron's ID track, the electron is removed. Jets within a cone of size $\Delta R = 0.2$ around an electron are removed, since a jet is always expected from clustering an electron's energy deposits in the calorimeter. Any electrons reconstructed within $\Delta R =\mathrm{min}(0.4,0.04+10\ \mathrm{\GeV}/\pt^{\mathrm{electron}})$ of the axis of any surviving jet 
are removed. Such electrons are likely to originate from semileptonic $b$- or $c$-hadron decays. If a jet is reconstructed within $\Delta R = 0.2$ of a muon and the jet has fewer than three associated tracks or the muon energy constitutes most of the jet energy then the jet is removed. Muons reconstructed within a cone of size $\Delta R ={\mathrm {min}}(0.4,0.04+10\ {\mathrm{\GeV}}/\pt^{\mathrm{muon}})$ around the jet axis of any surviving jet  
are removed. 
Jets that are reconstructed within a cone of size $\Delta R = 0.2$ around the axis of a hadronically decaying $\tau$-lepton are removed. 

\subsection{Event selection and categorisation}

The online event selection relies on either the \met\ or the single-charged-lepton triggers. Events passing the trigger selection and satisfying basic quality requirements are then categorised according to the charged lepton multiplicity, the vector boson's  transverse momentum, and jet multiplicity. Events are assigned to the 0-, 1- and 2-lepton channels depending on the number of charged leptons $\ell$, targeting the $ZH \to vv b\bar{b}$, $WH \to \ell \nu b \bar{b}$ and $ZH \to \ell\ell b\bar{b}$ signatures, respectively. 
Although $\tau$-leptons from vector-boson decays are not targeted explicitly, they pass the selection 
with reduced efficiency through leptonic decays of the $\tau$-lepton into muons and electrons.
All events are required to have at least two jets, and exactly two must pass the $b$-tagging requirement.
The Higgs boson candidate is reconstructed from the two $b$-tagged jets and the 
highest-\pt\ (leading) $b$-tagged jet is required to have $\pt>45$~\GeV. 

The analysis covers the phase space at large Higgs boson (and equivalently vector boson) transverse momentum, which has the highest signal-to-background ratio. For the same reason, events are categorised according to the reconstructed vector boson's transverse momentum \ptv. This observable corresponds to \met\ in the 0-lepton channel, to the size of the vectorial sum of \vecmet\ and the charged lepton's  transverse momentum in the 1-lepton channel, and the transverse momentum of the 2-lepton system in the 2-lepton channel. In the 0- and 1-lepton channels a single region is defined, with \ptv$>150$~\GeV. In the 2-lepton channel two regions are considered, $75$~\GeV~$<\ptv<150$~\GeV\ and $\ptv>150$~\GeV.

Events are further split into two categories according to jet multiplicity. In the 0- and 1-lepton channels, events are considered with exactly two or exactly three jets. Events with four or more jets are rejected in these channels to reduce the large background arising from $t\bar{t}$ production. In the 2-lepton channel, extra sensitivity is gained by accepting events with higher jet multiplicity due to the lower level of the \ttb\ background, thus the categories become either exactly two jets or three or more jets. For simplicity, these two selection categories are referred to as the 2- and 3-jet categories for all three lepton channels.

The event selection criteria for the three channels are detailed below and summarised in Table~\ref{tbl:evselTable}. 
The 1- and 2-lepton selections are both divided into two sub-channels depending on the flavour of the leptons: either electron or muon. There are small differences between these two sub-channels and these are mentioned when appropriate. The two sub-channels are merged to form the single 1- and 2-lepton channels used for the statistical analysis.
The statistical analysis uses eight signal regions (SRs) and six control regions (CRs). 
Multivariate discriminants are used as the main observables to extract the signal, as described in Section \ref{subsec:MVA}. 

The predicted cross-sections times branching ratios for \wzh\ with \wln, \zll, \znn,
and \hbb, as well as the acceptances in the three channels after full selection
are given in Table~\ref{tab:crosssection_acc}. The non-negligible acceptance for the $WH$ process in the 0-lepton channel is mostly due to events with hadronically decaying $\tau$-leptons produced in the $W$ decay, and the larger acceptance for the $gg\to ZH$ process with respect to $qq\to ZH$  is due to the harder \ptv\ distribution from the gluon-induced process.
\begin{landscape}
\begin{table}[p]
\centering
\caption{Summary of the event selection in the 0-, 1- and 2-lepton channels.}
\label{tbl:evselTable}
\begin{tabular}{l | c | c | c | c} \hline \hline
Selection & 0-lepton & \multicolumn{2}{c|}{1-lepton} & 2-lepton\\ 
               &                & $e$ sub-channel & $\mu$ sub-channel &   \\ \hline
Trigger & \met\ & Single lepton & \met\  & Single lepton \\
        \hline
Leptons & 0 \VHloose\ leptons & 1 tight electron & 1 medium muon  & 2 \VHloose\ leptons with $\pt>7$~\GeV \\
         &   with $\pt>7$~\GeV                    & $\pt>27$~\GeV  & $\pt>25$~\GeV   & $\ge$ 1 lepton with $\pt>27$~\GeV \\
        \hline
\met\ & $>$ 150~\textrm{\GeV} & $>$ 30~\textrm{\GeV} & -- & --  \\
\hline
\mll\ & -- & \multicolumn{2}{c|}{--}  & 81~\textrm{\GeV}~$<$ \mll\ $<$ 101~\textrm{\GeV} \\
\hline
Jets  & \multicolumn{3}{c|}{Exactly 2 or 3 jets} & Exactly 2 or $\geq$ 3 jets \\
\hline
Jet \pt &  \multicolumn{4}{c}{$>$ 20 \textrm{\GeV}} \\
\hline
$b$-jets & \multicolumn{4}{c}{Exactly 2 $b$-tagged jets} \\
\hline
Leading $b$-tagged jet \pt\  & \multicolumn{4}{c}{$>$ 45 \textrm{\GeV}} \\
\hline
$H_{\mathrm{T}}$ & $>$ 120 (2 jets), $>$150 \textrm{\GeV} (3 jets) & \multicolumn{2}{c|}{--}  & -- \\
\hline
$\textrm{min}[\Delta\phi(\vecmet,\vec{\mathrm{jets}}) ]$ & $> 20\ensuremath{^\circ}$ (2 jets), $> 30\ensuremath{^\circ}$ (3 jets) & \multicolumn{2}{c|}{--}  & -- \\ \hline
$\Delta\phi(\vecmet, \vec{bb})$ & $> 120\ensuremath{^\circ}$ & \multicolumn{2}{c|}{--}  & -- \\
\hline
$\Delta\phi(\vec{b}_1,\vec{b}_2)$ & $< 140\ensuremath{^\circ}$ & \multicolumn{2}{c|}{--}  & -- \\
\hline
$\Delta\phi(\vecmet,\vec{E}_{\mathrm{T},\mathrm{trk}}^{\mathrm{miss}})$ & $< 90\ensuremath{^\circ}$ & \multicolumn{2}{c|}{--}  & -- \\ \hline 
$\pt^V$ regions & \multicolumn{3}{c|}{$>150$~\GeV} & (75, 150]~\GeV, $>150$~\GeV \\ \hline\hline
Signal regions & $\checkmark$ &\multicolumn{2}{c|}{$\mbb \geq 75$~\GeV\ or $m_{\mathrm{top}}\leq225$~\GeV} & Same-flavour leptons \\
        &                   &            \multicolumn{2}{c|}{}         & Opposite-sign charge ($\mu\mu$ sub-channel)\\ \hline
Control regions & -- &\multicolumn{2}{c|}{$\mbb < 75$~\GeV\ and $m_{\mathrm{top}}>225$~\GeV} & Different-flavour leptons\\ \hline \hline
\end{tabular}
\end{table}
\end{landscape}

\begin{table}[tb!]
\begin{center}
\caption{The cross-section times branching ratio (B) and acceptance for the three channels 
at $\sqrt{s}=13$~\TeV. 
The $qq$- and $gg$-initiated $ZH$ processes are shown separately.  
The branching ratios are calculated considering only decays to muons and electrons for \zll, 
decays to all three lepton flavours for \wln\ and decays to all neutrino flavours for \znn. 
The acceptance is calculated as the fraction of events remaining in the combined signal and control
regions after the full event selection. 
\label{tab:crosssection_acc}}
\begin{tabular}{l|ccccc}
\hline\hline
\multicolumn{5}{c}{$m_H = 125$~\GeV\ at $\sqrt{s}=13$~\TeV} \\
\hline
\multirow{2}{*}{Process} & \multirow{2}{*}{Cross-section $\times$ B [fb]} & \multicolumn{3}{c}{Acceptance [\%]}\\
\cline{3-5}
&  & 0-lepton & 1-lepton & 2-lepton\\
\hline
$qq\to ZH\to\ell\ell b\bar{b}$  &  29.9  & $<$~0.1     &  $<$~0.1 & 7.0 \\ 
$gg\to ZH\to\ell\ell b\bar{b}$   &   4.8  & $<$~0.1      & $<$~0.1 & 15.7 \\
$qq\to WH\to\ell\nu b\bar{b}$  & 269.0  & 0.2 & 1.0 & -- \\
$qq\to ZH\to\nu\nu b\bar{b}$ &  89.1  & 1.9 & --        & -- \\
$gg\to ZH\to\nu\nu b\bar{b}$   &   14.3  & 3.5 & --        & -- \\
\hline\hline
\end{tabular}
\end{center}
\end{table}

\subsubsection{Zero-lepton selection}

The online event selection relies on an $E_{\mathrm{T}}^{\mathrm{miss}}$ trigger. The threshold for this trigger was 70~\GeV\ for the 2015 data, and it was initially raised to 90~\GeV\ and then to 110~\GeV\ during 2016.
In the offline analysis events are required to have no loose leptons and  \met > 150~\GeV. When compared to the offline selection, the \met\ trigger is fully efficient for $\met > 180$~\GeV, and it is $85-90$\% efficient at $\met=150$~\GeV, depending on the data taking period. The trigger efficiency is measured in $W$~+~jets and $t\bar{t}$ events in data using an orthogonal set of single-muon triggers; these measurements are utilised to determine data-over-simulation scaling factors, used to correct the simulation. The scaling factors are within 5\% of unity and parameterised as a function of \met .  A selection based on the scalar sum of the transverse momenta of the jets in the event, $H_{\mathrm{T}}$, is used to remove a marginal region of phase space in which the trigger efficiency exhibits a small dependence on the jet multiplicity. For 2-jet events the requirement is $H_{\mathrm{T}} > 120$~\GeV, and $H_{\mathrm{T}} > 150$~\GeV\ is required for 3-jet events.

In order to suppress the multi-jet background, which is mostly due to jets mismeasured in the calorimeters, four angular selection criteria are applied:
\begin{itemize}
\item $\Delta\phi(\vec{E}_{\mathrm{T}}^{\mathrm{miss}}, \vec{E}_{\mathrm{T},\mathrm{trk}}^{\mathrm{miss}}) < 90^\circ$,		
\item $\Delta\phi(\vec{b}_1,\vec{b}_2) < 140^\circ$,
\item $\Delta\phi(\vec{E}_{\mathrm{T}}^{\mathrm{miss}}, \vec{bb}) > 120^\circ$,		
\item $\min [ \Delta\phi(\vec{E}_{\mathrm{T}}^{\mathrm{miss}}, \vec{\mathrm{jets}}) ] > 20^\circ\ \mathrm{for\ 2\ jets},\ >30^\circ\ \mathrm{for\ 3\ jets}$.
\end{itemize}

Here $\Delta\phi(\vec{a},\vec{b})$ indicates the difference in azimuthal angle between objects $\vec{a}$ and $\vec{b}$; $\vec{b}_1$ and $\vec{b}_2$ are the two $b$-tagged jets forming the Higgs boson candidate's dijet system $\vec{bb}$; $\vec{E}_{\mathrm{T},\mathrm{trk}}^{\mathrm{miss}}$ is defined as the missing transverse momentum calculated from the negative vector sum of the transverse momenta of tracks reconstructed in the inner detector and identified as originating from the primary vertex. The final selection is a requirement on the azimuthal angle between the $\vec{E}_{\mathrm{T}}^{\mathrm{miss}}$ vector and the closest jet.

\subsubsection{One-lepton selection}
\label{sele1lept}

For the electron sub-channel, events are selected using a logical OR of single-electron triggers with \pt\ thresholds of 24~\GeV, 60~\GeV\ and 120~\GeV\ for the 2015 data and with increased thresholds of 26~\GeV, 60~\GeV\ and 140~\GeV\ in 2016. The lowest-threshold trigger in 2016 includes isolation and identification requirements that are looser than any of the isolation and identification requirements applied in the analysis. 
These requirements are removed or relaxed for the higher-threshold triggers.
The muon sub-channel uses the same \met\  triggers as the 0-lepton channel. Since muons are not included in the \met\ calculation at trigger level, in events where a muon is present  this trigger is in effect selecting events based on \ptv, and is therefore fully efficient for values of \ptv\ above 180~\GeV. This trigger is preferred because it has an overall signal efficiency (with respect to the offline selection) of 98\%, compared to $\sim 80\%$ efficiency for the combination of single-muon triggers, which is due to the limited muon trigger chamber coverage in the central $|\eta|$ region of the detector.  Events are required to contain exactly one tight electron with \pt\ above 27~\GeV\ (electron sub-channel) or one medium muon with \pt\ above 25~\GeV\ (muon sub-channel), and no additional loose leptons. In the electron sub-channel, where multi-jet production is a significant background, an additional selection of $E_{\mathrm{T}}^{\mathrm{miss}} > 30$~\GeV\ is applied.

Control regions enhanced in the $W$~+~HF background are defined for both the 2- and 3-jet categories. These are obtained by applying two additional selection requirements beyond the respective nominal selection criteria: $\mbb < 75$~\GeV\ and $m_{\mathrm{top}}>225$~\GeV. 
To calculate the reconstructed top quark mass, $m_{\text{top}}$, an estimate of the four-momentum of the neutrino from the $W$ boson decay is required. The vector $\vec{E}_{\mathrm{T}}^{\mathrm{miss}} $ is assumed to give an estimate of the neutrino's transverse momentum 
components and then $p^{\nu}_{{z}}$ can be determined up to a possible two-fold ambiguity by constraining the mass of the lepton-plus-neutrino system to be the $W$ boson mass.\footnote{In the case of negative discriminant in the quadratic equation, the $\vec{E}_{\mathrm{T}}^{\mathrm{miss}} $ vector is shifted such that the discriminant becomes zero.} The top quark is then reconstructed by considering the reconstructed $W$ boson and one of the two $b$-tagged jets. 
The combination of $b$-tagged jet and $p^{\nu}_{{z}}$ resulting in $m_{\text{top}}$ closest to 172.5~\GeV\ is selected.
The requirement on the reconstructed top quark mass  significantly reduces the contamination from $t\bar{t}$ and single-top-quark events in the $W$~+~HF CRs. The events in the control regions are removed from the corresponding signal regions. In the $W$~+~HF CRs, between 75\% and 78\% of the events are expected to be from $W$~+~HF production.

\subsubsection{Two-lepton selection}

Events are selected in the electron sub-channel using the same single-electron triggers as for the 1-lepton channel. For the muon sub-channel a logical OR of single-muon triggers with \pt\ thresholds of 20~\GeV\ and 40~\GeV\ is used for 2015 data, and 24--26~\GeV\ and 40--50~\GeV\ for 2016 data, with the increase of the thresholds applied to cope with the increasing instantaneous luminosity. The lowest-threshold triggers include an isolation requirement that is removed for the higher-threshold triggers. The trigger efficiency with respect to the offline selection ranges from 97\% to 99.5\% for the electron sub-channel and from 87\% to 90\% for the muon sub-channel, depending on the \ptv\ region.
To ensure that the trigger efficiency reached its plateau, the lepton that triggered the event is required to have $\pt>$ 27~\GeV.
Exactly two loose leptons of the same flavour are required. 
 In dimuon events, the two muons are required to have opposite-sign charge. This is not used in the electron sub-channel, where the charge misidentification rate is not negligible. The invariant mass of the dilepton system must be consistent with the $Z$ boson mass: $81$~\GeV~$< m_{\ell\ell} < 101$~\GeV. This requirement suppresses backgrounds with non-resonant lepton pairs, such as $t\bar{t}$ and multi-jet production.  

Control regions are defined to be very pure in $t\bar{t}$ and $Wt$ background by applying the nominal selection but requiring an $e\mu$ lepton flavour combination instead of $ee$ or $\mu\mu$, and with no opposite-charge requirement. The $t\bar{t}$ and $Wt$ events in these control regions are kinematically identical to those in the signal region, 
except for slight differences in acceptance between electrons and muons.
These regions are called $e\mu$ CR in the following. In the $e\mu$ CRs, more than 99\% of the events are expected to be from \ttbar\ and single top quark production, and between 88\% and 97\% from \ttbar\ production alone.

\subsection{Selection for the dijet-mass analysis}
\label{subsec:dijetmass_selection}

To validate the result of the multivariate analysis, a second analysis is performed where the multivariate discriminants are replaced by the dijet invariant mass of the two $b$-tagged jets, \mbb. 
This second analysis adopts the same objects and event selection criteria as described in Table~\ref{tbl:evselTable}, with the additional selection criteria shown in Table~\ref{tbl:evselTableCBA}. With respect to the \ptv\ regions described earlier, the events with $\ptv>150$~\GeV\ are further split into two categories: $150~\mathrm{\GeV}<\ptv \le 200$~\GeV\ and $\ptv>200$~\GeV. Events with $\ptv \le 150$~\GeV\ are rejected if  $\Delta R(\vec{b}_1,\vec{b}_2)>3.0$, where $\Delta R(\vec{b}_1, \vec{b}_2)$ is the separation of the two $b$-tagged jets in the $(\eta, \phi)$ plane. 
For $150~\mathrm{\GeV}<\ptv \le 200$~\GeV, the events are rejected if  $\Delta R(\vec{b}_1,\vec{b}_2)>1.8$. For \ptv$>200$~\GeV\ events are rejected if $\Delta R(\vec{b}_1,\vec{b}_2)>1.2$. 

In the 1-lepton channel, since the low \mbb\ range in the dijet mass spectrum provides sufficient information to constrain the \whf\ background normalisation, no dedicated \whf\ control region is defined. Also, a requirement on the $W$ boson's transverse mass \mtw$<120$~\GeV\ is used to suppress events from \ttbar\ background. The $W$ boson's transverse mass is defined as $m_{\mathrm{T}}^{W} = \sqrt{2p_{\mathrm{T}}^\ell\met (1-\cos(\Delta\phi(\vec{\ell},\vecmet)))}$,
where $p_{\mathrm{T}}^\ell$ is the lepton's  transverse momentum.

In the 2-lepton channel, the \ttbar\ background is suppressed thanks to the additional requirement  $\met/\sqrt{S_{\mathrm{T}}}<3.5\sqrt{\mathrm{\GeV}}$, where $S_{\mathrm{T}}$ is defined as the scalar sum of the transverse momenta of all jets and leptons in the event. Events with \ptv$>150$~\GeV\  in the  $e\mu$ CR are used inclusively in \ptv.
\begin{table}[pht]
\centering
\caption{Summary of the event selection criteria in the 0-, 1- and 2-lepton channels for the dijet-mass analysis, 
applied in addition to those described in Table~\ref{tbl:evselTable} for the multivariate analysis.
\label{tbl:evselTableCBA}}
\begin{tabular}{l | c | c | c} \hline \hline
\multicolumn{4}{c}{Channel} \\ \hline 
Selection & 0-lepton & 1-lepton & 2-lepton\\ \hline
\mtw\ & - & $<$ 120~\textrm{\GeV} & - \\
\hline
$\met/\sqrt{S_{\mathrm{T}}}$ & - & - & $<3.5\sqrt{\mathrm{\GeV}}$ \\ \hline  \hline 
\multicolumn{4}{c}{}\\\hline\hline
\multicolumn{4}{c}{$\pt^V$ regions} \\ \hline 
$\pt^V$ & (75, 150]~\GeV & (150, 200]~\GeV &(200, $\infty$)~\GeV \\
& (2-lepton only)&  & \\ \hline
$\Delta R(\vec{b}_1,\vec{b}_2)$ & $<$3.0 & $<$1.8 &$<$1.2 \\ \hline \hline
\end{tabular}
\end{table}

\section{Multivariate analysis}
\label{subsec:MVA}

Multivariate discriminants making use of boosted decision trees (BDTs) are constructed, trained and evaluated in each lepton channel and analysis region separately. 
Two versions of the BDTs, using the same input variables, are trained. The nominal version is designed to separate the $VH, H \to b\bar{b}$ signal from the sum of the expected background processes, and is referred to as BDT$_{VH}$. A second one, which is used to validate the analysis, aims at separating the $VZ, Z \to b\bar{b}$ diboson process from the sum of all other expected background processes (including $VH$), and is referred to as BDT$_{VZ}$ .

The input variables used for the BDTs are chosen in order to maximise the separation in the $VH$ search. Starting from the dijet mass ($\mbb$), additional variables describing the event kinematics and topology are
tried one at a time and the one yielding the best separation gain is added to the list. This procedure is repeated until
adding more variables results in a negligible performance gain.
The final selections of variables for
the different channels are listed in Table~\ref{tbl:MVAinputs}. 
The $b$-tagged jets are labelled
in decreasing $\pT$ as $b_1$ and $b_2$, and $|\Delta\eta(\vec{b}_1, \vec{b}_2)|$ is their separation in pseudorapidity. 
 In 3-jet events, the
third jet is labelled as $\mathrm{jet}_3$ and the mass of the 3-jet system is denoted $m_{bbj}$. The azimuthal angle between the vector boson and the system of $b$-tagged jets is denoted $\Delta\phi(\vec{V}, \vec{bb})$, and their pseudorapidity separation is denoted $|\Delta\eta(\vec{V}, \vec{bb})|$. In the 0-lepton channel, $m_{\mathrm{eff}}$ is defined as the scalar sum of the transverse momenta of all jets and $E_{\mathrm{T}}^{\mathrm{miss}}$ ($m_{\mathrm{eff}}=H_{\mathrm{T}}+E_{\mathrm{T}}^{\mathrm{miss}}$). In the 1-lepton channel, the angle 
between the lepton and the closest $b$-tagged jet in the transverse plane is denoted
$\mathrm{min}[\Delta\phi(\vec{\ell},\vec{b})]$. 
In  the 1-lepton channel, two variables are used to improve the rejection of the $t\bar{t}$ background: the rapidity difference between the $W$ and Higgs boson candidates, $|\Delta Y(\vec{V},\vec{bb})|$ and, assuming that the event is $t\bar{t}$, the reconstructed top quark mass, $m_{\text{top}}$. To construct the $|\Delta Y(\vec{V},\vec{bb})|$ variable,  the four-vector of the neutrino in the $W$ boson decay is estimated as explained in Section~\ref{sele1lept} for  $m_{\text{top}}$. 
The distributions of input variables of the BDTs are compared between data and simulation, 
and good agreement is found within the uncertainties.

The Toolkit for Multivariate Data Analysis, TMVA~\cite{TMVA}, is used to train the BDTs, with values 
of the training parameters similar to those described in Ref.~\cite{HIGG-2013-23}. 
In order to make use of the complete set of simulated MC events for the 
BDT training and 
evaluation in an unbiased way, 
the MC events are split into two samples of equal size, $A$ and $B$. 
The performance of the BDTs trained on sample $A$ ($B$) is evaluated with sample $B$ ($A$) in order
to avoid using identical events for both training and evaluation of the same BDT. 
Half of the data are analysed with the BDTs trained on sample $A$, and the other half with the BDTs 
trained on sample $B$.
At the end, the output distributions of the BDTs trained on samples $A$ and $B$ are
merged for both the simulated and data events.
A dedicated procedure is applied to transform the BDT output distributions
to obtain a smoother distribution for the background processes and finer binning
in the regions with the largest signal contribution, 
whilst ensuring that the statistical uncertainty of the simulated background is less than 20\% in each bin. 
The binning procedure is 
described in more detail in Ref.~\cite{HIGG-2013-23}.

\begin{table}[tb!]
\begin{center}
\caption{Variables used for the multivariate discriminant in each of the categories.}
\label{tbl:MVAinputs}
\begin{tabular}{l|ccc}
\hline\hline
Variable                                      &  0-lepton   & 1-lepton    & 2-lepton    \\
\hline                                      
\ptv                                          &     $\equiv \met $       & $\times$  & $\times$  \\
\met                                          &  $\times$ & $\times$  & $\times$  \\
$\pt^{{b}_1}$                                 &  $\times$ & $\times$  & $\times$  \\
$\pt^{{b}_2}$                                 &  $\times$ & $\times$  & $\times$  \\
$m_{{bb}}$                                    &  $\times$ & $\times$  & $\times$  \\
$\Delta R(\vec{b}_1,\vec{b}_2)$                      &  $\times$ & $\times$  & $\times$  \\
$|\Delta\eta(\vec{b}_1,\vec{b}_2)|$                  &  $\times$ &           &   \\ 
$\Delta\phi(\vec{V},\vec{bb})$                         &  $\times$ & $\times$  & $\times$  \\
$|\Delta\eta(\vec{V},\vec{bb})|$                       &             &             & $\times$  \\
$m_{\mathrm{eff}}$                               &  $\times$ &             &             \\
$\mathrm{min}[\Delta\phi(\vec{\ell},\vec{b})]$         &             & $\times$  &          \\
\mtw                                          &             & $\times$  &             \\
\mll                                          &             &             & $\times$  \\
\mtop                        &   & $\times$  &   \\
$|\Delta Y(\vec{V},\vec{bb})|$                        &   & $\times$  &   \\
\hline
  & \multicolumn{3}{c}{Only in 3-jet events} \\
\hline
$\pt^{\mathrm{jet}_3}$                          &  $\times$ & $\times$  & $\times$  \\
$m_{bbj}$                                     &  $\times$ & $\times$  & $\times$  \\
\hline\hline
\end{tabular}
\end{center}
\end{table}


\section{Estimation of the multi-jet background}
\label{sec:multijet}
The MC samples summarised in Section~\ref{sec:samples} are used to model background processes with 
$W$ or $Z$ boson decays into leptons;  these are defined as electroweak (EW) backgrounds in the following. 
Multi-jet backgrounds are produced with large cross-sections and thus, despite not providing genuine leptonic signatures, have the potential to contribute 
a non-negligible background component. In the following this background contribution is discussed channel by channel.

\subsection{0-lepton channel}

As described in Section~\ref{sec:selection}, specific criteria are applied in the event selection to suppress the multi-jet backgrounds. A data-driven method is used to estimate the residual contribution.  
After removing the selection applied to the $\min [ \Delta\phi(\vec{E}_{\mathrm{T}}^{\mathrm{miss}}, \vec{\mathrm{jets}}) ]$ variable, a fit to this distribution in the 3-jet category is performed to extract the multi-jet contribution 
while allowing the \ttbar\ and $Z$~+~jets background normalisations to float.
In multi-jet background events, a fake $E_{\mathrm{T}}^{\mathrm{miss}}$ can arise from a jet energy fluctuation, and it is expected that its direction is close to the direction of the poorly measured jet. Therefore, the $\min [ \Delta\phi(\vec{E}_{\mathrm{T}}^{\mathrm{miss}}, \vec{\mathrm{jets}}) ]$ variable is very effective in suppressing the multi-jet contribution, which is confined to low values of $x=\min [ \Delta\phi(\vec{E}_{\mathrm{T}}^{\mathrm{miss}}, \vec{\mathrm{jets}}) ]$ and is parameterised with a falling exponential ($\exp\left(- x/c\right)$). The parameter $c$ is determined in the fit itself, while the templates for the other backgrounds are taken directly from simulation. After the nominal selection criteria are applied, the residual multi-jet contamination within an $80$~\GeV~$<m_{bb}<160$~\GeV\ mass window is found to be $\sim 10$\% of the signal contribution and negligible
($<0.1$\%) with respect to the total background. The BDT distribution for the multi-jet background is estimated from the data at 
low $\min [ \Delta\phi(\vec{E}_{\mathrm{T}}^{\mathrm{miss}}, \vec{\mathrm{jets}}) ]$, and found to have a shape similar to the one expected for the sum of the remaining backgrounds. 
The small multi-jet contribution is therefore absorbed in the floating normalisation factors of the EW backgrounds in the global likelihood fit.
The same data-driven estimation technique cannot be used in the 2-jet region, where events at low values of $\min [ \Delta\phi(\vec{E}_{\mathrm{T}}^{\mathrm{miss}}, \vec{\mathrm{jets}}) ]$ are removed by the other selection requirements. A multi-jet \textsc{ Pythia8} MC sample generated with the A14 tune and NNPDF2.3LO PDFs is used to extrapolate the data-driven estimate from the 3- to the 2-jet region, with the extrapolation factor derived after removing any $b$-tagging requirement. The contribution in the 2-jet region is found to be negligible. Multi-jet production in the 0-lepton channel is therefore found to be a small enough background that it can be neglected in the global likelihood fit.

\subsection{1-lepton channel}

Both the electron and muon sub-channels have contributions from multi-jet events. The dominant contribution to this background stems from real muons or electrons from heavy-flavour hadrons that undergo semileptonic decays. In the electron sub-channel a second contribution arises 
from $\gamma \to e^{+}e^{-}$ conversions of photons produced in the decay of neutral pions in jets,
  or directly from $\pi^0$ Dalitz decays. Although those leptons are not expected to be isolated, a small but non-negligible fraction passes the lepton isolation requirements. This background is estimated separately in the electron and muon sub-channels, and in the 2- and 3-jet categories, using similar procedures.

In each signal region, a template fit to the $W$ boson candidate's  transverse  mass (\mtw) distribution is performed in order to extract the multi-jet yield. The variable \mtw\ is chosen as it offers the clearest discrimination between the multi-jet and EW processes. The template used for the multi-jet contribution is obtained from data in a control region after subtraction of the residual EW contribution, based on MC predictions, while the template for the EW contribution in the signal region is obtained directly from MC predictions. The control region is enriched in multi-jet events that are kinematically close to  the corresponding signal region but not overlapping with it, and is defined by applying the nominal selection but inverting the tight isolation requirement. To increase the statistical precision of the data-driven estimate, the number of required $b$-tags is reduced from two to one. The template fit determines the normalisation of the multi-jet contribution in the signal region, while the shape of the BDT discriminant (or of other relevant observables) is obtained analogously to the \mtw\ template. Both the normalisation and shape derived for the BDT discriminant are then used in the global likelihood fit.

Since the efficiency of the tight isolation requirement on multi-jet events depends in general on lepton kinematics, and on the composition of the multi-jet background, the control regions that are based on inverting such a requirement provide biased estimators for the multi-jet templates in the corresponding signal regions. The templates are therefore corrected for such a bias, by applying event-by-event extrapolation factors that depend on lepton \pt\ and $\eta$, and, in the electron sub-channel, also on the value of \met. These extrapolation factors are derived in additional control regions where the 2- and 3-jet requirements of the nominal selection are replaced by a 1-jet requirement, and the $b$-tagging requirement is removed. The extrapolation factors are computed as the
ratio of the number of events with an isolated lepton to the number of events with a non-isolated lepton, after removing the MC-predicted EW background contribution.

The estimate of the normalisations of the $W$~+~jet and top quark (\ttbar\ and single top quark) background contributions in the signal region provided by Monte Carlo simulations is subject to significant uncertainties. In addition, the \mtw\ distributions of the $W$~+~jet and top quark backgrounds are sufficiently different that a common normalisation factor induces a bias in the multi-jet estimate. The normalisation of these two backgrounds is therefore left free to be determined in the template fit used to extract the multi-jet contribution. In order to improve their relative separation, the fit to the \mtw\ distribution in the signal region is performed together with a fit to the overall yield in the corresponding $W$~+~HF control region. Furthermore, in order to improve the statistical precision in the determination of the $W$~+~jet and top quark background normalisation factors, the multi-jet template fit is performed simultaneously in the electron and muon sub-channels. This corresponds to performing separate fits for the two sub-channels, but with common $W$~+~jet and top quark background normalisation factors. 

The multi-jet contribution in the 2-jet region is found to be $4.8\%$ ($4.6\%$) of the total background contribution in the electron (muon) sub-channel, while in the 3-jet region it is found to be $0.3\%$ ($0.5\%$). These estimates are subject to sizeable systematic uncertainties, which are described in Section~\ref{sec:syst}.

\subsection{2-lepton channel}

Requiring two isolated leptons with a dilepton invariant mass compatible with that of the $Z$ boson strongly suppresses the contributions from multi-jet events. The residual contribution is estimated using a fit to the dilepton mass distribution in a sample of events where the two lepton candidates have the same charge. The fit model includes expected contributions from EW backgrounds from simulation and an exponential model for the multi-jet background. An estimate is then made of the fraction of the background in a mass window around the $Z$ boson peak in the signal region that could be attributed to multi-jet events based on the assumption that the numbers of opposite-charge and same-charge events are equal for the multi-jet background. Inside a mass window $81$~\GeV~$< m_{\ell\ell} < 101$~\GeV\ the fraction of the background in the signal region coming from multi-jet events is estimated to be 0.03\% and 0.2\% for the muon and electron sub-channels, respectively. The residual multi-jet contamination within a  $100$~\GeV~$<m_{bb}<140$~\GeV\ mass window is found to be $\sim 8$\% of the signal contribution, without an $m_{bb}$ resonant shape, and found to have a BDT shape similar to the one expected for the sum of the remaining backgrounds. 
The multi-jet contamination is also extracted in the $e\mu$ control region and found to be 0.3\% of the total background. The multi-jet contribution in the 2-lepton channel is thus small enough to have a negligible impact on the signal extraction and is therefore not included in the global likelihood fit.


\section{Systematic uncertainties}
\label{sec:syst}
The sources of systematic uncertainty can be broadly divided into four groups: those of experimental nature, those related to the modelling of the simulated backgrounds, those related to the multi-jet background estimation, and those associated with the Higgs boson signal simulation. The finite size of the simulated background samples is also an important source of systematic uncertainty, and, whenever possible, generator-level filters are employed to enhance the amount of simulated events in the phase-space region that is most relevant for the analysis.

\subsection{Experimental uncertainties}
\label{subsec:syst_exp}
The dominant experimental uncertainties originate from the flavour-tagging simulation-to-data efficiency correction factors, from the jet energy scale corrections and the modelling of the jet energy resolution.
Flavour-tagging simulation-to-data efficiency correction factors are derived~\autocite{PERF-2012-04} 
separately for $b$-jets, $c$-jets and light-flavour jets. All three correction factors depend on jet \pt\ (or \pt\ and $|\eta|$) and have uncertainties estimated from multiple sources. These are decomposed into uncorrelated components which are then treated independently, resulting in three uncertainties for $b$-jets and for $c$-jets, and five for light-flavour jets. 
The approximate size of the uncertainty in the tagging efficiency is 2\% for $b$-jets, 10\% for $c$-jets and 30\% for light jets. Additional uncertainties are considered in the extrapolation of the $b$-jet efficiency calibration above $\pt=300$~\GeV\ and in the misidentification of hadronically decaying $\tau$-leptons as $b$-jets. The uncertainties in the jet energy scale and resolution are based on their respective measurements in data~\autocite{Aaboud:2017jcu,PERF-2011-04}. The many sources of uncertainty in the jet energy scale correction are decomposed into 21 uncorrelated components which are treated as independent. An additional specific uncertainty is considered that affects the energy calibration of $b$- and $c$-jets.

Uncertainties in the reconstruction, identification, isolation and trigger efficiencies of muons~\autocite{PERF-2015-10} 
and electrons~\autocite{ATLAS-CONF-2016-024}, along with the uncertainty in their energy scale and resolution, are estimated based upon 13~\TeV\ data. These are found to have only a small impact on the result. 
The uncertainties in the energy scale and resolution of the jets and leptons are propagated
to the calculation of \met, which also has additional  uncertainties from the scale, resolution 
and efficiency of the tracks used to define the soft term~\autocite{ATL-PHYS-PUB-2015-023}, along with the modelling
of the underlying event. An uncertainty is assigned to the simulation-to-data \met\ trigger scale factors to account for the statistical uncertainty in the measured scale factors and differences between the scale factors determined from $W$~+~jets and $t\bar{t}$ events.
The uncertainty in the luminosity is 2.1\% for the 2015 data and 3.4\% for the 2016 data, resulting in an uncertainty of 3.2\% for the combined dataset. It is derived, following a methodology similar to that detailed in Ref.~\autocite{Aaboud:2016hhf}, 
from a preliminary calibration of the luminosity scale using $x$--$y$ beam-separation scans performed 
in 2015 and 2016. 
The average number of interactions per bunch crossing is rescaled by 9\% to improve the agreement between simulation with data, and an uncertainty, as large as the correction, is included.

\subsection{Simulated background uncertainties}

Modelling uncertainties are derived for the simulated backgrounds and broadly cover three areas: normalisation, acceptance differences that affect the relative normalisation between analysis regions with a common background normalisation, and the differential distributions of the most important kinematic variables. These uncertainties are derived either from particle-level comparisons between nominal and alternative samples 
using the RIVET~\cite{Buckley:2010ar} framework, or from comparisons to data in control regions. The particle-level comparisons are cross-checked with detector-level simulations whenever these are available, and good agreement is found. When acceptance uncertainties are estimated all the nominal and alternative samples are normalised using the same production cross-section. Such uncertainties are estimated by adding the differences between the nominal and alternative samples in quadrature. Shape uncertainties are considered in each of the analysis regions separately, with the samples scaled to have the same normalisation in each region. In this case, the uncertainty is taken from the alternative generator which has the largest shape difference compared to the nominal sample. Shape uncertainties are only derived for the \mbb\ and \ptv\ variables, as it was found that it is sufficient to only   consider the changes induced in these variables by an alternative generator to cover the overall shape variation of the BDT$_{VH}$ discriminant. The systematic uncertainties affecting the modelling of the background samples are reported in Tables~\ref{tab:bkg_systematics} and~\ref{tab:bkgdib_systematics}, and the specific details of how the uncertainties are estimated are provided below for each simulated background sample.

\begin{table}[p!]
\begin{small}
\begin{center}
	\caption
	{Summary of the systematic uncertainties in the background modelling for 
$Z$~+~jets, $W$~+~jets, \ttb, single top quark and multi-jet production. 
    An ``S'' symbol is used when only a shape uncertainty is assessed. 
    The regions for which the normalisations float independently are listed in brackets.
	\label{tab:bkg_systematics}}
\begin{tabular}{ l | c}
	\hline\hline
	\multicolumn{2}{c}{$Z$~+~jets}\\
	\hline
	$Z+ll$ normalisation & 18\% \\
	$Z+cl$ normalisation & 23\% \\
	$Z+bb$ normalisation & Floating (2-jet, 3-jet) \\
	$Z+bc$-to-$Z+bb$ ratio & 30 -- 40\% \\
	$Z+cc$-to-$Z+bb$ ratio & 13 -- 15\% \\
	$Z+bl$-to-$Z+bb$ ratio & 20 -- 25\% \\
	0-to-2 lepton ratio & 7\% \\
	\mbb, \ptv  & S \\
	\hline\hline
	\multicolumn{2}{c}{$W$~+~jets}\\
	\hline
	$W+ll$ normalisation & 32\% \\
         $W+cl$ normalisation & 37\% \\
         $W+bb$ normalisation & Floating (2-jet, 3-jet)\\
 	$W+bl$-to-$W+bb$ ratio & 26\% (0-lepton) and 23\% (1-lepton) \\
	$W+bc$-to-$W+bb$ ratio & 15\% (0-lepton) and 30\% (1-lepton)    \\
	$W+cc$-to-$W+bb$ ratio & 10\% (0-lepton) and 30\% (1-lepton) \\
	0-to-1 lepton ratio & 5\%  \\
	$W$~+~HF CR to SR ratio & 10\% (1-lepton) \\
	\mbb, \ptv  & S \\
	\hline\hline
	\multicolumn{2}{c}{\ttb\ (all are uncorrelated between the 0+1 and 2-lepton channels)}\\
	\hline
	\ttb\ normalisation & Floating (0+1 lepton, 2-lepton 2-jet, 2-lepton 3-jet) \\
	0-to-1 lepton ratio & 8\% \\
	2-to-3-jet ratio & 9\% (0+1 lepton only) \\
	$W$~+~HF CR to SR ratio & 25\% \\
	\mbb, \ptv & S \\
	\hline\hline
	\multicolumn{2}{c}{Single top quark}\\	
	\hline
	Cross-section & 4.6\% ($s$-channel), 4.4\% ($t$-channel), 6.2\% ($Wt$) \\
	Acceptance 2-jet & 17\% ($t$-channel), 35\% ($Wt$) \\
	Acceptance 3-jet & 20\% ($t$-channel), 41\% ($Wt$) \\
        \mbb, \ptv & S ($t$-channel, $Wt$) \\
	\hline\hline
	\multicolumn{2}{c}{Multi-jet (1-lepton)}\\	
	\hline
		Normalisation & 60 -- 100\% (2-jet), 100 -- 400\% (3-jet) \\
		BDT template & S \\
	\hline\hline
\end{tabular}
	\end{center}
	\end{small}
	\end{table}

\begin{table}[ht!]
\begin{small}
\begin{center}
\caption
	{Summary of the systematic uncertainties in the background modelling for diboson production. 
	``PS/UE'' indicates parton shower~/ underlying event.
    An ``S'' symbol is used when only a shape uncertainty is assessed.
    When determining the $(W/Z)Z$ diboson production signal strength, 
    the normalisation uncertainties in $ZZ$ and $WZ$ production are removed. 
	\label{tab:bkgdib_systematics}}
\begin{tabular}{ l | c}
	\hline\hline
	\multicolumn{2}{c}{$ZZ$}\\	
	\hline
        Normalisation & 20\% \\
        0-to-2 lepton ratio & 6\% \\
       Acceptance from scale variations (var.) & 10 -- 18\% (Stewart--Tackmann jet binning method)\\
       Acceptance from PS/UE var. for 2 or more jets & 5.6\% (0-lepton), 5.8\% (2-lepton)\\
        Acceptance from PS/UE var. for  3  jets & 7.3\% (0-lepton), 3.1\% (2-lepton)\\
       	\mbb, \ptv,  from  scale var.& S (correlated with $WZ$ uncertainties)\\
       	\mbb, \ptv, from PS/UE var.& S (correlated with $WZ$ uncertainties)\\
       	\mbb,  from matrix-element var.& S (correlated with $WZ$ uncertainties)\\
    \hline\hline
	\multicolumn{2}{c}{$WZ$}\\	
	\hline
        Normalisation & 26\% \\
        0-to-1 lepton ratio & 11\% \\
        Acceptance from scale var. & 13 -- 21\%   (Stewart--Tackmann jet binning method)\\
        Acceptance from PS/UE var. for 2 or more jets & 3.9\%\\
        Acceptance from PS/UE var. for  3  jets & 11\% \\
       	\mbb, \ptv,  from  scale var.& S (correlated with $ZZ$ uncertainties)\\
       	\mbb, \ptv, from PS/UE var.& S (correlated with $ZZ$ uncertainties)\\
       	\mbb,  from matrix-element var.& S (correlated with $ZZ$ uncertainties)\\
	\hline\hline
	\multicolumn{2}{c}{$WW$} \\	
	\hline
    	Normalisation & 25\% \\
	\hline\hline
\end{tabular}
	\end{center}
	\end{small}
	\end{table}

\textbf{\vjets\ production} The \vjets\ backgrounds are subdivided into three different components based upon the jet flavour labels of the two $b$-tagged jets in the event. The main background contributions ($V+bb$, $V+bc$, $V+bl$ and $V+cc$) are jointly considered as the \vhf\ background. Their overall normalisation, separately in the 2- and 3-jet categories, is free to float in the global likelihood fit, as detailed in Section~\ref{sec:fit}. The remaining flavour components, \vcl\ and \vll, make up less than $\sim1$\% of the background in each analysis region, so only uncertainties in the normalisation of these backgrounds are included. 

Acceptance uncertainties are estimated for the relative normalisations of the different regions that share a common floating normalisation parameter. In the case of the \whf\ background, this includes the uncertainties in the ratio of the event yield in the 0-lepton channel to that in the 1-lepton channel and, in the 1-lepton channel, in the ratio of the event yield in the \whf\ control region to that in the signal region. For the \zhf\ background, there is an uncertainty in the ratio of the event yield in the 0-lepton channel to that in the 2-lepton channel. These ratio uncertainties act as effective extrapolation uncertainties from one region to another. 

Uncertainties are also estimated in the relative normalisation of the four heavy-flavour components that make up the \vhf\ background. These are taken as uncertainties in the $bc$, $cc$ and $bl$ yields compared to the dominant $bb$ yield and are estimated separately for the 0- and 1-lepton channels in the case of \whf\ and separately for the 0-lepton, 2-lepton 2-jet and 2-lepton 3-jet regions in the case of \zhf.

The normalisation and acceptance uncertainties are all calculated by adding the differences between the nominal \textsc{Sherpa 2.2.1} sample and its associated systematic variations in quadrature, including a variation of (i) the renormalisation scale by factors of 0.5 and 2; (ii) the factorisation scale by factors of 0.5 and 2; (iii) the CKKW merging scale from 30~\GeV\ to 15~\GeV; (iv) the parton-shower/resummation scale by factors of 0.5 and 2. In addition, the difference between the \textsc{Sherpa 2.2.1} nominal sample and an alternative sample produced with a different matrix-element generator is added in quadrature to the rest to yield the total uncertainty. The alternative sample is produced with \textsc{Madgraph5}\_a\textsc{MC@NLO~v2.2.2}~\cite{Alwall:2011uj}, with up to four extra partons at LO, and interfaced to \textsc{Pythia 8.212}; the A14 tune is used together with the \textsc{NNPDF2.3LO} PDF set.
 
Uncertainties in the shapes of the \mbb\ and \ptv\ distributions are estimated for \zhf\ by comparing the \zjets\ background to data in signal-depleted regions with a very high \zjets\ purity, specifically the 1- and 2-tag regions of the 2-lepton channel, with the \mbb\ region around the Higgs boson mass excluded in the 2-tag case. In order to remove most of the residual $t\bar{t}$ contamination, a selection requirement is made on $\met / \sqrt{S_{\mathrm{T}}}<3.5$~$\sqrt{\mathrm{\GeV}}$ as done for the dijet-mass analysis. 

For the \whf\ background, due to the limited number of events in the dedicated control region, shape uncertainties are based on the same systematic uncertainty sources as for the normalisation and acceptance uncertainties; in all event categories, since scale variations are found to have a minor effect on the shapes of the distributions, the systematic uncertainties are dominated by the comparison of the nominal \textsc{Sherpa 2.2.1} sample with \textsc{Madgraph5}\_a\textsc{MC@NLO~v2.2.2}.

\textbf{\ttb\ production} Uncertainties are derived from comparing the nominal sample (\textsc{Powheg+Pythia8}) to alternative samples with different parton-shower generation (\textsc{Powheg+Herwig7}~\autocite{Bahr:2008pv,Bellm:2015jjp}), matrix-element generation (\textsc{Madgraph5}\_a\textsc{MC@NLO+Pythia8}) and settings of the nominal generator designed to increase or decrease the amount of radiation.
Due to the significantly different regions of phase space probed, the \ttb\ background in the 0- and 1-lepton channels (jointly referred to as 0+1 lepton in the following) is considered independently from the \ttb\ background in the 2-lepton channel; different overall floating normalisation factors are considered, and acceptance uncertainties are derived separately and taken as uncorrelated between the 0+1 and 2-lepton channels. For the 0+1 lepton channels, uncertainties are considered in the normalisation ratios of the 3-jet and 2-jet regions, the $W$~+~HF control region and signal region, and the 1-lepton and 0-lepton channels. These uncertainties are estimated by comparing the difference between the ratios of the yields in the two regions under consideration in the alternative \ttb\ samples and those measured in the nominal sample. The differences between the nominal and each of the alternative samples are summed in quadrature to provide an overall uncertainty. For the 2-lepton channel, the normalisations in the 2- and 3-jet regions are both left floating, and are effectively determined in their respective $e\mu$ control regions. 
Uncertainties in the shapes of the \ptv\ and \mbb\ distributions are estimated in the 0+1 and 2-lepton channels separately. The difference between the nominal sample and \textsc{Madgraph5}\_a\textsc{MC@NLO} provides by far the largest variation, and is therefore considered as a systematic uncertainty in the shapes of these distributions.

\textbf{Single top quark production} In the $Wt$ and $t$-channels, uncertainties are derived in the normalisation, acceptance and shapes of the \mbb\ and \ptv\ distributions. The $s$-channel only has a normalisation uncertainty derived as its contribution is negligible overall. \\
For the $t$-channel, the nominal samples (\textsc{Powheg+Pythia6}) are compared to alternative samples, which are similar to those used in the \ttb\ case using different parton-shower generation (\textsc{Powheg+Herwig++}), and matrix-element generation (\textsc{Madgraph5}\_a\textsc{MC@NLO+Herwig++}).
For the $Wt$ channel, uncertainties related to  the interference between the $Wt$ and $t\bar{t}$ production processes are assessed by using a diagram subtraction scheme instead of the nominal diagram removal scheme~\autocite{Frixione:2008yi,Re:2010bp}. For both the $t$- and $Wt$-channels, the settings of the nominal generator are varied so as to maximise or minimise the amount of radiation. The normalisation uncertainties take into account variations of the renormalisation and factorisation scales, $\alphas$ and PDFs. Uncertainties in the acceptance in both the 2- and 3-jet regions are derived by comparing the alternative generators and summing the differences with respect to the nominal sample in quadrature. Shape uncertainties are derived for the \mbb\ and \ptv\ distributions. These uncertainties cover all the differences in the shapes of the kinematic distributions investigated by comparing nominal and alternative samples.	

\textbf{Diboson production} The diboson backgrounds are composed of three distinct processes, $WZ$, $WW$ and $ZZ$ production. Given the small contribution from $WW$ production ($<0.1\%$ of the total background) only a normalisation uncertainty is assigned. The more important contributions from the $WZ$ and $ZZ$ backgrounds have uncertainties derived for the overall normalisation, the relative acceptance between regions and for the \mbb\ and \ptv\ shapes. Uncertainties are derived by comparing the nominal sample (\textsc{Sherpa 2.2.1}) to the alternative samples with varied factorisation, renormalisation and resummation scales, and using the Stewart--Tackmann method~\autocite{Stewart:2011cf} to estimate scale variation uncertainties for the acceptance in the jet multiplicity categories. 
Additional uncertainties in the overall acceptance, in the relative acceptance across jet multiplicities and in the shape of the \mbb\ and \ptv\ distributions are estimated in the parton-shower and underlying-event model.
These are estimated by considering the difference between  \textsc{Powheg+Pythia8} and \textsc{Powheg+Herwig++}, as well as changes in the \textsc{Pythia8} parton-shower tune. The envelope of the two effects is considered to define these uncertainties.
A systematic uncertainty in the shape of the \mbb\ distribution 
 results from the comparison of \textsc{Sherpa 2.2.1} and \textsc{Powheg+Pythia8}.
This changes the shape of the \mbb\ distribution for values in the range 100 -- 130~\GeV\ by 10 -- 20\%. Acceptance uncertainties are derived for the ratio of 0-to-1 lepton channels and for the ratio of the 2-to-3 jet regions for $WZ$ production. In the $ZZ$ production case the acceptance uncertainties are derived for the ratio of the 0-to-2 lepton channels and of the 2-to-3 jet regions. Uncertainties in the acceptance and \mbb\ or \ptv\ shapes of the diboson background due to PDF and $\alphas$ variations were found to have a negligible impact.
 
\subsection{Multi-jet background uncertainties}
The multi-jet background in the 1-lepton channel is estimated from data as outlined in Section~\ref{sec:multijet}. 
Systematic uncertainties can have an impact on the multi-jet estimates in two ways: either changing the \mtw\ distributions used in the multi-jet template fits, therefore impacting the extracted multi-jet normalisations, or directly changing the multi-jet BDT distributions used in the global likelihood fit. Several sources of uncertainty are considered, uncorrelated between the electron and muon sub-channels. The respective variations are added in quadrature for the normalisations, or considered as separate shape uncertainties.  The variations are obtained by changing the definition of the multi-jet control region (2 versus 1 $b$-tag, more stringent isolation requirements, a different single-electron trigger to probe a potential trigger bias in the isolation requirements);  removing the bias correction that makes use of the \pt-, $\eta$-, and \met- dependent extrapolation factors derived in the (1-jet, 0 $b$-tag) region; varying the normalisation of the contamination from the top ($t\bar{t}$ and $Wt$) and \vjets\ processes in the multi-jet control region. In addition, the following sources of uncertainty that only have an impact on the multi-jet normalisation, are considered: use of the \met\ variable instead of \mtw\ for the multi-jet template fit and, for the electron sub-channel only, the inclusion of the $\met<30$~\GeV\ region, which significantly enhances the multi-jet contribution in the template fit. 

 \subsection{Signal uncertainties} 
The signal samples are normalised using their inclusive cross-sections, as described in Section~\ref{sec:samples}, and an additional scale factor computed using the \textsc{Hawk} generator is applied as a function of \ptv\ to correct for the sizeable impact of the NLO (EW) corrections to the \ptv\ distributions. The systematic uncertainties that affect the modelling of the signal are summarised in Table~\ref{tab:sig_systematics}.

Uncertainties in the calculations of the $VH$ production cross-sections and the $H\to\bbbar$ branching ratio are assigned following the recommendations of the LHC Higgs Cross Section working group \cite{Dittmaier:2011ti,Dittmaier:2012vm,Heinemeyer:2013tqa,Harlander:2013mla,Brein:2012ne}. 
The uncertainties in the overall $VH$  production cross-section from missing higher-order terms in the QCD perturbative expansion are obtained by varying the renormalisation scale $\mu_{\mathrm{R}}$ and factorisation scale $\mu_{\mathrm{F}}$ independently, from $1/3$ to $3$ times their original value. The PDF+$\alphas$ uncertainty in the overall $VH$ production cross-section is calculated from the 68\% CL interval using the PDF4LHC15\_nnlo\_mc PDF set. The latest recommendations of the LHC Higgs working group~\cite{deFlorian:2016spz} do not distinguish between uncertainties in $qq\to ZH$ production and $gg\to ZH$ production. To obtain the scale uncertainties separately for these two processes, it is assumed that the uncertainty in $qq\to ZH$ production is identical to the  uncertainty in $WH$ production. The $gg\to ZH$ production uncertainty is then derived such that the sum in quadrature of the $qq\to ZH$ and $gg\to ZH$ production uncertainties (considering their respective production cross-sections) amount to the scale uncertainty in the overall $ZH$ production. 
Since the PDF+$\alphas$ uncertainty is larger for $WH$ production than $ZH$ production, the method used for the scale uncertainty cannot be used for this uncertainty. 
The PDF+$\alphas$ uncertainty in the $gg\to ZH$ production is taken from previous recommendations~\cite{Heinemeyer:2013tqa} and the uncertainty in the $qq\to ZH$ production is taken from the latest recommendation~\cite{deFlorian:2016spz}.

Another systematic uncertainty in the overall $VH$ cross-section originates from 
missing higher-order electroweak corrections. 
This is estimated as the maximum variation among three quantities: the maximum size expected for the missing NNLO EW effects (1\%), the size of the NLO EW correction and the uncertainty in the photon-induced cross-section relative to the total $(W/Z)H$ cross-section described in Section \ref{sec:samples}.
The systematic uncertainty in the $H\to b\bar{b}$ branching ratio is 1.7\%~\autocite{Djouadi:1997yw}. This uncertainty takes into account missing higher-order QCD and EW corrections as well as uncertainties in the $b$-quark mass and in the value of $\alphas$.

Acceptance and shape systematic uncertainties are derived to account for missing higher-order QCD and EW corrections, for PDF+$\alphas$ uncertainty, and for variations of the parton-shower and underlying-event models.   
Uncertainties in the acceptance and in the shape of the \mbb\ and \ptv\ distributions, originating from missing higher-order terms in QCD, are estimated by comparing the nominal samples to those generated with weights corresponding to varied factorisation and renormalisation scales applied. The Stewart--Tackmann method is used to assign scale variation uncertainties in the acceptance in the jet multiplicity categories.
Uncertainties due to the parton-shower and underlying-event models are estimated by considering the difference between  \textsc{Powheg+MiNLO+Pythia8} and \textsc{Powheg+MiNLO+Herwig7}, as well as changes in the \textsc{Pythia8} parton-shower tune. The latter effect is assessed in events generated with \textsc{Madgraph5}\_a\textsc{MC@NLO} and showered with \textsc{Pythia8}, using the A14 tune and its variations. The envelope of the two effects is considered to define uncertainties separately in the overall acceptance, in the relative acceptance across jet multiplicities and in the shape of the \mbb\ and \ptv\ distributions.
The PDF+$\alphas$ uncertainty in the acceptance between regions and in the \mbb\ and \ptv\ shapes is estimated applying the PDF4LHC15\_30 PDF set and its uncertainties, according to the PDF4LHC recommendations~\autocite{Butterworth:2015oua}. 

\begin{table}[ht!]
\begin{small}
\begin{center}
	\caption
	{Summary of the systematic uncertainties in the signal modelling. ``PS/UE'' indicates parton shower~/ underlying event.
    An ``S'' symbol is used when only a shape uncertainty is assessed.
	\label{tab:sig_systematics}}
\begin{tabular}{ l | c}
	\hline\hline
    \multicolumn{2}{c}{Signal}\\
    \hline
    Cross-section (scale) & 0.7\% ($qq$), 27\% ($gg$) \\
    Cross-section (PDF) & 1.9\% ($qq\rightarrow WH$), 1.6\% ($qq\rightarrow ZH$), 5\% ($gg$) \\
    Branching ratio & 1.7 \% \\
    Acceptance from scale variations (var.) & 2.5 -- 8.8\%   (Stewart--Tackmann jet binning method)\\
    Acceptance from PS/UE var. for 2 or more jets & 10 -- 14\% (depending on lepton channel)\\
    Acceptance from PS/UE var. for  3  jets & 13\% \\
    Acceptance from PDF+$\alphas$ var. & 0.5 -- 1.3\% \\
    \mbb, \ptv, from scale var.& S \\
    \mbb, \ptv, from PS/UE var.& S \\
    \mbb, \ptv, from PDF+$\alphas$ var. & S \\ 
    \ptv\ from NLO EW correction & S \\
	\hline\hline
\end{tabular}
	\end{center}
	\end{small}
	\end{table}


\section{Statistical analysis}
\label{sec:fit}
\subsection{Analysis of the 13 \TeV\ data}
\label{sec:statAnalysis13TeV}

A statistical fitting procedure 
based on the Roostats framework~\cite{Moneta:2010pm,Verkerke:2003ir} 
is used to extract the strength of the Higgs boson signal from the data.

The signal strength is a parameter, $\mu$, that multiplies the SM
Higgs boson production cross-section times branching ratio into \bb.
A binned likelihood function is constructed as the product of
Poisson probability terms over the bins of the input distributions
involving the numbers of data events and the expected
signal and background yields, taking into account the effects of the floating 
background normalisations and the systematic uncertainties.

\begin{table}[htbp]
\begin{center}
\caption[Regions used in likelihood fit]
{The distributions used in the global likelihood fit for the signal regions (SR) and  
control regions (CR) for all the categories in each channel, for the nominal multivariate analysis. 
\label{tab:13tevregions}}
\advance\leftskip -0.2cm
\begin{tabular}{   l | c | c | c | c | c  }
\hline\hline
\multirow{4}{*}{Channel} & \multirow{4}{*}{SR/CR} & \multicolumn{4}{c}{Categories} \\
  \cline{3-6}
 & & \multicolumn{2}{c|}{$75$~\GeV~$<\ptv < 150$~\GeV}  & \multicolumn{2}{c}{$\ptv > 150$~\GeV} \\
 \cline{3-6}
 & & 2 jets & 3 jets & 2 jets & 3 jets \\
\hline \hline
0-lepton & SR &  - & - & BDT & BDT  \\ 
1-lepton & SR &  - & - & BDT & BDT  \\ 
2-lepton & SR & BDT  & BDT & BDT & BDT \\ 
\hline
1-lepton & $W$~+~HF CR & - & - & Yield & Yield \\ 
2-lepton & $e\mu$ CR & \mbb & \mbb & Yield &\mbb \\ 
\hline\hline
\end{tabular}
\end{center}
\end{table}

The different regions entering the likelihood fit are summarised in Table~\ref{tab:13tevregions}.
The primary inputs to the 
global fit are the BDT$_{VH}$ discriminants in the
eight 2-$b$-tag signal regions 
defined by the three lepton channels, up to two \ptv\ intervals and the two jet multiplicity categories. Additional inputs are the event yields in the two $W$~+~HF control regions in the 1-lepton channel subdivided into the two number-of-jet categories, and the \mbb\ distributions or the event yields for the four  $e\mu$ control regions defined by the two \ptv\ intervals and the two number-of-jet categories.
The electron and muon sub-channels are combined in the fit.
Altogether, there are 141 bins in the 14 regions used in the global fit.
In addition to the global fit with all channels combined, separate 0-, 1- and 2-lepton channel fits are 
performed, where only the analysis regions specific to a single channel are considered and 
a channel-specific signal strength is obtained.

The effect of systematic uncertainties in the signal and background predictions
is described by nuisance parameters (NPs), $\vec{\theta}$, which
are constrained by Gaussian or log-normal probability density functions, the latter being 
used for normalisation uncertainties to prevent
normalisation factors from becoming negative in the fit.
The expected
numbers of signal and background events in each bin are functions of $\mu$ and $\vec{\theta}$.
For each NP, the prior is added as a penalty term to the likelihood, 
$\mathcal{L} (\mu,\vec{\theta})$, which decreases
as soon as the nuisance parameter $\theta$ is shifted away from its nominal value. 
The statistical uncertainties of background predictions from simulation 
are included through one nuisance parameter per bin, using the Beeston--Barlow 
technique~\autocite{Barlow:1993dm}.

The test statistic $q_\mu$ is constructed from 
the profile likelihood ratio 
$$q_\mu = - 2 \ln \Lambda_\mu \mathrm{~~with~~} \Lambda_\mu =
\mathcal{L} (\mu,\hat{\hat{\vec{\theta}}}_\mu)/\mathcal{L} (\hat{\mu},\hat{\vec{\theta}}),$$ 
where $\hat{\mu}$ and $\hat{\vec{\theta}}$ are the parameters that maximise the
likelihood, and 
$\hat{\hat{\vec{\theta}}}_\mu$ are the nuisance parameter values that maximise the
likelihood for a given $\mu$.
To measure the compatibility of the background-only hypothesis with the observed
data, the test statistic used is 
$q_0 = -2\ln\Lambda_0$.
The results are presented in terms of the probability $p_0$ of the background-only hypothesis,
and the best-fit signal strength value $\hat\mu$ with its associated uncertainty $\sigma_\mu$.
The fitted $\hat\mu$ value is obtained  by maximising the likelihood function with respect to all parameters. 
The uncertainty $\sigma_\mu$ is obtained from the variation of $q_\mu$ by one unit.
Expected results are obtained in the same way as the observed results by replacing the data
in each input bin by the prediction from simulation with all NPs set to their best-fit values,
as obtained from the fit to the data, except for the signal strength parameter, which is kept at its 
nominal value.

The data have sufficient statistical power to constrain the largest background normalisation NPs, 
which are left free to be determined in the fit without having priors. This applies to the \ttb, \whf\ and \zhf\ processes.
The corresponding normalisation factors expressed with respect to 
their expected nominal value 
and resulting from the global fit to the 13~\TeV\ data, are shown in Table~\ref{tab:resultscalefactorsMVA}. 
As stated in Section~\ref{sec:syst}, the \ttb\ background is normalised independently for the 
2-lepton channel and for the 0- and 1-lepton channels. In the 2-lepton channel, the \ttb\
background is almost entirely due to events in which both top quarks decay into $(W\to\ell\nu)b$ 
(dileptonic decays) with all final-state objects detected (apart from the neutrinos). 
In the 0- and 1-lepton channels, it is in part due to dileptonic decays with one or two of the leptons 
(often a $\tau$-lepton) undetected, and in part due to cases where one of the top quarks decays into 
$(W\to q\overline{q}')b$ (semileptonic decays) with at least one undetected light- or $c$-quark jet. 
Furthermore, the \ptv\ range probed is different in the 0- and 1-lepton channels: 
$\ptv>150$~\GeV\ in contrast to $\ptv>75$~\GeV\ in the 2-lepton channel. 
For the \zhf\ and \whf\ backgrounds, the data have enough statistical power to constrain the normalisations in 
the 2-jet and in the 3-jet categories independently. The normalisation factors for these backgrounds 
can deviate significantly from one due to the large theoretical uncertainty in the cross-sections 
of the contributing processes. 

\begin{table}[tb!]
\begin{center}
\caption{Factors applied to 
the nominal normalisations of the \ttb, \whf\, and \zhf\ backgrounds,
as obtained from the global fit to the 13~\TeV\ data for the nominal 
multivariate analysis, used to extract the Higgs boson signal.
The errors include the statistical and systematic uncertainties. 
\label{tab:resultscalefactorsMVA}}
\begin{tabular}{l|c}
\hline\hline
Process & Normalisation factor \\
\hline
\ttb\ 0- and 1-lepton  & $0.90 \pm 0.08$ \\ 
\ttb\ 2-lepton 2-jet  & $0.97 \pm 0.09$ \\ 
\ttb\ 2-lepton 3-jet  & $1.04 \pm 0.06$ \\ 
\whf\ 2-jet &  $1.22 \pm 0.14$ \\
\whf\ 3-jet &  $1.27 \pm 0.14$ \\
\zhf\ 2-jet &  $1.30 \pm 0.10$ \\
\zhf\ 3-jet &  $1.22 \pm 0.09$ \\
\hline\hline
\end{tabular}
\end{center}
\end{table}

The systematic uncertainties are encoded in variations of the nominal BDT$_{VH}$ 
or \mbb\ templates, and of the nominal yields across analysis categories, 
for each up-and-down ($\pm 1\sigma$) variation.
The limited size of the MC samples for some simulated background processes in some regions can cause large local fluctuations in templates of systematic variations.
When the impact of a systematic variation translates into a reweighting of the nominal 
template, no statistical fluctuations are expected beyond those already present in the
nominal template. This is the case, for instance, for the $b$-tagging uncertainties. For those, no specific action is taken.
On the other hand, when a systematic variation may introduce changes in the events selected, 
as is the case for instance with the JES uncertainties, additional statistical fluctuations
may be introduced, which affect the templates of systematic variations.
In such cases, a smoothing procedure is applied to each systematic-variation template
in each region. 
To reduce the complexity of the fit, systematic uncertainties that have a negligible impact on the final 
results are pruned away, region by region. Studies were performed to verify that the smoothing 
and pruning procedures do not induce any bias in the result. More details about the smoothing and pruning procedures 
can be found in Ref.~\autocite{HIGG-2013-23}.

In order to understand the effect of systematic uncertainties on the final results, the breakdown of the contributions to the uncertainties in $\hat\mu$ is reported in Table~\ref{tab:syst_unc}. The individual sources of systematic uncertainty detailed in Section~\ref{sec:syst} are combined into categories. To assess the contribution of a category to the total systematic uncertainty, all NPs associated with the uncertainties within the category are fixed to their fitted values and the fit is repeated. The difference in quadrature between the uncertainties for $\hat\mu$ from this fit and from the nominal fit 
provides an estimate of the systematic uncertainty attached to the considered category of uncertainties.
As shown in the table, the systematic uncertainties for the modelling of the signal play a dominant role, 
followed by the uncertainty due to the limited size of the simulated samples, the modelling of the backgrounds and the $b$-jet tagging uncertainty.

\begin{table}[tb]
\caption{Breakdown of the contributions to the uncertainties in $\hat\mu$. 
The sum in quadrature of the systematic uncertainties attached to the 
categories differs from the total systematic uncertainty due to correlations. 
The $b$-tagging extrapolation uncertainty refers to the extrapolation of the $b$-jet 
calibration above $\pt=300$~\GeV.
\label{tab:syst_unc}}
\begin{center}
\begin{tabular}{ l | l  c }
\hline\hline
\multicolumn{2}{l}{Source of uncertainty} & $\sigma_\mu$ \\
\hline
\multicolumn{2}{l}{Total} & 0.39 \\
\multicolumn{2}{l}{Statistical} & 0.24 \\
\multicolumn{2}{l}{Systematic} & 0.31 \\
\hline
\multicolumn{3}{l}{Experimental uncertainties}\\
\hline
\multicolumn{2}{l}{Jets} & 0.03 \\
\multicolumn{2}{l}{\met} & 0.03 \\
\multicolumn{2}{l}{Leptons} & 0.01 \\
\multicolumn{3}{l}{} \\
\multirow{3}{*}{$b$-tagging~~~} & $b$-jets & 0.09 \\
& $c$-jets & 0.04 \\
& light jets & 0.04 \\
& extrapolation & 0.01 \\
\multicolumn{3}{l}{} \\
\multicolumn{2}{l}{Pile-up} & 0.01 \\
\multicolumn{2}{l}{Luminosity} & 0.04 \\
\hline
\multicolumn{3}{l}{Theoretical and modelling uncertainties}\\
\hline
\multicolumn{2}{l}{Signal} & 0.17 \\
\multicolumn{3}{l}{} \\
\multicolumn{2}{l}{Floating normalisations} & 0.07 \\
\multicolumn{2}{l}{$Z$~+~jets} & 0.07 \\
\multicolumn{2}{l}{$W$~+~jets} & 0.07 \\
\multicolumn{2}{l}{\ttb} & 0.07 \\
\multicolumn{2}{l}{Single top quark} & 0.08 \\
\multicolumn{2}{l}{Diboson} & 0.02 \\
\multicolumn{2}{l}{Multijet} & 0.02 \\
\multicolumn{3}{l}{} \\
\multicolumn{2}{l}{MC statistical} & 0.13 \\
\hline\hline
\end{tabular}
\end{center}
\end{table}

\subsection{Dijet-mass analysis}

In the dijet-mass analysis, the BDT$_{VH}$ discriminant is replaced by the \mbb\ variable 
as the main input used in the global fit, and the number of signal regions is increased from 
eight to fourteen, as a consequence of splitting the event categories with \ptv$>150$~\GeV\ 
in two 
in each of the three lepton channels. 
The different regions entering the likelihood fit are summarised in Table~\ref{tab:13tevregions_cutbased}.
Altogether, for the dijet-mass analysis, there are 283 bins in the 18 regions used in the global fit. 
\begin{table}[h!tbp]
\begin{center}
\caption[Regions used in likelihood fit for the dijet-mass analysis]
{The distributions used in the global likelihood fit for the dijet-mass analysis, 
for the signal regions (SR) and control regions (CR), for all the categories in each channel. 
The two regions marked with~$^*$ ($^\dagger$) are merged into a single region, to reduce statistical uncertainties.
\label{tab:13tevregions_cutbased}}
\advance\leftskip -0.2cm
\begin{tabular}{   l | c | c | c | c | c | c | c }
\hline\hline
\multirow{4}{*}{Channel} & \multirow{4}{*}{SR/CR} & \multicolumn{6}{c}{Categories} \\
  \cline{3-8}
 & & \multicolumn{2}{c|}{$75$~\GeV$<\ptv<$150~\GeV}  & \multicolumn{2}{c|}{$150$~\GeV$< \ptv <$200~\GeV} & \multicolumn{2}{c}{$\ptv > 200$~\GeV} \\
 \cline{3-8}
 & & \ \ \ \ \ 2 jets\ \ \ \ \  & 3 jets & \ \ \ \ \ 2 jets\ \ \ \ \  & 3 jets & 2 jets & 3 jets\\
\hline \hline
0-lepton & SR &  -       & -        & \mbb & \mbb & \mbb & \mbb  \\ 
1-lepton & SR plus $W$~+~HF CR &  -       & -        & \mbb & \mbb & \mbb & \mbb  \\ 
2-lepton & SR & \mbb  & \mbb & \mbb & \mbb & \mbb & \mbb \\ 
\hline
2 lepton & $e\mu$ CR & \mbb & \mbb & Yield$^*$ &\mbb$^\dagger$ & Yield$^*$ & \mbb$^\dagger$ \\ 
\hline\hline
\end{tabular}
\end{center}
\end{table}

\subsection{Diboson analysis}

The diboson analysis targets diboson production with a $Z$ boson decaying into a pair of $b$-quarks
and produced in association with either a $W$ or $Z$ boson. This process has a signature
that is similar to the one considered in this analysis,
and therefore provides an important validation of the $VH$ result. 
The cross-section is about nine times 
larger than for the SM Higgs boson with a mass of 125~\GeV, 
the \mbb\ distribution peaks at lower values, and 
the \ptbb\ spectrum is softer. 
The multivariate discriminant BDT$_{VZ}$ is used to extract the diboson 
signal.
In the diboson-analysis fits, 
the normalisation of the diboson contributions 
is allowed to vary with a multiplicative scale factor $\mu_{VZ}$ 
with respect to the SM prediction, 
except for the small contribution from $WW$ production,
which is treated as a background and constrained within its uncertainty. 
The overall normalisation uncertainties for the $WZ$ and $ZZ$ processes 
are removed, while 
all other systematic uncertainties are kept identical to those in the nominal 
fit used to extract the Higgs boson signal.
A SM Higgs boson with $m_H = 125$~\GeV\ is included as a background, with
a production cross-section at the SM value with an uncertainty of $50$\%.
The diboson and Higgs boson BDTs provide sufficient separation between the 
$VZ$ and $VH$ processes that they only have a weak direct correlation ($<1\%$) 
in their results.

\subsection{Combination with Run 1 data}

The statistical analysis of the 13 \TeV\ data is combined with the results of the data recorded at 7~\TeV\ and 8~\TeV, reported in Ref.~\autocite{HIGG-2013-23}.  
No change is implemented in the analysis of the 7~\TeV\ and 8~\TeV\ data, but several studies were carried out on the correlation and 
compatibility of the 13~\TeV\ results and the 7~\TeV\ and 8~\TeV\ results.
Studies on the correlation of the experimental systematic uncertainties between the 7~\TeV, 8~\TeV\ and 13~\TeV\ analyses were performed for the dominant uncertainties. 

The changes in the detector layout (inclusion of the IBL), in the tagging discriminating variable, in the used working points, in the $b$-tagging calibration analyses, and in the way the discriminating variable is used in the analysis support the choice of assuming a negligible correlation in the experimental systematic uncertainties affecting the $b$-tagging across datasets. Nevertheless, even correlating the leading systematic uncertainties for the $b$-jet efficiencies measured in data affects the combined measurement of $\mu$ by less than $5\%$, and has a negligible impact on its uncertainty. 
Different correlation schemes for the jet energy scale uncertainties were tested, with no significant impact on the combined result observed: a \textit{weak} correlation scheme was finally adopted, where only the $b$-jet-specific jet energy scale uncertainty is correlated across the 7~\TeV, 8~\TeV\ and 13~\TeV\ analyses.

Studying the impact of potential correlations in the modelling of the background processes is difficult, due to the changes in centre-of-mass energy, Monte Carlo generators, object and event selection, and in the software tools used for simulation, reconstruction and analysis. However, the potential impact of 
underestimating or omitting correlations is limited by the fact that each of these modelling systematic uncertainties only constitutes a fraction of the total uncertainty, and, furthermore, that this fraction in most cases varies with the centre-of-mass energy following variations in cross-section and acceptance. To evaluate the maximum potential effect of these correlations, a $\chi^2$-combination of the two measurements, the signal strengths from the Run~1 and Run~2 datasets, was performed and studied as a function of different linear correlation coefficients, expressing the degree of correlation between the two measurements. These coefficients were chosen to correspond to different correlation schemes, from uncorrelated to fully correlated, between the \ttbar, \zhf, and \whf\ normalisations and systematic shape variations across the two datasets, and they were computed based on the assumed correlation and the relative contribution of a specific uncertainty to the total uncertainty for $\mu$. In all cases considered, the impact on the combined signal strength was found to be smaller than 1\%, while the effect on the signal strength uncertainty was found to be smaller than 4\%. 

As a result of these studies, among the experimental uncertainties, only the $b$-jet-specific jet energy scale uncertainty is correlated across the 7~\TeV, 8~\TeV\ and 13 \TeV\ datasets for the combined results. For the Higgs boson signal, theory uncertainties in the overall cross-section, in the $H\to\bbbar$ branching ratio and in the \ptv-dependent NLO EW corrections, are correlated across the different datasets.


\section{Results}
\label{sec:result}
The results of the Higgs boson search and diboson analysis are reported below. In the following the fitted signal strength parameters are denoted $\mu$ and $\mu_{VZ}$ rather than $\hat\mu$ and $\hat\mu_{VZ}$.

\subsection{Results of the SM Higgs boson search at $\sqrt{s}=13$~\TeV}

\begin{figure}[hp]
  \centering
  \begin{tabular}{cccc}
    \includegraphics[width=0.4\linewidth]{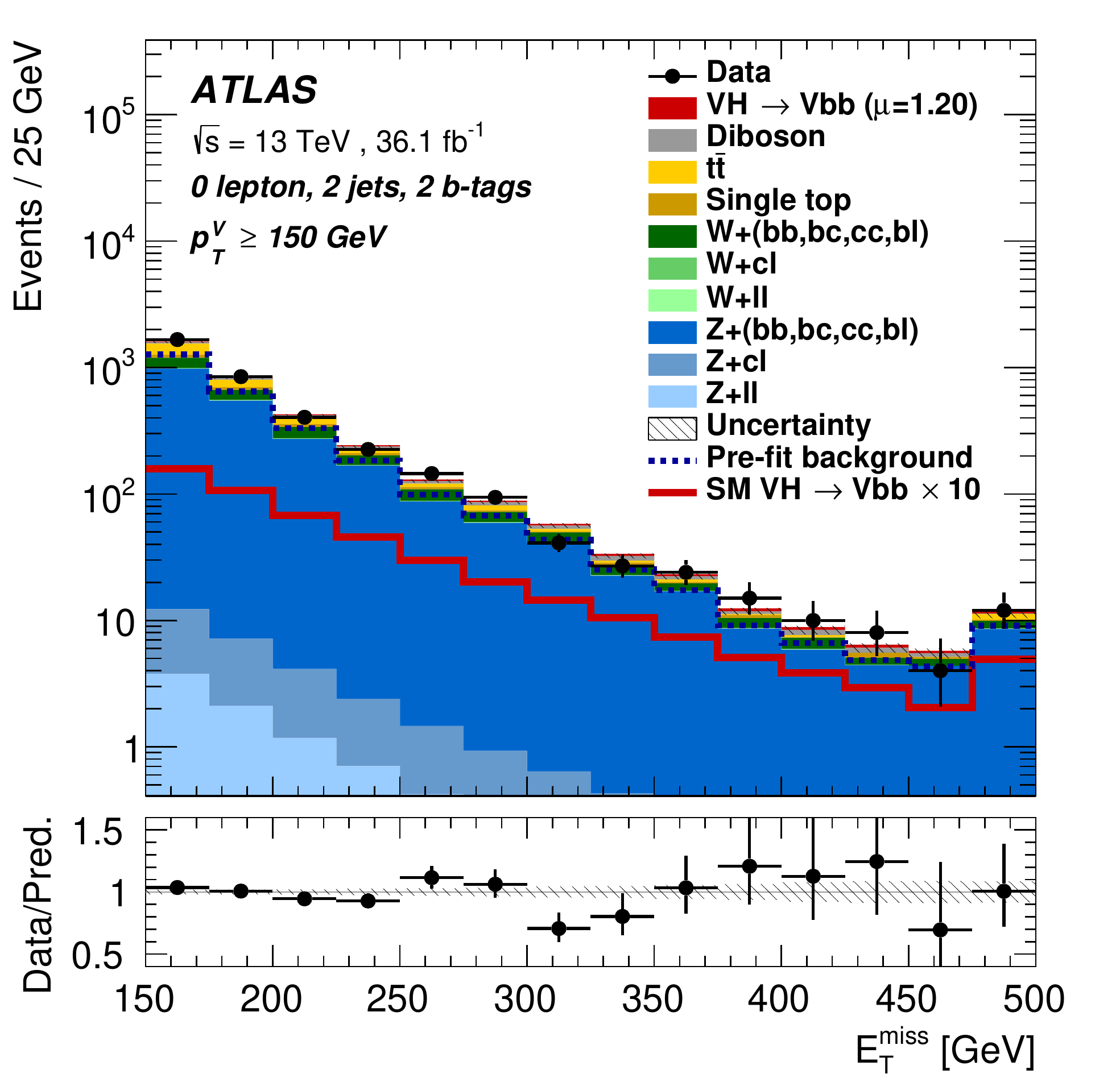}
    \includegraphics[width=0.4\linewidth]{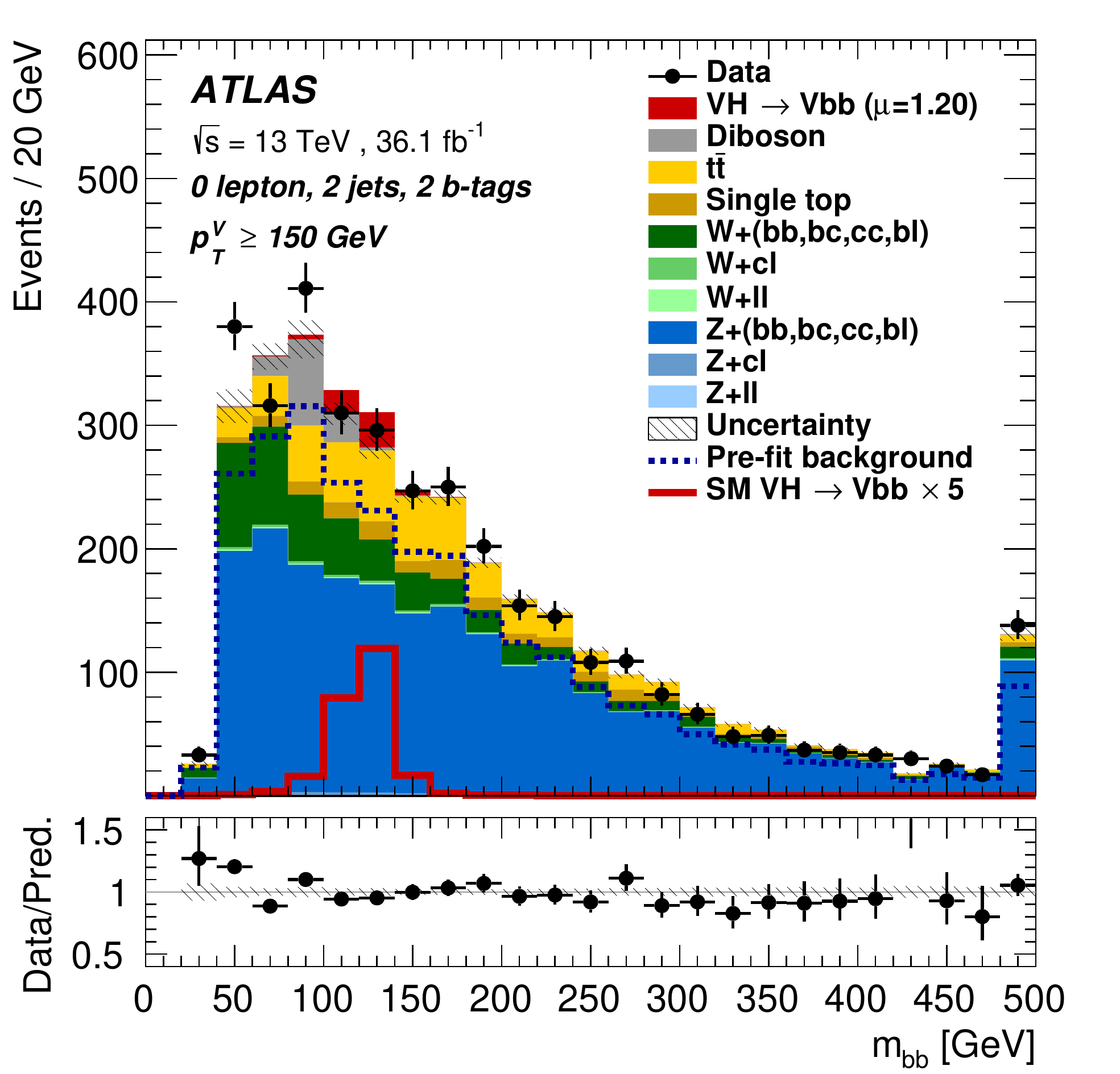}\\
    \includegraphics[width=0.4\linewidth]{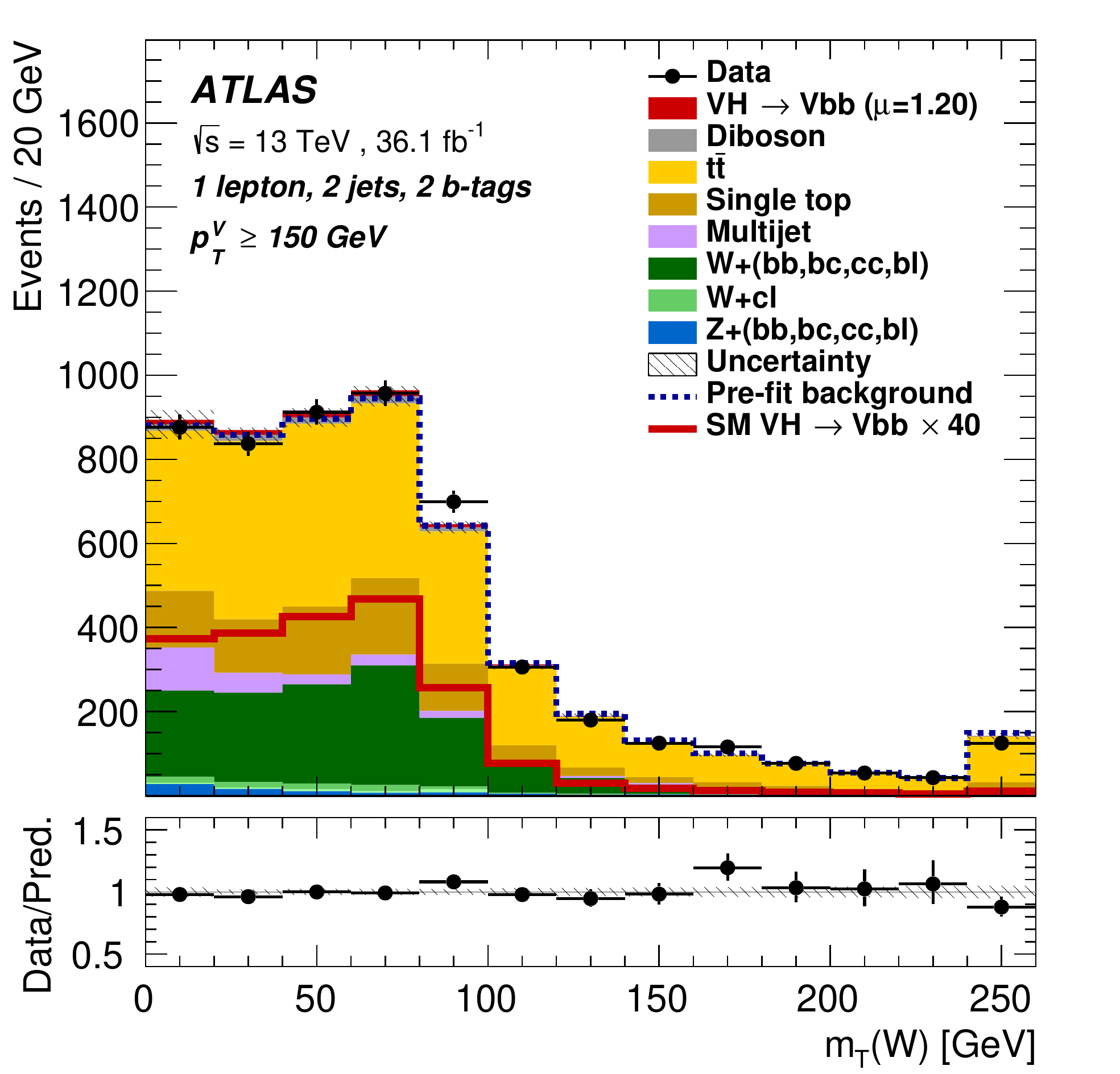}
    \includegraphics[width=0.4\linewidth]{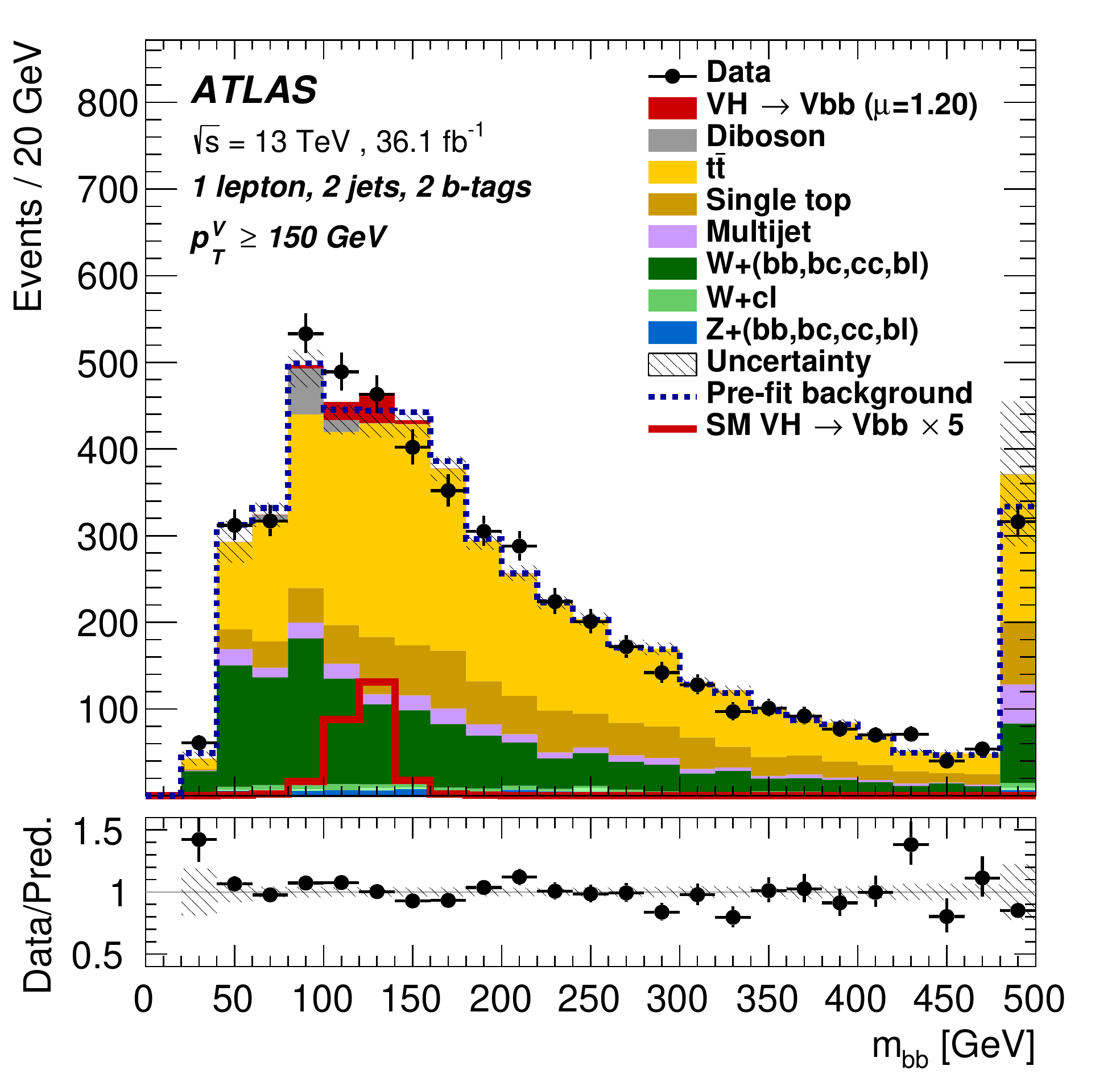}\\
    \includegraphics[width=0.4\linewidth]{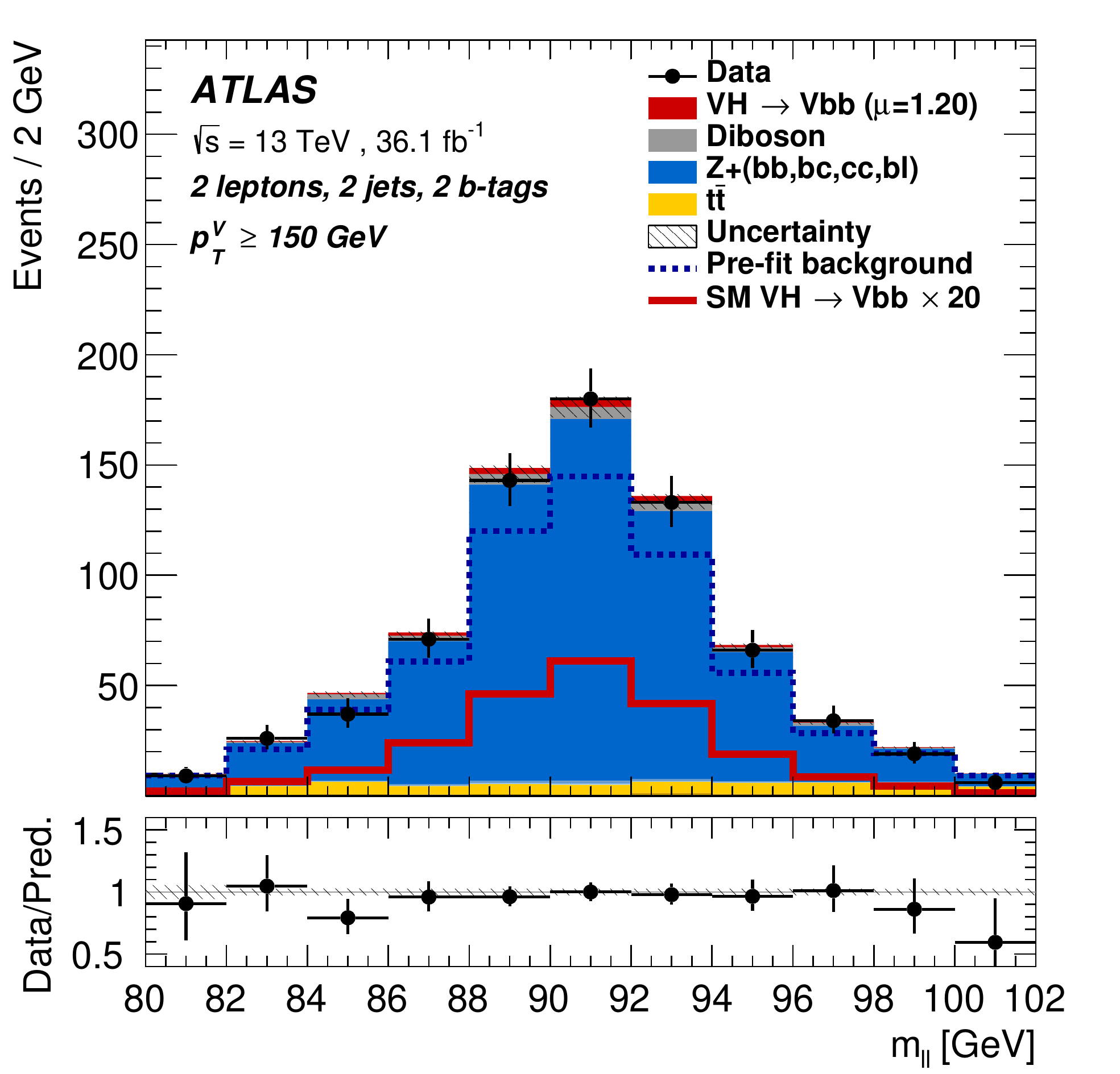}
    \includegraphics[width=0.4\linewidth]{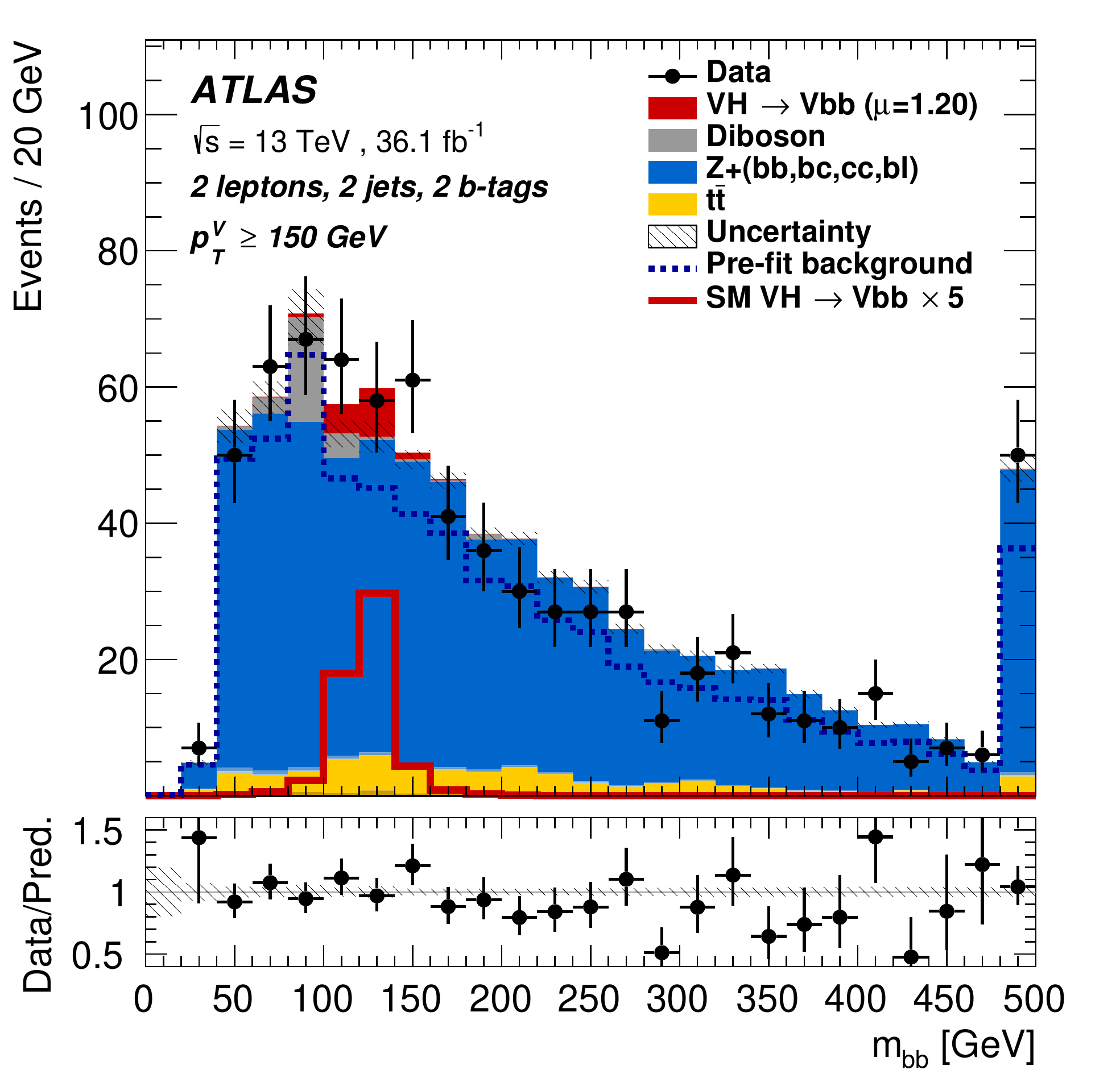}\\
  \end{tabular}
    \caption{The post-fit distributions for \met~(top left), \mtw~(middle left), \mll~(bottom left) and \mbb~(right) in the 0-lepton (top), 1-lepton (middle) and 2-lepton (bottom) channels for 2-jet, 2-$b$-tag events in the high \ptv\ region.  The background contributions after the global likelihood fit are shown as filled histograms. The Higgs boson signal ($\mh = 125$~\GeV) is shown as a filled histogram on top of 
the fitted backgrounds normalised to the signal yield extracted from data ($\mu=1.20$),
and unstacked as an unfilled histogram, scaled by the factor indicated in the legend. 
The entries in overflow are included in the last bin.
The dashed histogram shows the total background as expected from the pre-fit 
MC simulation. The size of the combined statistical and systematic uncertainty for the
sum of the fitted signal and background is indicated by the hatched band. The ratio
of the data to the sum of the fitted signal and background is shown in the lower panel. }
    \label{fig:kinematics}
\end{figure}

\begin{figure}[htbp]
  \centering
  \begin{tabular}{cccc}
\includegraphics[width=0.4\linewidth]{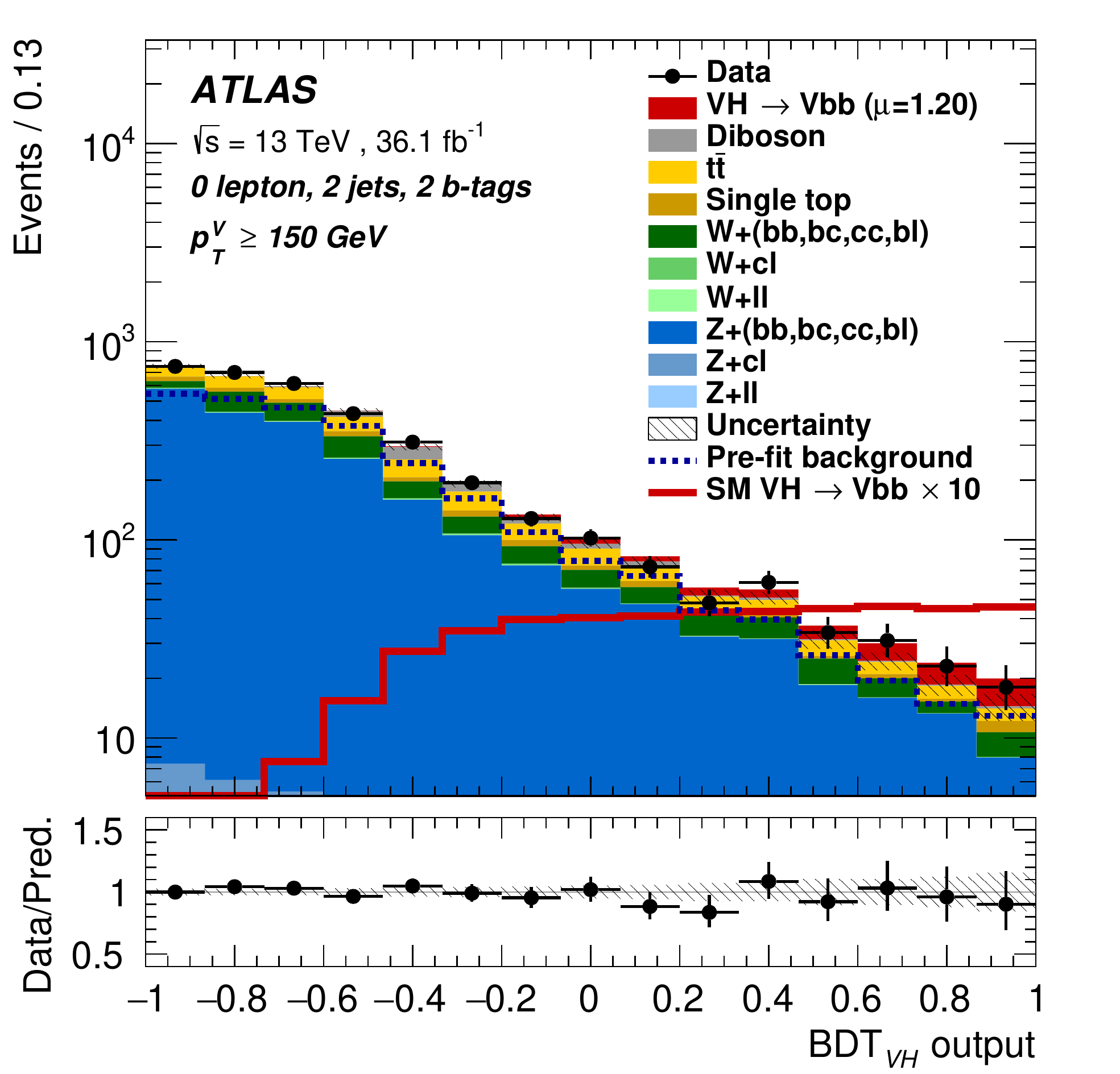}
\includegraphics[width=0.4\linewidth]{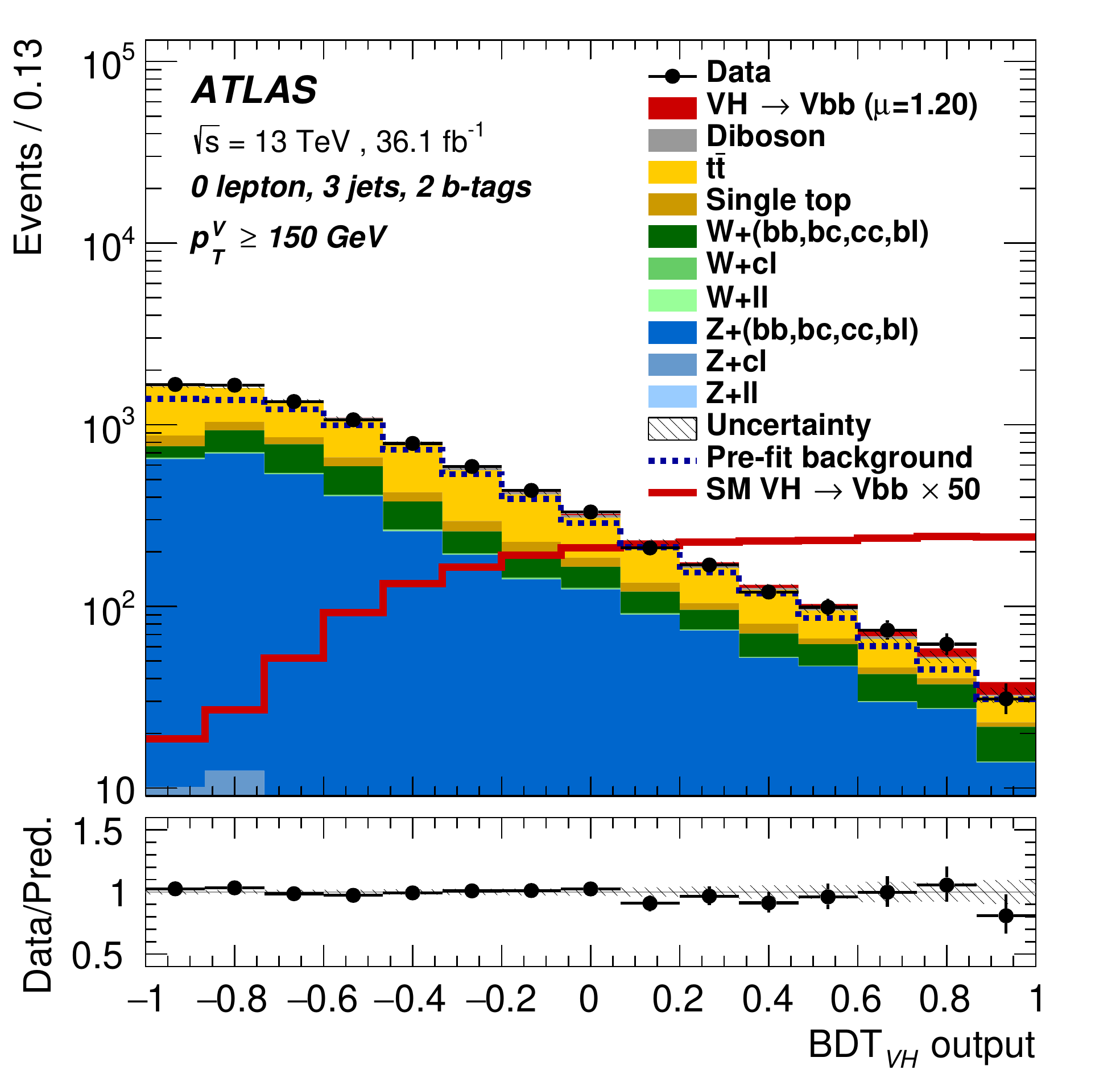} \\
\includegraphics[width=0.4\linewidth]{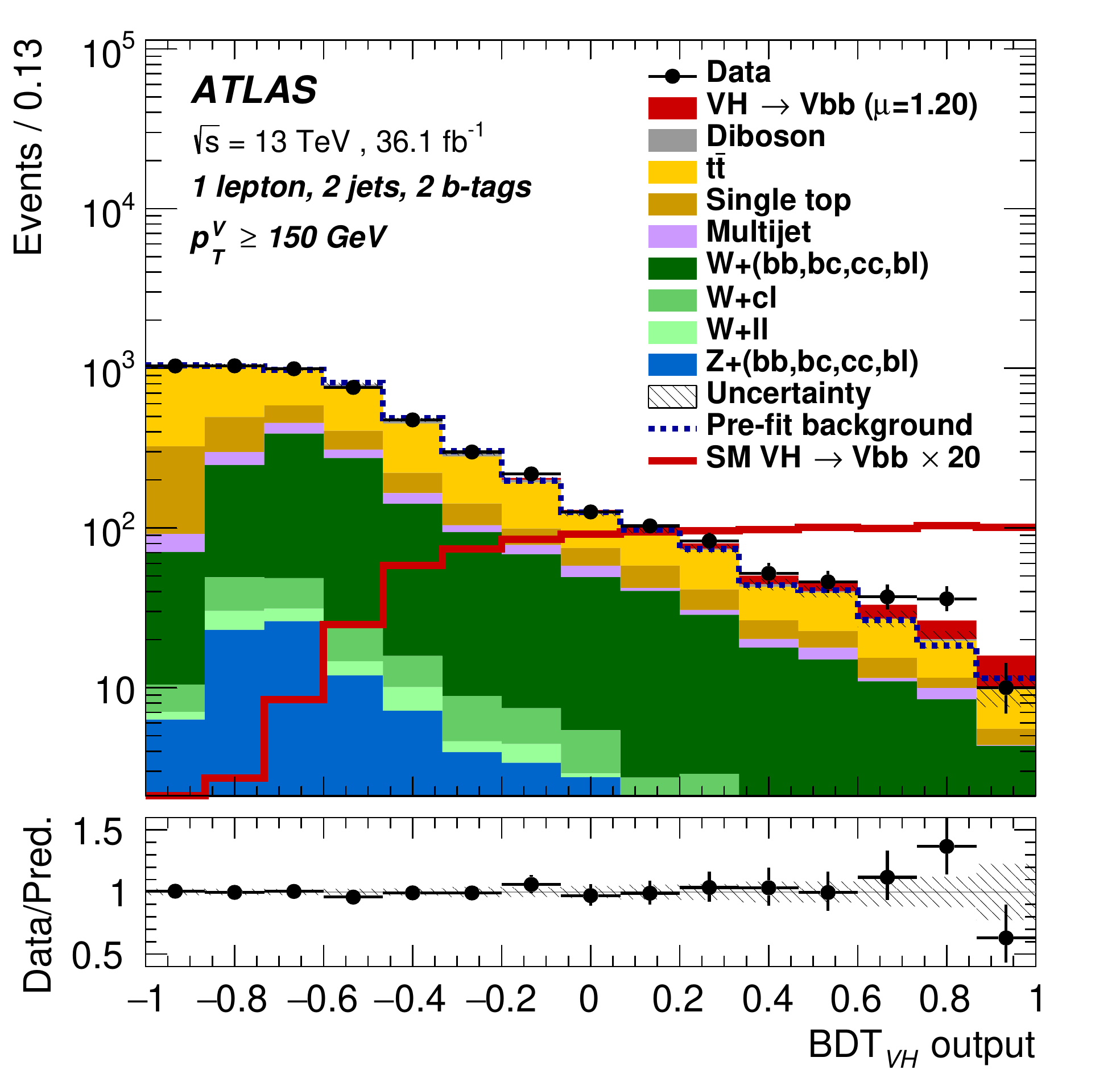}
\includegraphics[width=0.4\linewidth]{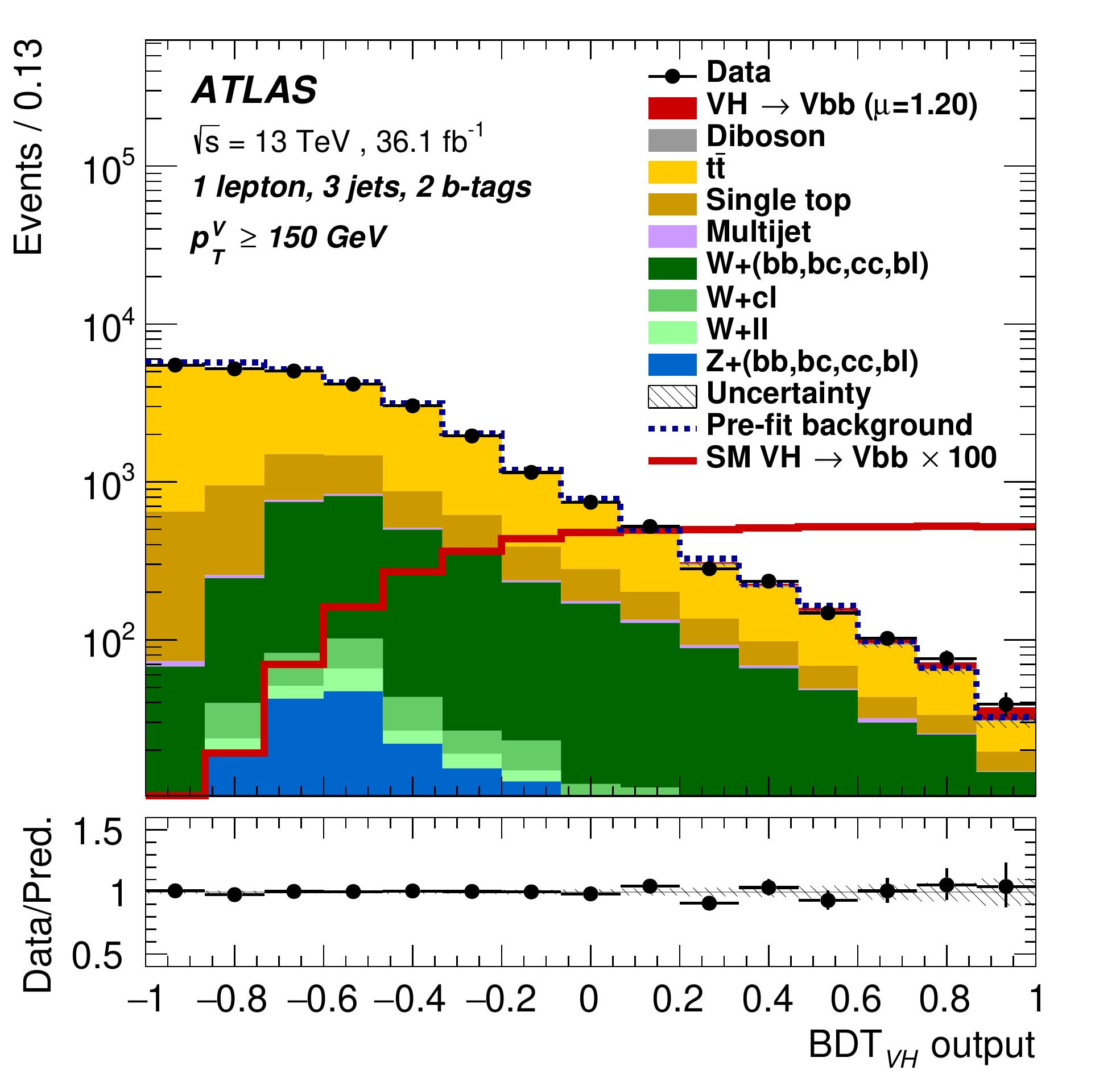} \\
\includegraphics[width=0.4\linewidth]{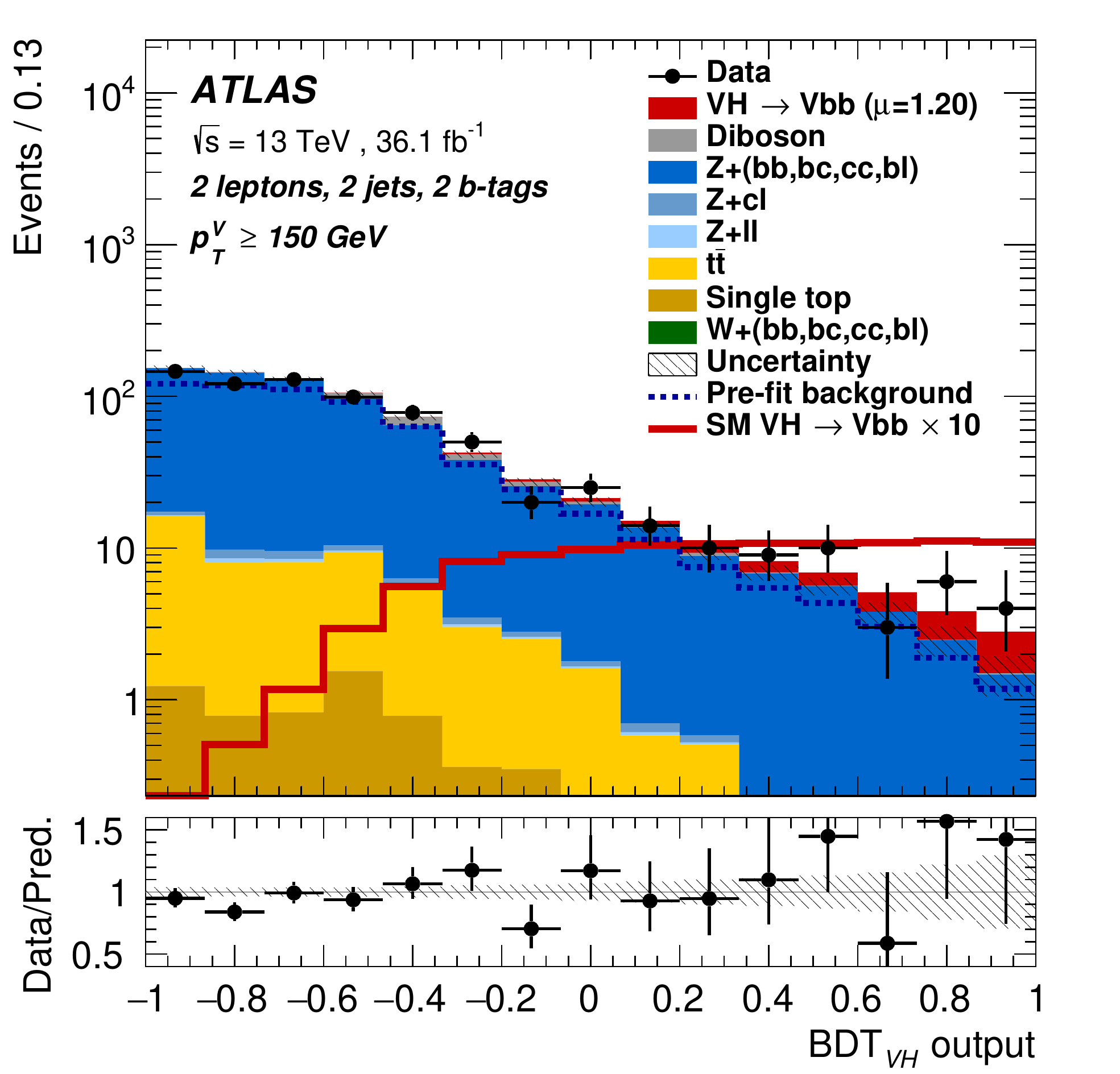}
\includegraphics[width=0.4\linewidth]{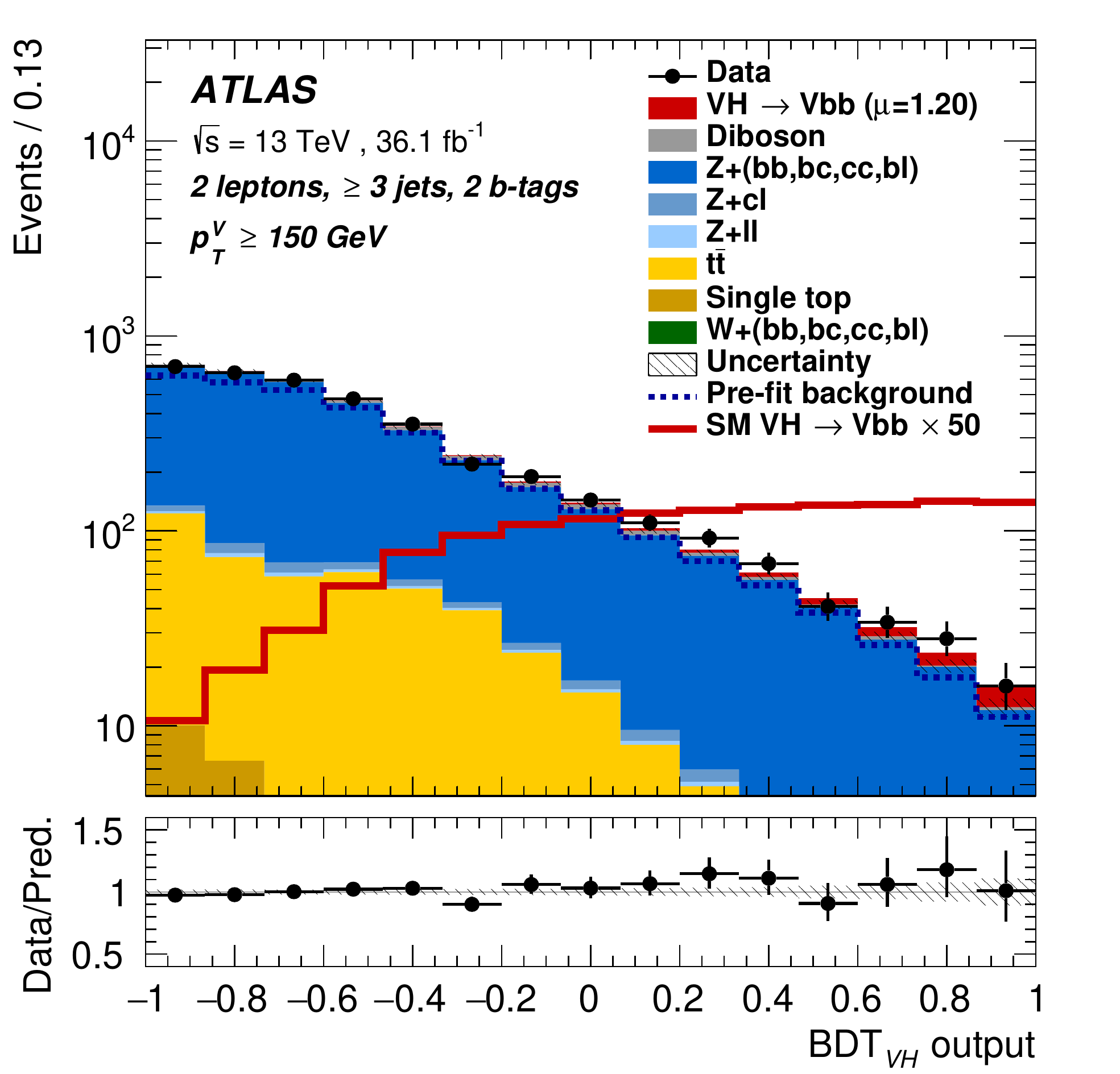} \\
  \end{tabular}
 \caption{The BDT$_{VH}$ output post-fit distributions in the 0-lepton~(top), 1-lepton~(middle) and 2-lepton~(bottom) channel for 2-$b$-tag events, in the 2-jet~(left) and exactly 3-jet (or $\ge 3$ jets for the 2-lepton case) (right) categories in the high \ptv\ region. The background contributions after the global likelihood fit are shown as filled histograms. The Higgs boson signal ($\mh = 125$~\GeV) is shown as a filled histogram on top of 
the fitted backgrounds normalised to the signal yield extracted from data ($\mu=1.20$), 
and unstacked as an unfilled histogram, scaled by the factor indicated in the legend. 
The dashed histogram shows the total background as expected from the pre-fit 
MC simulation. The size of the combined statistical and systematic uncertainty for the
sum of the fitted signal and background is indicated by the hatched band. The ratio
of the data to the sum of the fitted signal and background is shown in the lower panel. 
 \label{fig:mva_output}}
\end{figure}

Figure~\ref{fig:kinematics} shows a selection of characteristic post-fit distributions for each of the lepton channels, 
while Figure~\ref{fig:mva_output} shows the BDT output distributions in the most sensitive (high-\ptv) categories. 
The background prediction in all post-fit distributions is obtained by normalising the backgrounds and setting the systematic uncertainties according to the values of the floating normalisations and nuisance parameters obtained in the signal extraction fit. The post-global likelihood fit signal and background yields are shown in Table~\ref{tbl:yieldTable} for all the analysis regions. 

\begin{sidewaystable}[phtb]
\centering
\caption{The fitted Higgs boson signal and background yields for each signal region category in each channel after the full selection of the multivariate analysis. 
The yields are normalised by the results of the global likelihood fit. 
All systematic uncertainties are included in the indicated uncertainties. 
An entry of ``--'' indicates that a specific background component is negligible in a certain region, or that no simulated events are left after the analysis selection.\label{tbl:yieldTable}} 
\begin{tabular}{l| r@{\,$\pm$\,}l| r@{\,$\pm$\,}l| r@{\,$\pm$\,}l| r@{\,$\pm$\,}l| r@{\,$\pm$\,}l| r@{\,$\pm$\,}l| r@{\,$\pm$\,}l| r@{\,$\pm$\,}l }
\hline\hline
 \multirow{2}{*}{ Signal regions } & \multicolumn{4}{c|}{0-lepton}  & \multicolumn{4}{c|}{1-lepton}  & \multicolumn{8}{c}{2-lepton}  \\
\cline{2-17}
& \multicolumn{4}{c|}{$p_\mathrm{T}^V > 150$~\GeV, 2-$b$-tag }  & \multicolumn{4}{c|}{$p_\mathrm{T}^V > 150$~\GeV, 2-$b$-tag }  & \multicolumn{4}{c|}{$75$~\GeV~$< p_\mathrm{T}^V < 150$~\GeV, 2-$b$-tag }  & \multicolumn{4}{c}{$p_\mathrm{T}^V > 150$~\GeV, 2-$b$-tag }  \\
\hline
Sample & \multicolumn{2}{c|}{2-jet} & \multicolumn{2}{c|}{3-jet} & \multicolumn{2}{c|}{2-jet} & \multicolumn{2}{c|}{3-jet} & \multicolumn{2}{c|}{2-jet} & \multicolumn{2}{c|}{$\geq$3-jet} & \multicolumn{2}{c|}{2-jet} & \multicolumn{2}{c}{$\geq$3-jet} \\
\hline
$Z+ll$ &  9.0 & 5.1  &  15.5 & 8.1  &  \multicolumn{2}{c|}{$<1$}  & \multicolumn{2}{c|}{--} &  9.2 & 5.4  &  35 & 19  &  1.9 & 1.1  &  16.4 & 9.3  \\
$Z+cl$ &  21.4 & 7.7  &  42 & 14  &  2.2 & 0.1  &  4.2 & 0.1  &  25.3 & 9.5  &  105 & 39  &  5.3 & 1.9  &  46 & 17  \\
$Z$~+~HF &  2198 & 84  &  3270 & 170  &  86.5 & 6.1  &  186 & 13  &  3449 & 79  &  8270 & 150  &  651 & 20  &  3052 & 66  \\
$W+ll$ &  9.8 & 5.6  &  17.9 & 9.9  &  22 & 10  &  47 & 22  &  \multicolumn{2}{c|}{$<1$}  &  \multicolumn{2}{c|}{$<1$}  &  \multicolumn{2}{c|}{$<1$}  &  \multicolumn{2}{c}{$<1$}  \\
$W+cl$ &  19.9 & 8.8  &  41 & 18  &  70 & 27  &  138 & 53  &  \multicolumn{2}{c|}{$<1$}  &  \multicolumn{2}{c|}{$<1$}  &  \multicolumn{2}{c|}{$<1$}  &  \multicolumn{2}{c}{$<1$}  \\
$W$~+~HF &  460 & 51  &  1120 & 120  &  1280 & 160  &  3140 & 420  &  3.0 & 0.4  &  5.9 & 0.7  &  \multicolumn{2}{c|}{$<1$}  &  2.2 & 0.2  \\
Single top quark &  145 & 22  &  536 & 98  &  830 & 120  &  3700 & 670  &  53 & 16  &  134 & 46  &  5.9 & 1.9  &  30 & 10  \\
$t \bar t$ &  463 & 42  &  3390 & 200  &  2650 & 170  &  20640 & 680  &  1453 & 46  &  4904 & 91  &  49.6 & 2.9  &  430 & 22  \\
Diboson &  116 & 26  &  119 & 36  &  79 & 23  &  135 & 47  &  73 & 19  &  149 & 32  &  24.4 & 6.2  &  87 & 19  \\
Multi-jet $e$ sub-ch. & \multicolumn{2}{c|}{--} & \multicolumn{2}{c|}{--} &  102 & 66  &  27 & 68  & \multicolumn{2}{c|}{--} & \multicolumn{2}{c|}{--} & \multicolumn{2}{c|}{--} & \multicolumn{2}{c}{--} \\
Multi-jet $\mu$ sub-ch. & \multicolumn{2}{c|}{--} & \multicolumn{2}{c|}{--} &  133 & 99  &  90 & 130  & \multicolumn{2}{c|}{--} & \multicolumn{2}{c|}{--} & \multicolumn{2}{c|}{--} & \multicolumn{2}{c}{--} \\
\hline
Total bkg. &  3443 & 57  &  8560 & 91  &  5255 & 80  &  28110 & 170  &  5065 & 66  &  13600 & 110  &  738 & 19  &  3664 & 56  \\
\hline
Signal (fit) &  58 & 17  &  60 & 19  &  63 & 19  &  65 & 21  &  25.6 & 7.8  &  46 & 15  &  13.6 & 4.1  &  35 & 11  \\
\hline
Data & \multicolumn{2}{c|}{ 3520 }  & \multicolumn{2}{c|}{ 8634 }  & \multicolumn{2}{c|}{ 5307 }  & \multicolumn{2}{c|}{ 28168 }  & \multicolumn{2}{c|}{ 5113 }  & \multicolumn{2}{c|}{ 13640 }  & \multicolumn{2}{c|}{ 724 }  & \multicolumn{2}{c}{ 3708 }  \\
\hline\hline
\end{tabular}
\end{sidewaystable}

s

For the tested Higgs boson mass of 125~\GeV, when all lepton channels are combined, the probability $p_0$ of obtaining from background alone a result at least as signal-like as the observation is 0.019\%. In the presence of a Higgs boson with that mass and the SM signal strength, the expected $p_0$ value is 0.12\%. The observation corresponds to an excess with a significance of 3.5 standard deviations, to be compared to an expectation of 3.0 standard deviations. Table~\ref{tab:results} shows the $p_0$ and significance values for separate lepton channel fits and for the combined global fit.

For all channels combined the fitted value of the signal strength parameter is 
\begin{eqnarray*}
\mu = 1.20 ^{+0.24}_{-0.23} \mathrm{(stat.)} ^{+0.34}_{-0.28} \mathrm{(syst.)}.
\end{eqnarray*}

\begin{figure}[tbhp!]
\begin{center}
\includegraphics[width=0.7\textwidth]{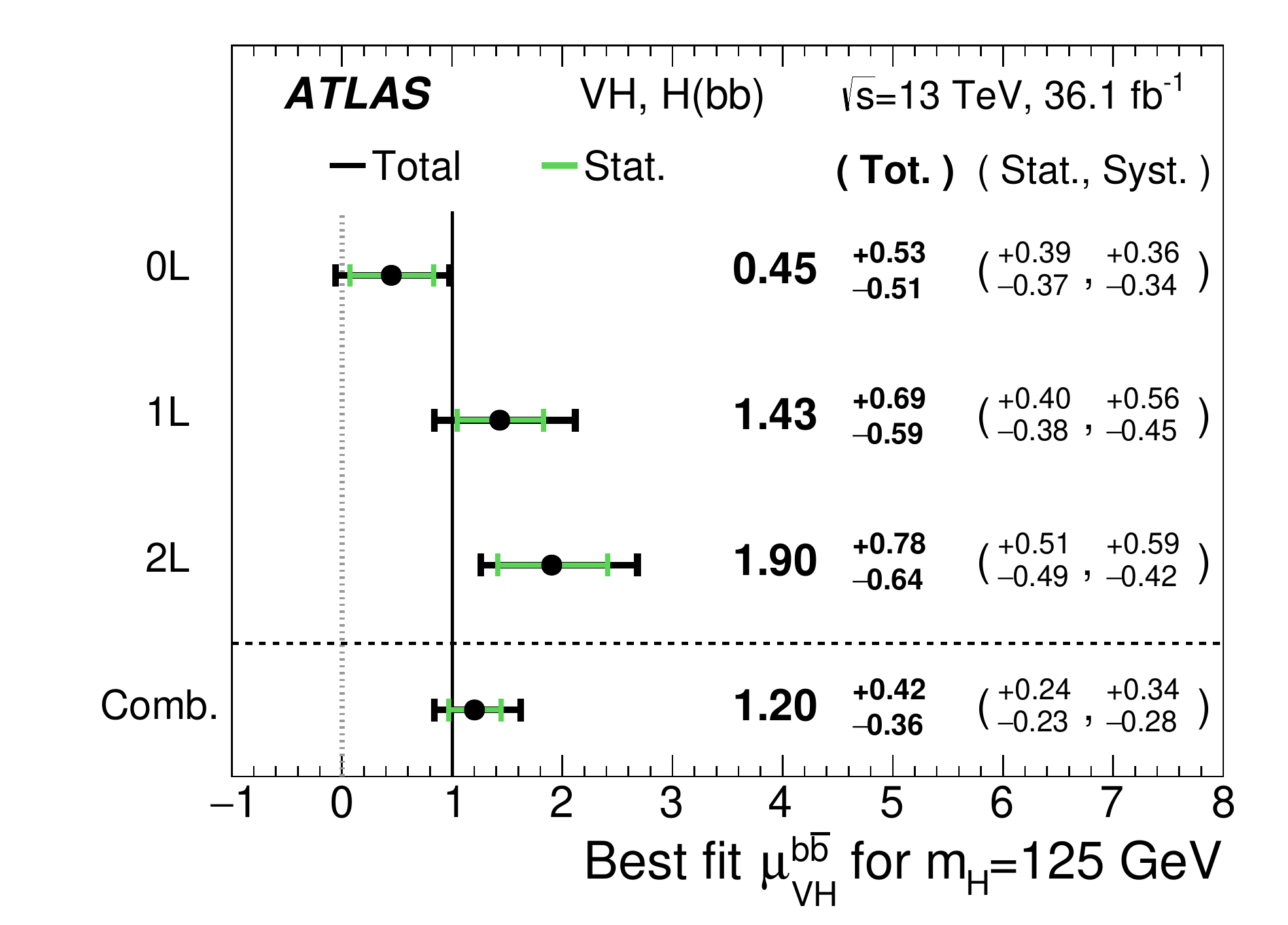}
\end{center}
\caption{The fitted values of the Higgs boson signal strength parameter $\mu$ for $\mh=125$~\GeV\ for the 0-, 1- and 2-lepton channels
and their combination. The individual $\mu$ values for the lepton channels	
are obtained from a simultaneous fit with the signal strength parameter for each of the
lepton channels floating independently. The compatibility of the individual signal strengths is 10\%. \label{fig:mu-higgs-b}}
\end{figure}

\begin{figure}
\begin{center}
\includegraphics[width=0.7\textwidth]{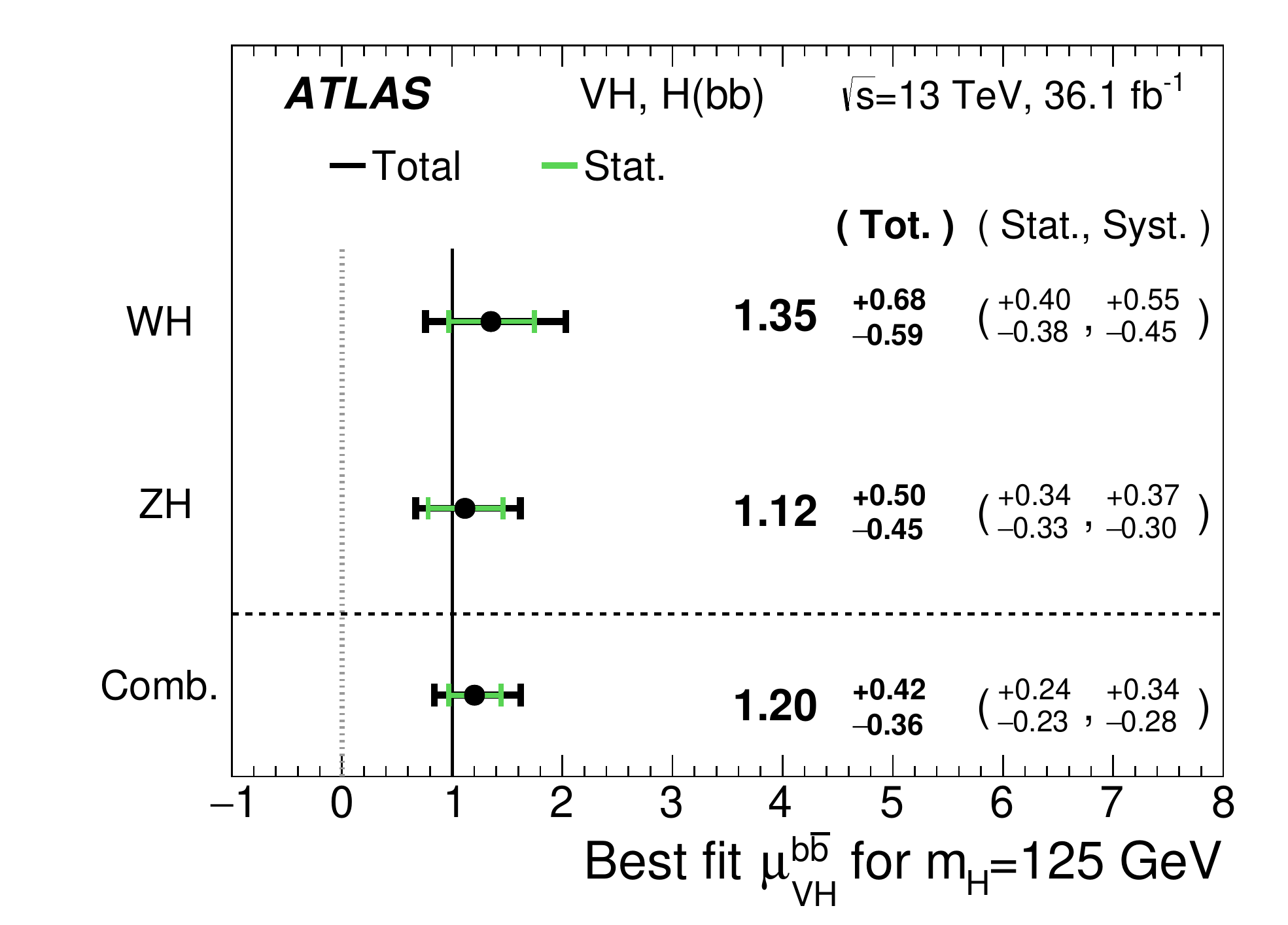}
\end{center}
\caption{The fitted values of the Higgs boson signal strength parameter $\mu$ for $\mh=125$~\GeV\ for the $WH$ and $ZH$ processes
and their combination. The individual $\mu$ values for the $(W/Z)H$ processes
are obtained from a simultaneous fit with the signal strength for each of the $WH$ and $ZH$ processes floating independently. 
The compatibility of the individual signal strengths is 75\%.\label{fig:mu-higgs-c}}
\end{figure}

Combined fits are also performed with floating signal strength parameters separately for
(i) the three lepton channels, or (ii) the $WH$ and $ZH$ production processes, but leaving all other NPs with the same correlation scheme as for the nominal result. The results of these fits are shown in Figures~\ref{fig:mu-higgs-b} and \ref{fig:mu-higgs-c}. The compatibility of the signal strength parameters measured in the three lepton channels is 10\%. 
The $WH$ and $ZH$ production modes are observed with a significance of 2.4 and 2.6 standard deviations, respectively. The linear correlation term between the signal strengths related to the $WH$ and the $ZH$ production modes is 0.6\%.
Assuming that the observed signal is due to the SM Higgs boson, with corresponding model-dependent extrapolation corrections to the inclusive phase space, the signal strengths can be interpreted as measurements of the $WH$ and $ZH$ production cross-sections times the $H \to b\bar{b}$ branching ratio. After removing the theoretical uncertainties for the production cross-sections and branching ratio, these are determined to be
\begin{eqnarray*}
 \sigma\left(WH\right) \times \mathrm{B}(H \rightarrow b\bar{b}) & = & 1.08^{+0.54}_{-0.47}~\mathrm{pb}, \\
 \sigma\left(ZH\right) \times \mathrm{B}(H \rightarrow b\bar{b}) & = & 0.57^{+0.26}_{-0.23}~\mathrm{pb}, 
\end{eqnarray*}
compared to expectations of 0.80~pb and 0.51~pb~\autocite{deFlorian:2016spz}, respectively. The cross-section for the sum of the $WH$ and $ZH$ production modes is determined to be $ \sigma\left(VH\right) \times \mathrm{B}(H \rightarrow b\bar{b})=1.58^{+0.55}_{-0.47}$~pb, compared to an expectation of 1.31~pb.\footnote{The cross-section for the sum of the $WH$ and $ZH$ production modes is obtained from a fit where both production modes are described by a common signal strength parameter. As a result of this, the total cross-section is not equal to the sum of the cross-sections measured for the separate production modes.} The uncertainties in the quoted theory predictions are negligible compared to the present experimental precision.

\begin{figure}[thb!]
\begin{center}
\includegraphics[width=0.7\textwidth]{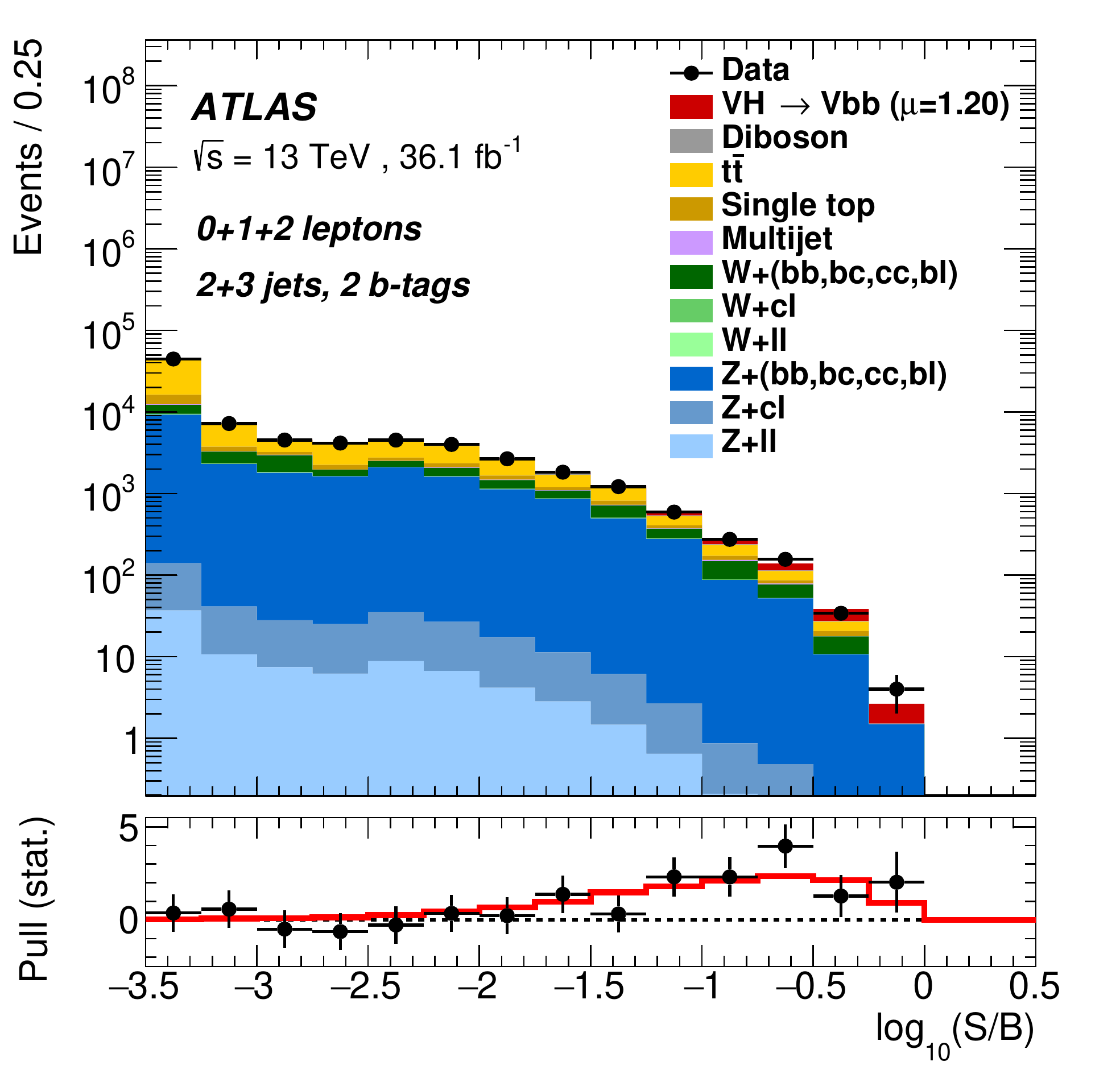}
\end{center}
\caption{Event yields as a function of $\log(S/B)$ for data, background
and a Higgs boson signal with $\mh=125$~\GeV. 
Final-discriminant bins in all  regions 
are combined into bins of $\log(S/B)$,
with the fitted signal being $S$ and the fitted background $B$.
The Higgs boson signal contribution is shown after rescaling the SM cross-section 
 according to the value of the signal strength parameter extracted from data ($\mu=1.20$).
 The pull (residual divided by its uncertainty) of the data with respect to the background-only prediction is also shown with statistical uncertainties only.
The full line indicates the pull of the prediction for signal ($\mu=1.20$) and background with respect to the background-only prediction.
}
\label{fig:logstob}
\end{figure}

Figure~\ref{fig:logstob} shows the data, background and signal yields, where final-discriminant bins in all regions are combined into bins of $\log(S/B)$. Here, $S$ and $B$ are the fitted signal and background yields, respectively. 
Details of the fitted values of the signal and of the various background
components in the four bins with the highest $S/B$ ratio in Figure~\protect\ref{fig:logstob} are provided in Table~\ref{tab:details}. 

{\renewcommand{\arraystretch}{1.3}
\begin{table}[phtb]
\caption{The expected and observed $p_{0}$ and significance values for the individual lepton channels and their combination using the 13 \TeV\ dataset. The expected values are evaluated assuming a SM Higgs boson with a mass of 125~\GeV.
\label{tab:results}}
\begin{center}
\begin{tabular}{l|cc|cc}
\hline\hline
\multirow{2}{*}{Dataset} & \multicolumn{2}{c|}{$p_{0}$} & \multicolumn{2}{c}{Significance} \\
\cline{2-5}
         & Exp. & Obs.                  & Exp. & Obs. \\ 
\hline
0-lepton                                   & 4.2\% & 30\% & $1.7$ & $0.5$   \\
1-lepton                                     & 3.5\% & 1.1\%  & $1.8$ & $2.3$\\
2-lepton                                     & 3.1\% & 0.019\%  & $1.9$ & $3.6$\\
\hline
Combined                                     & 0.12\% & 0.019\% & $3.0$ & $3.5$ \\
\hline\hline
\end{tabular}
\end{center}
\end{table}
}

\begin{table}[hpbt!]
\caption{The numbers of fitted signal and background events and the observed 
numbers of events in the four highest $S/B$ bins of Figure~\protect\ref{fig:logstob}. 
An entry of ``--'' indicates that a specific background component is negligible in a certain bin, or that no simulated events are left after the analysis selection.
\label{tab:details}}
\begin{center}
		\begin{tabular}{l|cccc}
		\hline\hline
Process           & Bin 11  & Bin 12  & Bin 13 & Bin 14 \\
\hline
Data             & 274     & 156     & 34     & 4      \\
Signal (fit)              & 32.4  & 25.0  & 11.1 & 1.1  \\
Total Background & 238.3 & 113.7 & 27.3 & 1.5\\
\hline
$Z+ll$          & 0.2   & 0.1   & $<0.1$  & $<0.1$  \\
$Z+cl$         & 0.7   & 0.4   &$<0.1$  & $<0.1$  \\
$Z$~+~HF             & 86.1  & 51.3  & 10.5 & 1.5  \\
$W+ll$          & 0.20   & 0.1   & $<0.1$  & --  \\
$W+cl$         & 1.6   & 0.2   & $<0.1$  & --  \\
$W$~+~HF             & 58.9  & 24.5  & 6.9 & --  \\
Single top quark       & 19.2  & 7.6   & 2.9  & --  \\
\ttbar\       & 61.3  & 25.7  & 6.2  & --  \\
Diboson          & 4.7   & 1.7  & 0.4  & $<0.1$  \\
Multi-jet $e$ sub-ch.      & 0.1   & --   & --  & --  \\
Multi-jet $\mu$ sub-ch.      & 5.2   & 2.0   & $<0.1$  & --  \\
\hline\hline
		\end{tabular}
\end{center}
\end{table}

\FloatBarrier

\subsection{Results of the dijet-mass analysis}
\label{subsec:dijetmassanalysis}

\begin{figure}[tp]
  \centering
  \begin{tabular}{cccc}
\includegraphics[width=0.33\linewidth]{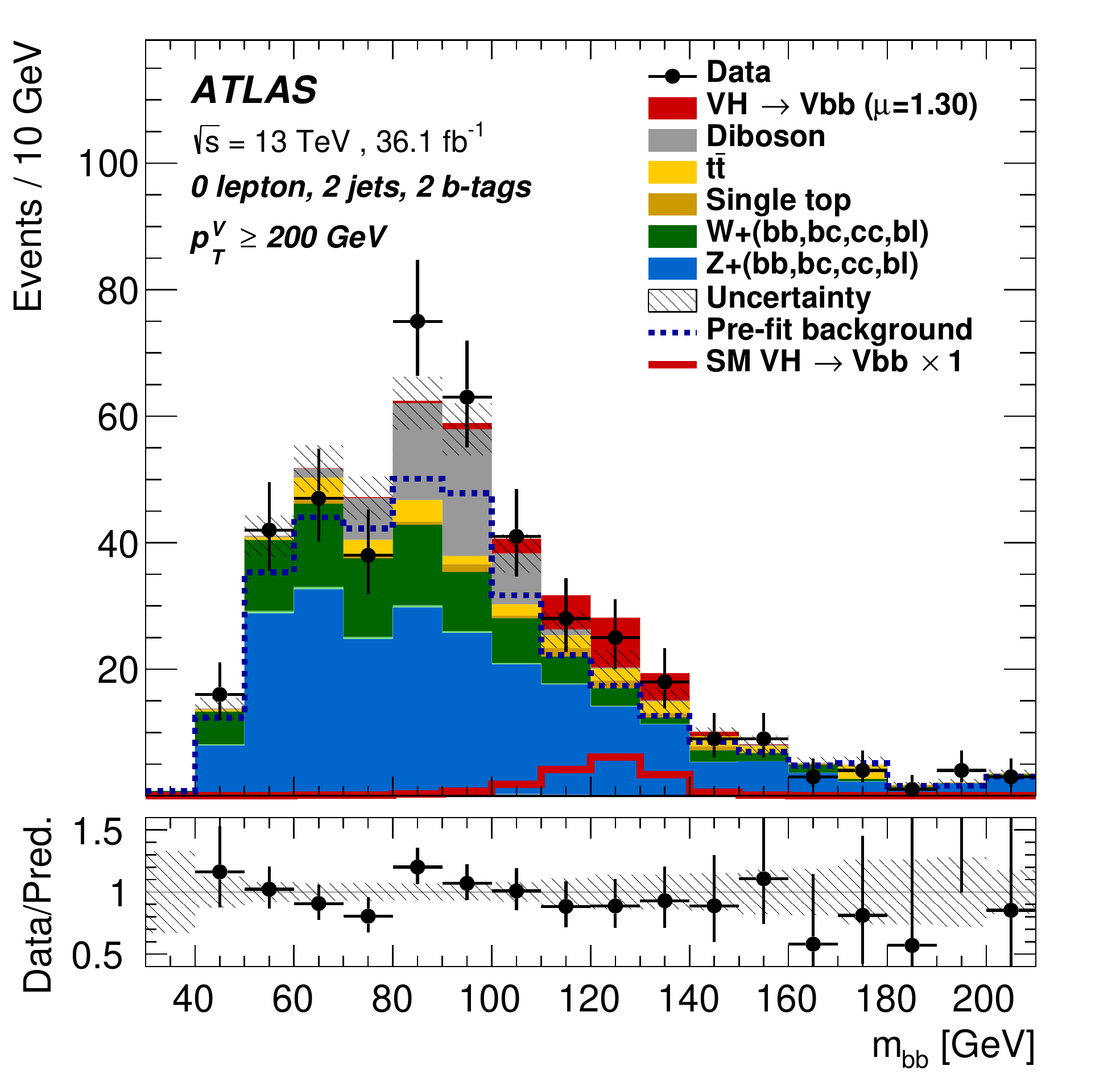}
\includegraphics[width=0.33\linewidth]{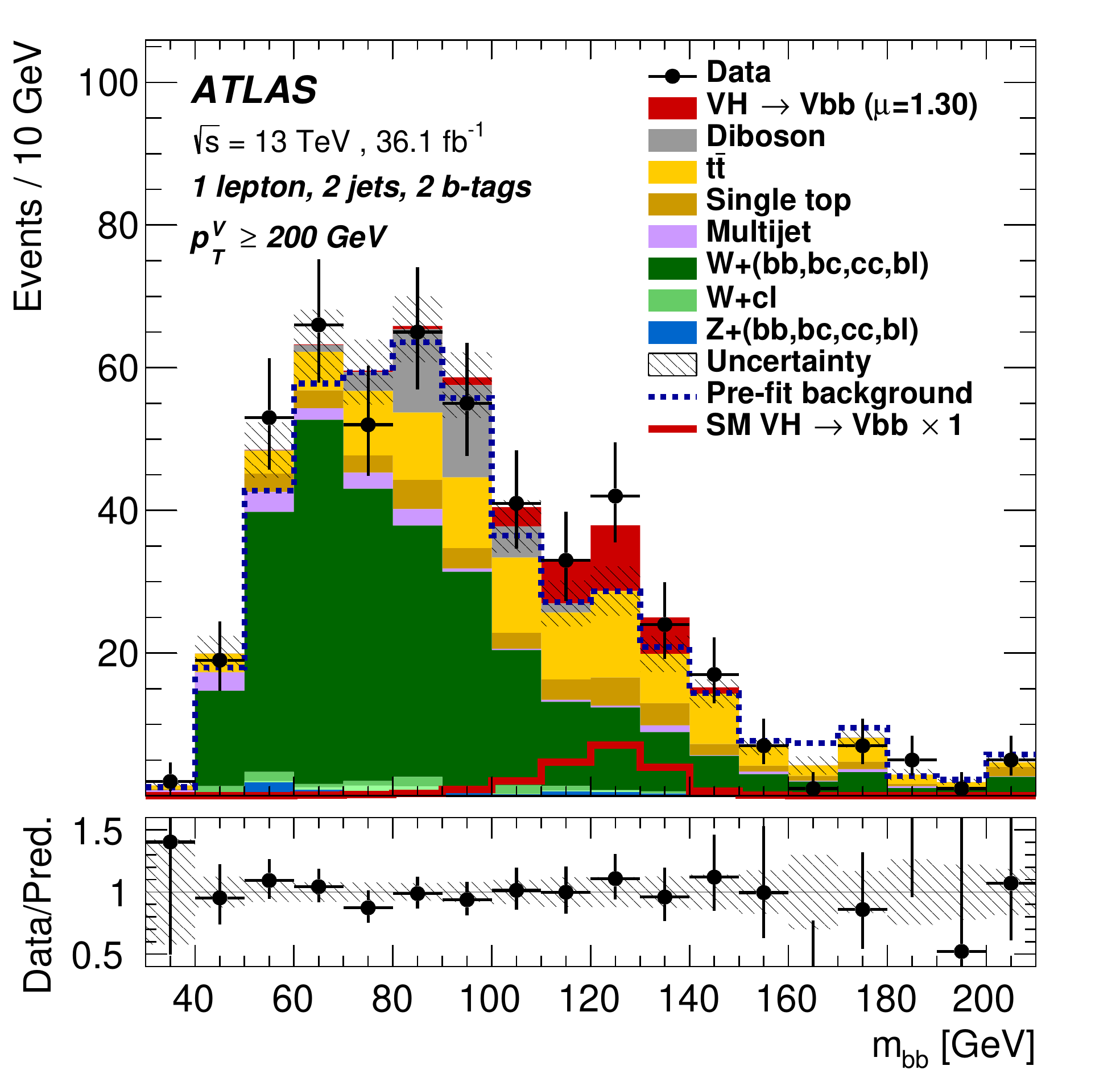} 
\includegraphics[width=0.33\linewidth]{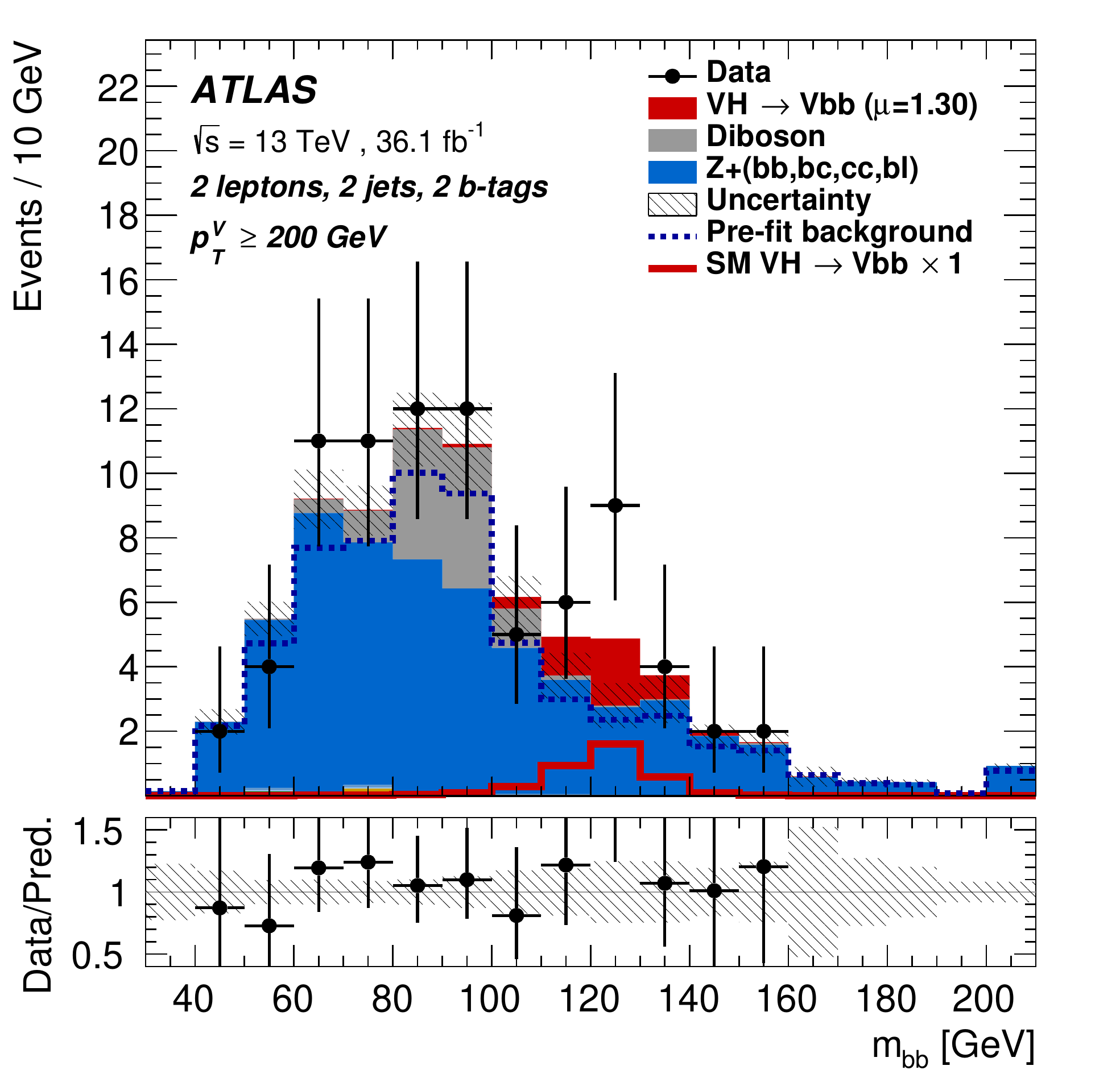}
  \end{tabular}
 \caption{The \mbb\ distributions in the 0-lepton~(left), 1-lepton~(middle) and 2-lepton~(right) channels for 2-$b$-tag events, in the 2-jet categories for $\ptv> 200$ \GeV. The background contributions after the global likelihood fit are shown as filled histograms. The Higgs boson signal ($\mh = 125$~\GeV) is shown as a filled histogram on top of 
the fitted backgrounds normalised to the signal yield extracted from data ($\mu=1.30$), 
and unstacked as an unfilled histogram, scaled by the factor indicated in the legend. 
The dashed histogram shows the total background as expected from the pre-fit 
MC simulation. The size of the combined statistical and systematic uncertainty for the
sum of the fitted signal and background is indicated by the hatched band. The ratio
of the data to the sum of the fitted signal and background is shown in the lower panel. 
 \label{fig:cba_output}}
\end{figure}

\begin{figure}[bp!]
\begin{center}
\includegraphics[width=0.49\textwidth]{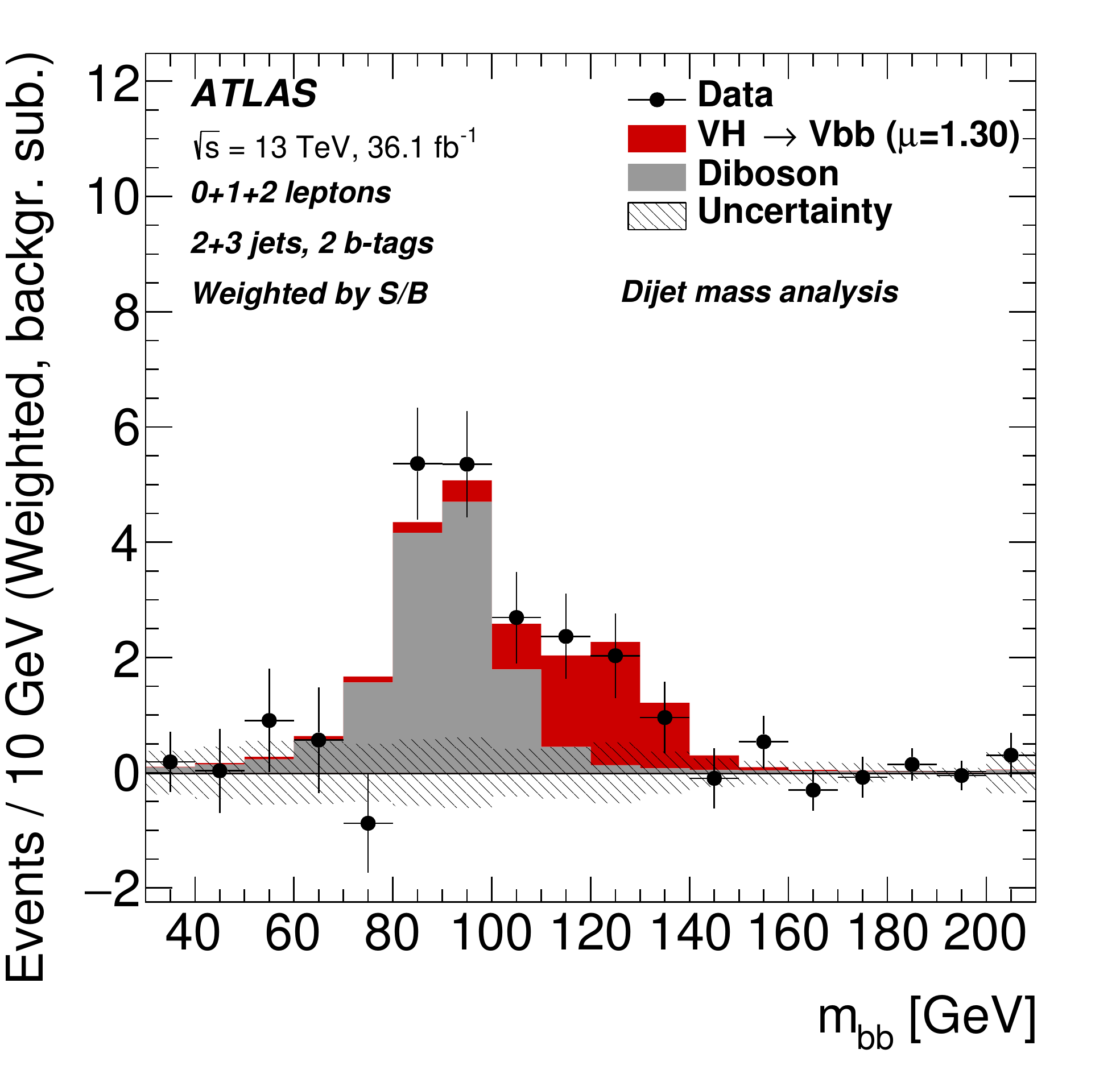}
\end{center}
\caption{The distribution of \mbb\ in data after subtraction of all backgrounds
except for the $WZ$ and $ZZ$ diboson processes, 
as obtained with the dijet-mass analysis. 
The contributions from all lepton channels, \ptv\ intervals and 
number-of-jets  categories are summed weighted by 
their respective value of the ratio of fitted Higgs boson signal and background.
The expected contribution of the associated $WH$ and $ZH$ production of a SM Higgs boson with $\mh=125$~\GeV\
is shown scaled by the measured combined signal strength ($\mu = 1.30$).
The size of the combined statistical and systematic uncertainty 
for the fitted background is indicated by the hatched band.
 \label{fig:mbb_cutbased}}       
\end{figure}

The distributions of \mbb\ in the dijet-mass analysis are shown in Figure~\ref{fig:cba_output} for the 2-jet category and the most sensitive analysis regions with $\ptv> 200$~\GeV\ for the 0-, 1- and 2-lepton channels separately. The \mbb\ distribution for all channels and regions summed, weighted by their respective value of the ratio of fitted Higgs boson signal and background yields, and after subtraction of all backgrounds except for the $(W/Z)Z$ diboson processes, is shown in Figure~\ref{fig:mbb_cutbased}. The data and the sum of expected signal and backgrounds are found to be in good agreement.

For all channels combined the fitted value of the signal strength parameter is 
\begin{eqnarray*}
\mu = 1.30 ^{+0.28}_{-0.27} \mathrm{(stat.)} ^{+0.37}_{-0.29} \mathrm{(syst.)}, 
\end{eqnarray*}
in good agreement with the result of the multivariate analysis.  
The observed excess has a significance of 3.5 standard deviations,
in comparison to an expectation of 2.8 standard deviations. 
Good agreement is also found in the values of signal strength parameters in the individual channels 
for the dijet-mass analysis compared to those for the multivariate analysis, 
with the largest difference between the respective central values of the two analyses being within 15\%.

\subsection{Results of the diboson analysis}

\begin{figure}[tbhp!]
\begin{center}
\includegraphics[width=0.7\textwidth]{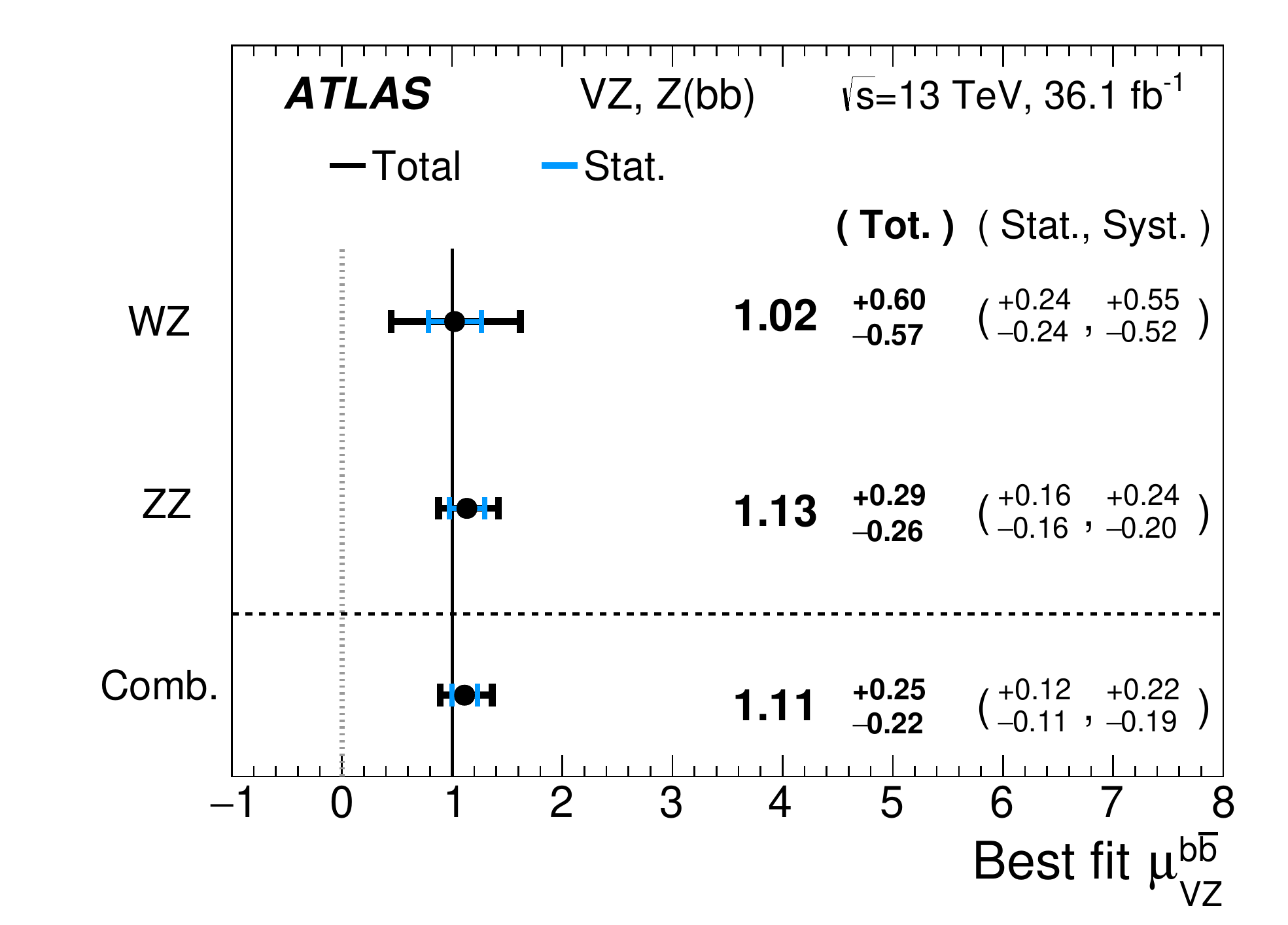}
\end{center}
\caption{The fitted values of the $VZ$  signal strength parameter $\mu_{VZ}$ for the $WZ$ and $ZZ$ processes
and their combination. The individual $\mu_{VZ}$ values for the $(W/Z)Z$ processes
are obtained from a simultaneous fit with the signal strength parameters for each of the $WZ$ and $ZZ$ processes floating independently.
The compatibility of the individual signal strengths is 88\%
\label{fig:mu-higgs-c-dib}}
\end{figure}

\begin{figure}[thb!]
\begin{center}
\includegraphics[width=0.7\textwidth]{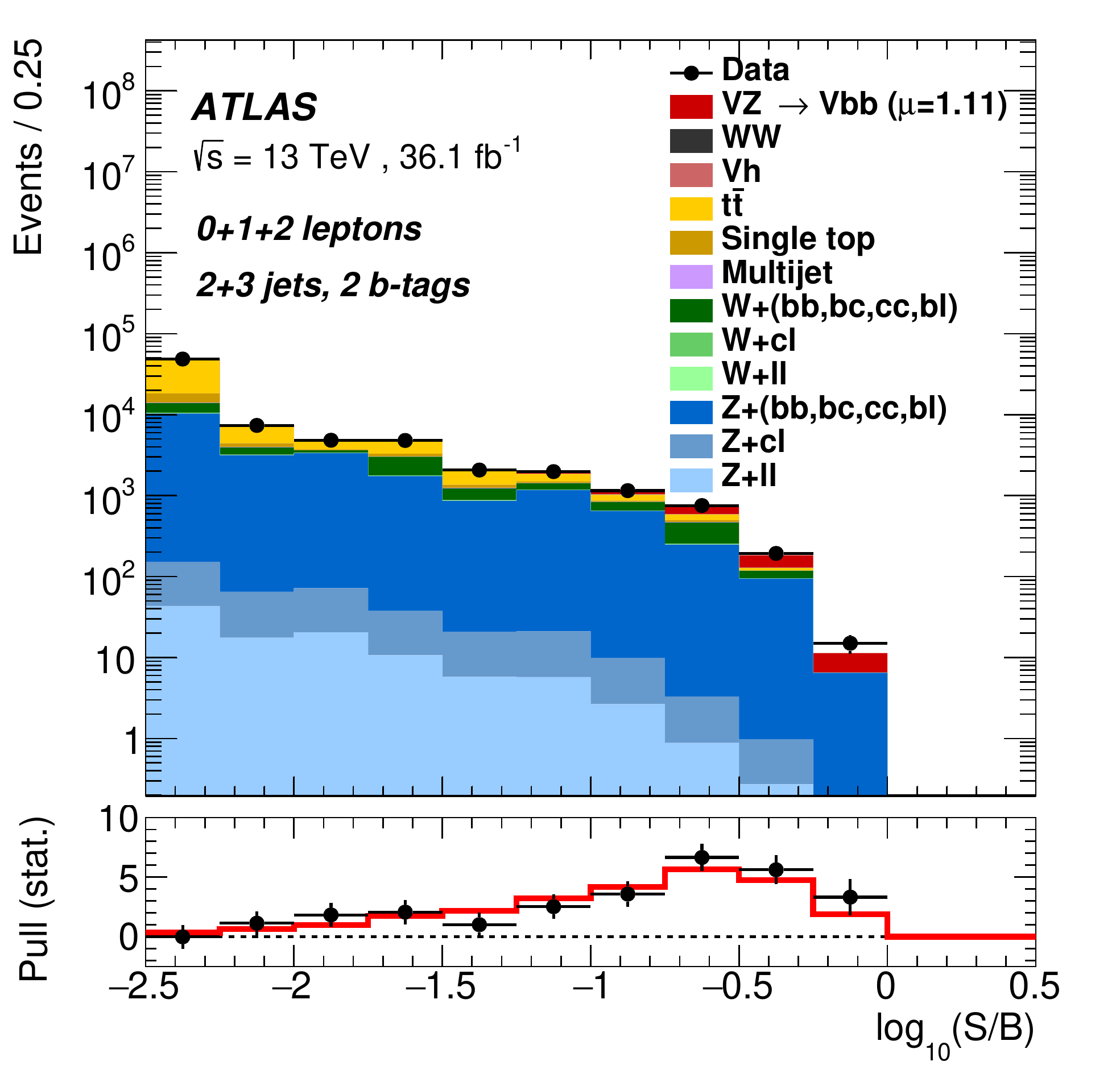}
\end{center}
\caption{Event yields as a function of $\log(S/B)$ for data, background
and $VZ$ processes. 
Final-discriminant bins in all  regions 
are combined into bins of $\log(S/B)$, 
with the fitted signal being $S$ and the fitted background $B$.
The $VZ$  contribution is shown after rescaling the SM cross-section 
 according to the value of signal strength extracted from data ($\mu=1.11$). 
The pull (residual divided by its uncertainty) of the data with respect to the background-only prediction is also shown with statistical uncertainties only.
The full line indicates the pull of the prediction for signal ($\mu=1.11$) and background with respect to the background-only prediction.
}
\label{fig:logstob-dib}
\end{figure}

The measurement of $VZ$ production based on the multivariate analysis described in Section~\ref{sec:fit} returns a value of signal strength 
\begin{eqnarray*}
\mu_{VZ} = 1.11 ^{+ 0.12}_{-0.11} \mathrm{(stat.)} ^{+ 0.22}_{-0.19} \mathrm{(syst.)}, 
\end{eqnarray*}
in good agreement with the Standard Model prediction. The $VZ$ signal is observed with a significance of 5.8 standard deviations, to be compared to an expected significance of~5.3 standard deviations.  Analogously to the $VH$ signal, fits are also performed with separate signal strength parameters for the $WZ$ and $ZZ$ production modes, and the results are shown in Figure~\ref{fig:mu-higgs-c-dib}.
Figure~\ref{fig:logstob-dib} shows the data, background and $VZ$ signal yields, where final-discriminant bins in all  regions are combined into bins of $\log(S/B)$.
Here, $S$ and $B$ are the fitted signal and background yields, respectively. 

Diboson production is also measured using the dijet-mass analysis. 
The $VZ$ signal yield is determined in the fit while 
the Higgs boson signal yield is kept fixed to the Standard Model prediction within 50\% uncertainty. 
The extracted signal strength is
\begin{eqnarray*}
\mu_{VZ} = 1.01 \pm 0.12 \mathrm{(stat.)} ^{+ 0.20}_{-0.17} \mathrm{(syst.)},
\end{eqnarray*} 
again in good agreement with the Standard Model prediction.

\FloatBarrier

\subsection{Results of the combination with Run 1}

The combination of the Run 1 and Run 2 analyses is used to estimate the combined probability $p_0$ of obtaining from a background-only experiment 
a signal at least as large as the one observed, to measure the combined signal strength, and to check the compatibility of the results from the two datasets.

\begin{figure}[htb!p]
\begin{center}
\includegraphics[width=0.7\textwidth]{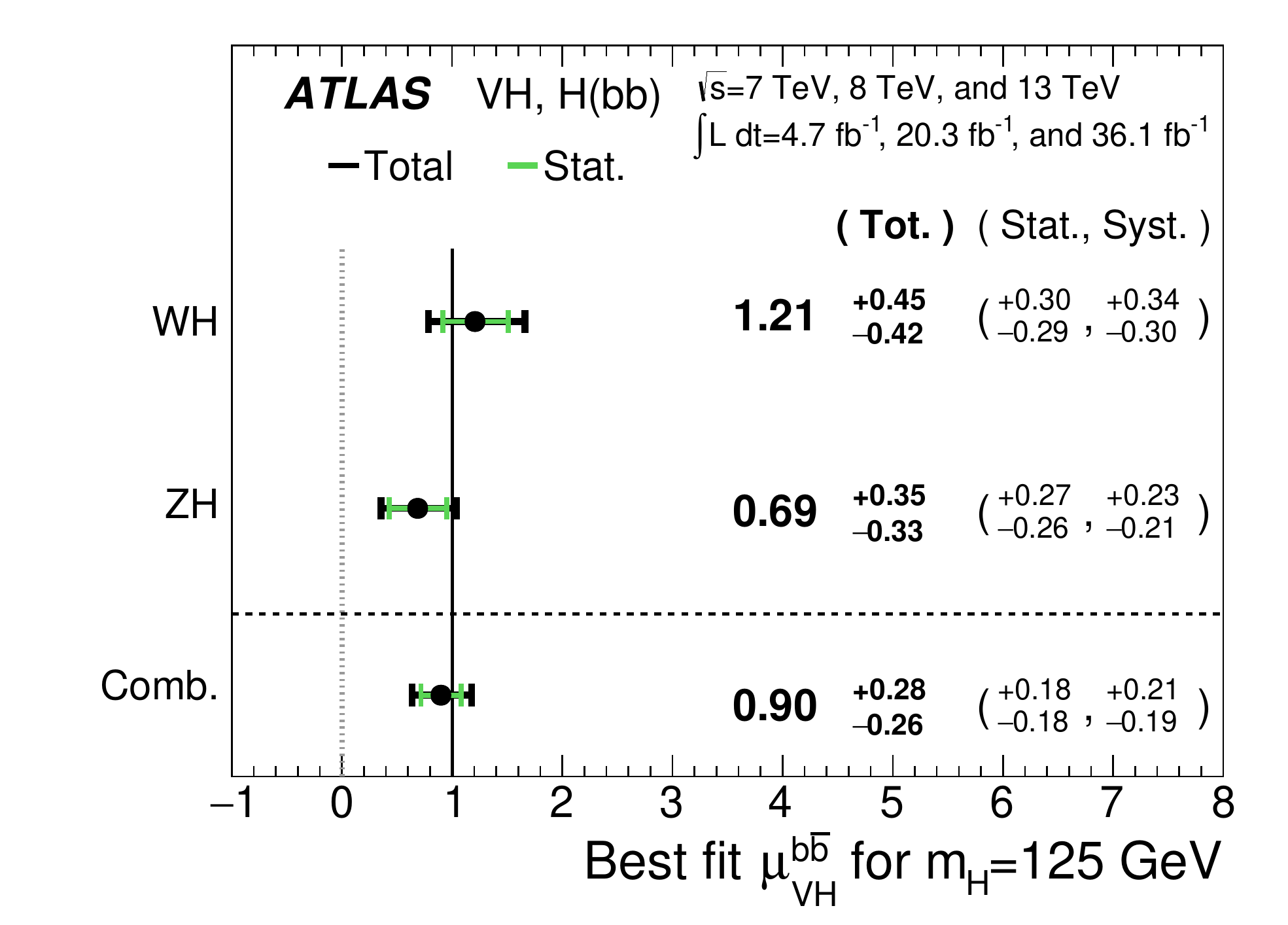}
\end{center}
\caption{The fitted values of the Higgs boson signal strength parameter $\mu$ for $\mh=125$~\GeV\ for the $WH$ and $ZH$ processes
and their combination with the 7~\TeV, 8~\TeV\ and 13~\TeV\ datasets combined. The individual $\mu$ values for the $(W/Z)H$ processes
are obtained from a simultaneous fit with the signal strength parameters for each of the $WH$ and $ZH$ processes floating independently. 
The compatibility of the individual signal strengths is 34\%. \label{fig:mu-higgs-c-comb}}
\end{figure}

\begin{figure}[htb!p]
\begin{center}
\includegraphics[width=0.7\textwidth]{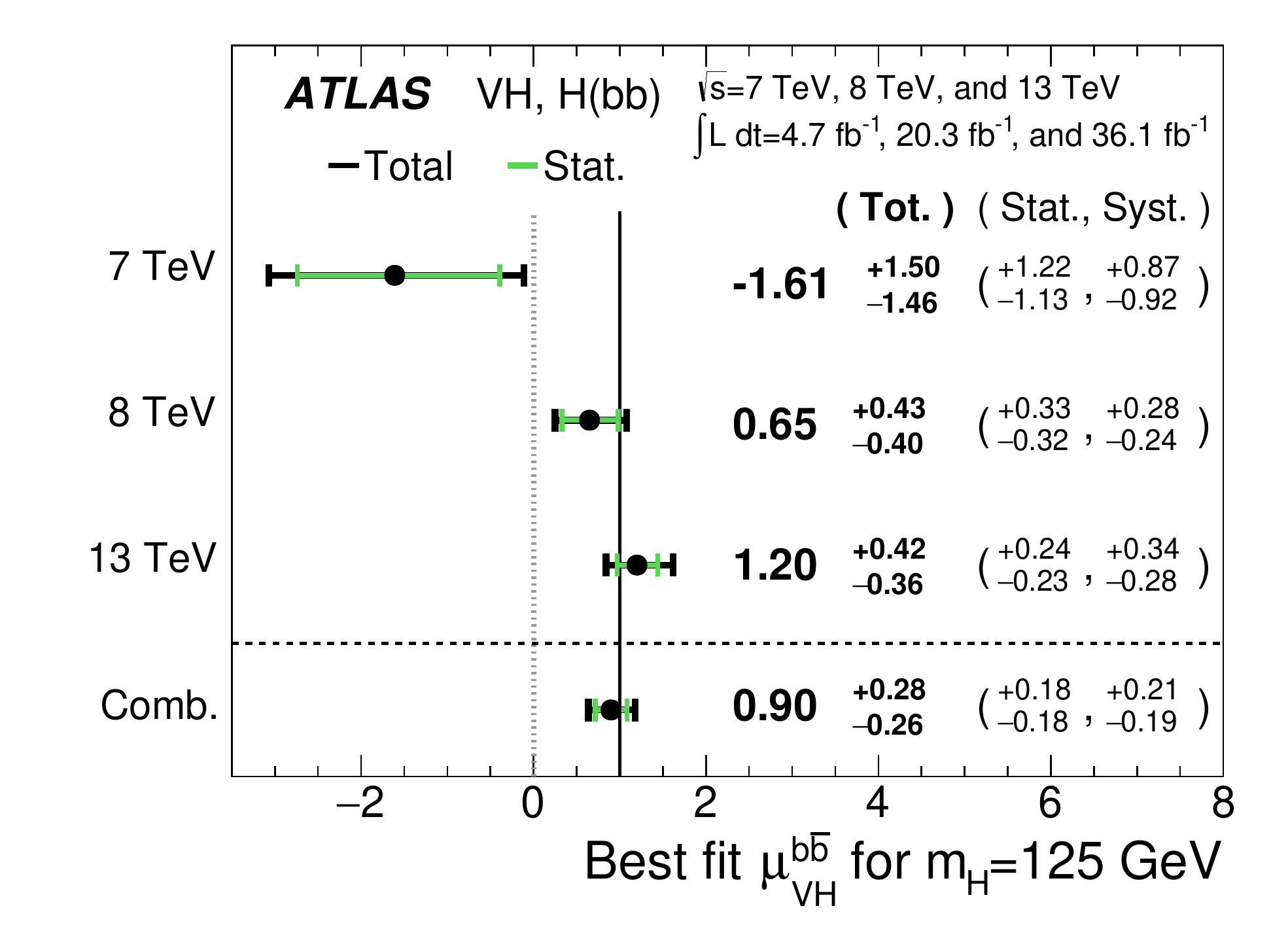}
\end{center}
\caption{The fitted values of the Higgs boson signal strength parameter $\mu$ for $\mh=125$~\GeV\ separately for the 7~\TeV, 8~\TeV\ and 13 \TeV\ datasets and their combination. 
\label{fig:mu-higgs-b-comb}}
\end{figure}

\begin{figure}[thb!]
\begin{center}
\includegraphics[width=0.7\textwidth]{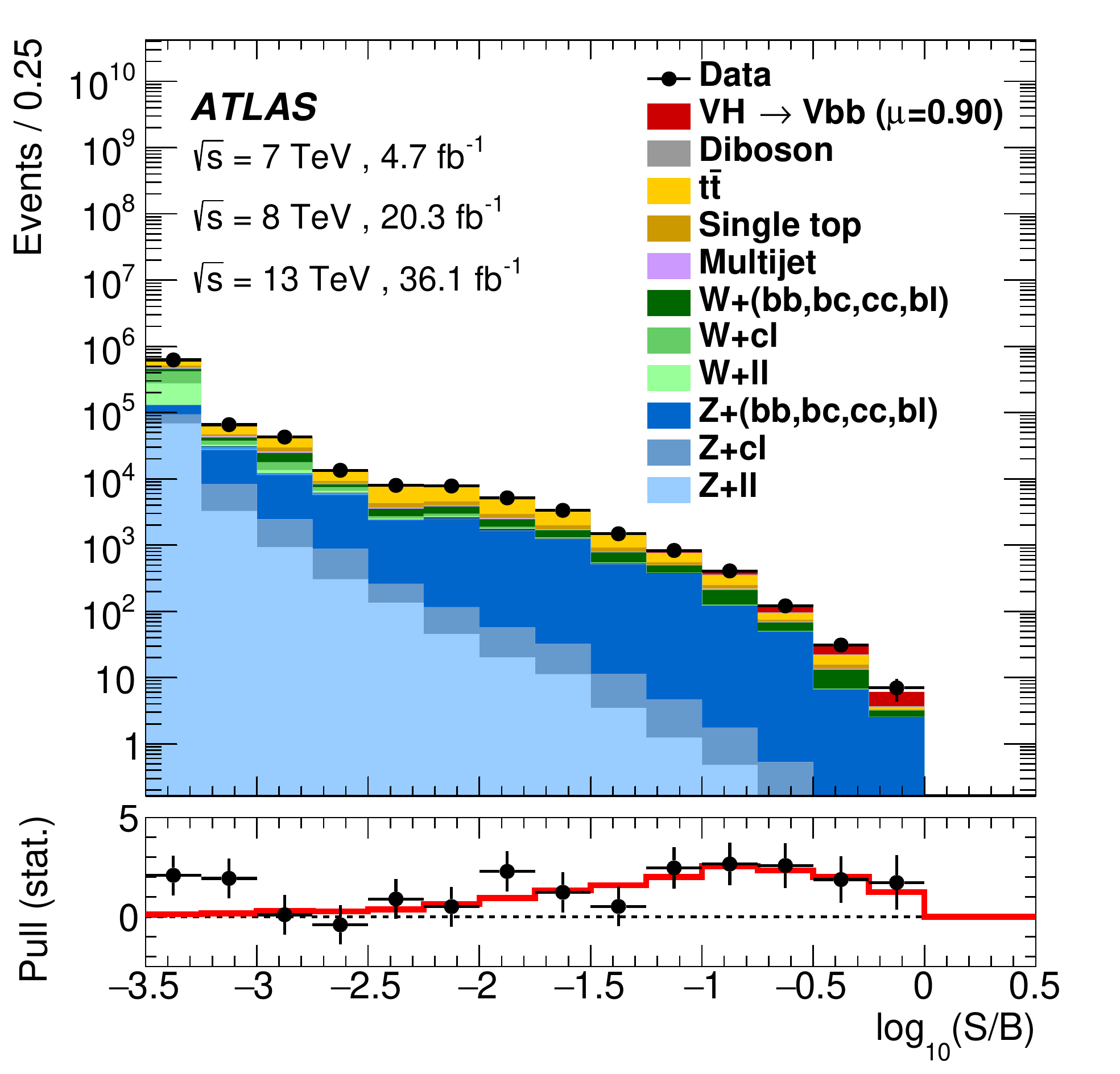}
\end{center}
\caption{Event yields as a function of $\log(S/B)$ for data, background
and Higgs boson signal with $\mh=125$~\GeV, for the 7~\TeV, 8~\TeV\ and 13~\TeV\ datasets combined. 
Final-discriminant bins in all  regions 
are combined into bins of $\log(S/B)$, with the fitted signal being $S$ and the fitted background $B$, for the 7~\TeV, 8~\TeV\ and 13~\TeV\ datasets combined.
The Higgs boson signal contribution is shown after rescaling the SM cross-section 
according to the value of signal strength extracted from data ($\mu=0.90$).
The pull (residual divided by its uncertainty) of the data with respect to the background-only prediction is also shown with statistical uncertainties only.
The full line indicates the pull of the prediction for signal ($\mu=0.90$) and background with respect to the background-only prediction.
}
\label{fig:logstob-comb}
\end{figure}

For the tested Higgs boson mass of 125 \GeV, the observed $p_0$ value 
is 0.018\%.  In the presence of a Higgs boson with that mass and the SM signal strength, the expected $p_0$ value is $3\times 10^{-5}$. The observation corresponds to an excess with a significance of 3.6 standard deviations,
to be compared to an expectation of 4.0 standard deviations. 
For all channels combined the fitted value of the signal strength parameter is 
\begin{eqnarray*}
\mu = 0.90 \pm 0.18 \mathrm{(stat.)} ^{+0.21}_{-0.19} \mathrm{(syst.)}. 
\end{eqnarray*}
The Run 1 and Run 2 analyses each contribute three measurements, corresponding to  the three lepton channels, yielding a total of six measurements. Their compatibility is estimated to be 7\%. Fits are also performed with the signal strength parameters floated independently for the $WH$ and $ZH$ production processes, and for Run 1 and Run 2. The compatibility of the signal strengths for the $WH$ and $ZH$ production processes is 34\%, and the results of this fit are shown in Figure~\ref{fig:mu-higgs-c-comb}.  
The compatibility of the signal strength parameters measured in Run 1 with those measured in Run 2 is 21\%. 
Figure ~\ref{fig:mu-higgs-b-comb} shows the signal strengths as measured  separately for the 7~\TeV, 8~\TeV\ and 13~\TeV\ datasets and their combination.
Figure~\ref{fig:logstob-comb} shows the data, background and signal yields, where final-discriminant bins in all  regions are combined into bins of $\log(S/B)$.
Here, $S$ and $B$ are the fitted signal and background yields, respectively. 


\FloatBarrier

\section{Conclusion}
\label{sec:conclusion}
Evidence for a Standard Model Higgs boson decaying into a $b\bar{b}$ pair and produced in association with a $W$ or $Z$ boson is presented, using data collected by the ATLAS experiment in proton--proton collisions from Run~2 of the Large Hadron Collider. This dataset corresponds to an integrated luminosity of 36.1~\fb collected at a centre-of-mass energy of $\sqrt{s} = $13~\TeV. 
An excess over the expected background is observed, with a significance of 3.5 standard deviations compared to an expectation of 3.0. The measured signal strength with respect to the SM prediction for $m_H = 125$~\GeV\ is found to be 
$\mu = 1.20 ^{+0.24}_{-0.23} \mathrm{(stat.)} ^{+0.34}_{-0.28} \mathrm{(syst.)}$. 

The analysis procedure adopted to extract the Higgs boson signal is also used to measure the yield of $(W/Z)Z$ production with $Z \rightarrow b\bar{b}$, where the ratio of the observed yield to that expected in the Standard Model is found to be $1.11  ^{+0.12}_{-0.11} \mathrm{(stat.)} ^{+ 0.22}_{-0.19} \mathrm{(syst.)}$.

The result of the search for the Standard Model Higgs boson based on Run 2 data is combined with previous results based on the full Run 1 dataset collected at centre-of-mass energies of 7~\TeV\ and 8~\TeV. An excess over the expected Standard Model background is observed, with a significance of 3.6 standard deviations compared to an expectation of 4.0. The measured signal strength with respect to the SM  expectation is found to be 
$\mu = 0.90 \pm 0.18 \mathrm{(stat.)} ^{+0.21}_{-0.19} \mathrm{(syst.)}$. 
Assuming the Standard Model production strength, the result is consistent with the value of the Yukawa coupling to bottom quarks in the Standard Model.


\section*{Acknowledgements}


We thank CERN for the very successful operation of the LHC, as well as the
support staff from our institutions without whom ATLAS could not be
operated efficiently.

We acknowledge the support of ANPCyT, Argentina; YerPhI, Armenia; ARC, Australia; BMWFW and FWF, Austria; ANAS, Azerbaijan; SSTC, Belarus; CNPq and FAPESP, Brazil; NSERC, NRC and CFI, Canada; CERN; CONICYT, Chile; CAS, MOST and NSFC, China; COLCIENCIAS, Colombia; MSMT CR, MPO CR and VSC CR, Czech Republic; DNRF and DNSRC, Denmark; IN2P3-CNRS, CEA-DSM/IRFU, France; SRNSF, Georgia; BMBF, HGF, and MPG, Germany; GSRT, Greece; RGC, Hong Kong SAR, China; ISF, I-CORE and Benoziyo Center, Israel; INFN, Italy; MEXT and JSPS, Japan; CNRST, Morocco; NWO, Netherlands; RCN, Norway; MNiSW and NCN, Poland; FCT, Portugal; MNE/IFA, Romania; MES of Russia and NRC KI, Russian Federation; JINR; MESTD, Serbia; MSSR, Slovakia; ARRS and MIZ\v{S}, Slovenia; DST/NRF, South Africa; MINECO, Spain; SRC and Wallenberg Foundation, Sweden; SERI, SNSF and Cantons of Bern and Geneva, Switzerland; MOST, Taiwan; TAEK, Turkey; STFC, United Kingdom; DOE and NSF, United States of America. In addition, individual groups and members have received support from BCKDF, the Canada Council, CANARIE, CRC, Compute Canada, FQRNT, and the Ontario Innovation Trust, Canada; EPLANET, ERC, ERDF, FP7, Horizon 2020 and Marie Sk{\l}odowska-Curie Actions, European Union; Investissements d'Avenir Labex and Idex, ANR, R{\'e}gion Auvergne and Fondation Partager le Savoir, France; DFG and AvH Foundation, Germany; Herakleitos, Thales and Aristeia programmes co-financed by EU-ESF and the Greek NSRF; BSF, GIF and Minerva, Israel; BRF, Norway; CERCA Programme Generalitat de Catalunya, Generalitat Valenciana, Spain; the Royal Society and Leverhulme Trust, United Kingdom.

The crucial computing support from all WLCG partners is acknowledged gratefully, in particular from CERN, the ATLAS Tier-1 facilities at TRIUMF (Canada), NDGF (Denmark, Norway, Sweden), CC-IN2P3 (France), KIT/GridKA (Germany), INFN-CNAF (Italy), NL-T1 (Netherlands), PIC (Spain), ASGC (Taiwan), RAL (UK) and BNL (USA), the Tier-2 facilities worldwide and large non-WLCG resource providers. Major contributors of computing resources are listed in Ref.~\cite{ATL-GEN-PUB-2016-002}.





\printbibliography

\newpage






\FloatBarrier
\newpage 

\begin{flushleft}
{\Large The ATLAS Collaboration}

\bigskip

M.~Aaboud$^\textrm{\scriptsize 137d}$,
G.~Aad$^\textrm{\scriptsize 88}$,
B.~Abbott$^\textrm{\scriptsize 115}$,
O.~Abdinov$^\textrm{\scriptsize 12}$$^{,*}$,
B.~Abeloos$^\textrm{\scriptsize 119}$,
S.H.~Abidi$^\textrm{\scriptsize 161}$,
O.S.~AbouZeid$^\textrm{\scriptsize 139}$,
N.L.~Abraham$^\textrm{\scriptsize 151}$,
H.~Abramowicz$^\textrm{\scriptsize 155}$,
H.~Abreu$^\textrm{\scriptsize 154}$
% ...
\bigskip
\\
% ...
$^{12}$ Institute of Physics, Azerbaijan Academy of Sciences, Baku, Azerbaijan\\
% ...
$^{88}$ CPPM, Aix-Marseille Universit{\'e} and CNRS/IN2P3, Marseille, France\\
% ...
$^{115}$ Homer L. Dodge Department of Physics and Astronomy, University of Oklahoma, Norman OK, United States of America\\
% ...
$^{119}$ LAL, Univ. Paris-Sud, CNRS/IN2P3, Universit{\'e} Paris-Saclay, Orsay, France\\
% ...
$^{137}$ $^{(a)}$ Facult{\'e} des Sciences Ain Chock, R{\'e}seau Universitaire de Physique des Hautes Energies - Universit{\'e} Hassan II, Casablanca; $^{(b)}$ Centre National de l'Energie des Sciences Techniques Nucleaires, Rabat; $^{(c)}$ Facult{\'e} des Sciences Semlalia, Universit{\'e} Cadi Ayyad, LPHEA-Marrakech; $^{(d)}$ Facult{\'e} des Sciences, Universit{\'e} Mohamed Premier and LPTPM, Oujda; $^{(e)}$ Facult{\'e} des sciences, Universit{\'e} Mohammed V, Rabat, Morocco\\
% ...
$^{139}$ Santa Cruz Institute for Particle Physics, University of California Santa Cruz, Santa Cruz CA, United States of America\\
% ...
$^{151}$ Department of Physics and Astronomy, University of Sussex, Brighton, United Kingdom\\
% ...
$^{154}$ Department of Physics, Technion: Israel Institute of Technology, Haifa, Israel\\
$^{155}$ Raymond and Beverly Sackler School of Physics and Astronomy, Tel Aviv University, Tel Aviv, Israel\\
% ...
$^{161}$ Department of Physics, University of Toronto, Toronto ON, Canada\\
% ...
$^{a}$ Also at Department of Physics, King's College London, London, United Kingdom\\
$^{b}$ Also at Institute of Physics, Azerbaijan Academy of Sciences, Baku, Azerbaijan\\
$^{c}$ Also at Novosibirsk State University, Novosibirsk, Russia\\
$^{d}$ Also at TRIUMF, Vancouver BC, Canada\\
$^{e}$ Also at Department of Physics {\&} Astronomy, University of Louisville, Louisville, KY, United States of America\\
% ...
$^{*}$ Deceased
\end{flushleft}


\end{document}